\documentclass[reprint, prx, aps, notitlepage, nofootinbib,longbibliography,floatfix, superscriptaddress]{revtex4-2}

\usepackage[utf8]{inputenc}
\usepackage{float}
\usepackage{graphicx}  
\usepackage{physics}
\usepackage{esint}
\usepackage{amssymb}  
\usepackage{mathtools}
\usepackage{amsmath}
\usepackage{amsthm}
\usepackage{mathrsfs}
\usepackage[colorlinks=true,linkcolor=blue,citecolor=blue,urlcolor=cyan,plainpages=false,pdfpagelabels]{hyperref}
\usepackage{color,xcolor,colortbl}
\usepackage{bm}
\usepackage[most]{tcolorbox}
\usepackage{enumitem}
\usepackage[normalem]{ulem}

\newcommand{\be}{\nopagebreak[3]\begin{equation}}
\newcommand{\ee}{\end{equation}}
\newcommand{\bfig}{\nopagebreak[3]\begin{figure}}
\newcommand{\efig}{\end{figure}}
\newcommand{\bea}{\nopagebreak[3]\begin{eqnarray}}
\newcommand{\eea}{\end{eqnarray}}
\newcommand{\e}{\text{e}}
\newcommand{\im}{\text{i}}

\numberwithin{footnote}{section}
\numberwithin{equation}{section}

\begin{document}

\title{Symplectic circuits, entanglement, and stimulated Hawking radiation in analog gravity}

\author{Anthony J. Brady}\email{ajbrady4123@arizona.edu}
\affiliation{Department of Electrical and Computer Engineering, University of Arizona, Tucson, Arizona 85721, USA}
\author{Ivan Agullo}
\email{agullo@lsu.edu}
\affiliation{Department of Physics and Astronomy, Louisiana State University, Baton Rouge, LA 70803, U.S.A.
}
\author{Dimitrios Kranas}
\email{dkrana1@lsu.edu}
\affiliation{Department of Physics and Astronomy, Louisiana State University, Baton Rouge, LA 70803, U.S.A.
}

\begin{abstract}
We introduce a convenient set of analytical tools (the Gaussian formalism) and diagrams (symplectic circuits) to analyze multi-mode scattering events in analog gravity, such as pair-creation a l\'a Hawking by  black hole and white hole analog event horizons. The diagrams prove to be valuable ansatzes for the scattering dynamics, especially in settings where direct analytic results are not straightforward and one must instead rely on numerical simulations. We use these tools to investigate entanglement generation in single- and multi-horizon scenarios, in particular when the Hawking process is stimulated with classical (e.g., thermal noise) and non-classical (e.g., single-mode squeezed vacuum) input states---demonstrating, for instance, that initial squeezing can enhance the production of entanglement and  overcome the deleterious effects that initial thermal fluctuations have on the output entanglement. To make further contact with practical matters, we examine how attenuation degrades quantum correlations between Hawking pairs. The techniques that we employ are generally applicable to analog gravity setups of (Gaussian) bosonic quantum systems, such as analog horizons produced in optical analogs and in Bose-Einstein condensates, and should be of great utility in these domains. We show the applicability of these techniques by putting them in action for an optical system containing a pair white-black hole analog, extending our previous analysis of [Phys. Rev. Lett. {\bf 128}, 091301 (2022)].
\end{abstract}

\date{\today}

\maketitle

\tableofcontents


\section{Introduction}\label{sec:intro}
In the 1970's, Stephen Hawking discovered that black holes are not completely black but, instead,  spontaneously emit thermal radiation as  black-bodies~\cite{Hawking74,Hawking75,Hawking:1976ra}---now famously referred to as Hawking radiation. This result originated from applying quantum field theory on the background space-time of a collapsing, massive body.  Since quantum mechanics of closed quantum-systems relies on unitary evolution, as is the case here, this presents an apparent contradiction: how can the final, outgoing quantum-state of the Hawking radiation be thermal---and therefore a mixed (as opposed to pure) quantum state---when the initial quantum state was pure (vacuum, in the case of spontaneous emission)? The contradiction is resolved\footnote{Though leads to a true puzzle (the information-loss paradox~\cite{Hawking:1976ra}) once the evaporation of the black hole is considered. We do not discuss this true puzzle here.} by observing one intriguing fact that follows from a more detailed analysis: there exists a counter-part gas of blackbody radiation falling into the black hole, which is entangled with the thermal outgoing Hawking radiation, thus purifying the full quantum state and avoiding the apparent contradiction~\cite{israel1976thermo}. The quantum entanglement between the outgoing Hawking radiation and the in-falling partner-radiation is a crucial feature of the process, highlighting the genuine `quantumness' of the effect.

The spontaneous generation of Hawking radiation (and the entangled partner-radiation) by an event horizon---a process which we shall generally refer to as the \textit{Hawking process} or \textit{Hawking effect}---was originally thought to be a mysterious property of black holes alone, but analog-gravity systems have changed that perspective entirely (e.g., in fluids~\cite{unruh81,Michel:2014zsa}, optical systems \cite{demircan11TRANSISTOR,rubino2012soliton,petev2013blackbody,finazzi13,Belgiorno:2014ana,linder16,Bermudez:2016hbl,Belgiorno:2017glw,jacquet20emission,Jacquet:2020jpj,rosenberg2020optical,Aguero-Santacruz:2020krw}, and Bose-Einstein condensates~\cite{cirac2000,Parentani09,macher2009,Finazzi:2010yq,Finazzi:2011jd,Finazzi:2012iu,Busch:2013gna,busch14,Michel:2016tog,nambu21}), revealing that the formation of an analog causal-barrier (event horizon) generically generates entangled Hawking quanta spontaneously from the vacuum with thermal properties~\cite{Visser2003,novello02ArtificialBHs,barcelo11,barcelo2019analogue,jacquet20}. This has stimulated a tremendous amount of research aimed at discovering physical systems which support such analog event-horizons, potentially leading to detectable signatures of Hawking quanta---and their entanglement---in the lab~\cite{philbin08, weinfurtner2011,euve2016, steinhauer2016,de19BEC,drori19,kolobov2021BEC}. 

Many detailed theoretical analyses have been done, primarily focusing on the microphysics of the Hawking process in a multitude of physically distinct platforms. From our perspective, it would be beneficial to have a set of universal tools and simple intuitive diagrams, which are applicable to various microphysical settings and which, more importantly, highlights the essential physics arising therein. Such may be especially relevant for the multi-mode scattering scenarios that are ever-present in generic dispersive media. Moreover, as experimental efforts strive to observe the Hawking effect and entanglement in the lab, some theoretical handle on practical considerations---such as, e.g., noise and system inefficiencies---would be useful when modeling such systems. Addressing these matters, in a comprehensive yet efficient way, is one goal of this paper.

A crucial aspect of the Hawking process---and one that we pay particular attention to---is the entanglement generated therefrom, whether that be from the spontaneous or stimulated Hawking effect. From our viewpoint, it is unclear from previous works what the stimulated Hawking process implies about the quantumness of the effect, since such is commonly regarded as having purely classical origin. We are thus motivated to clarify what is quantum and what is not in the stimulated Hawking process. We do so by evaluating entanglement in a precise and operationally-meaningful manner (via the logarithmic negativity~\cite{peres96, plenio05}), which allows us to describe in detail the harmful role that thermal noise and ineffiencies play, as well as show how some stimulating radiation---such as a single-mode squeezed vacuum---can actually enhance entanglement and even help to vanquish the detrimental effects of thermal noise.

When appropriate, we compare the logarithmic negativity to another indicator of entanglement prolifically utilized in the analog-gravity literature (such as a particular Cauchy-Schwarz inequality originally introduced in Refs.~\cite{Busch:2013gna,busch14}), and point out subtle but important distinctions between the two when making physically meaningful statements about the entanglement within the system. In particular, we discuss how the explanatory power of such Cauchy-Schwarz inequalities are limited and can lead to false conclusions about entanglement if they are pushed beyond their range of applicability. We highlight this point with a few simple examples relevant to practical realizations of the analog Hawking effect. Portions of this article contain a detailed and extended analysis of the ideas  presented succinctly in \cite{agullo2022prl}. 

This paper is organized as follows. In Section~\ref{sec:prelims}, we summarize the tools we use throughout this article to deal with Gaussian states of linear systems. The tone of this section is pedagogical. We defend the power and usefulness of these techniques, and the visual tools associated with them, to describe the physics behind the Hawking effect in analog systems. In Section \ref{sec:primer}, we show the utility of these tools by applying them to the Hawing effect in gravitationally-produced, spherically symmetric black holes. Section \ref{sec:recipe_lim} describes how to  build symplectic circuits to describe the Hawking effect. Section \ref{sec:recipe_lim} applies these tools to a second example: the Hawking effect in  optical systems containing a pair white-black hole, and uses them to analyze in some detail the production of entanglement in this multi-mode system. This section also includes a detailed comparison with numerical simulations, in order to evaluate the accuracy of the proposed symplectic circuit to describe the system.   Section \ref{sec:nonclassical_assist} discusses  ideas to amplify the generation of entanglement in the Hawking process, in particular   the use of quantum inputs such as single-mode squeezed states. We also evaluate the way ambient noise and losses affect the entanglement in the final state. Section \ref{sec:conclude} contains an overview of the main results in this article and discusses them from a broader perspective. The appendices contain material of a more technical nature helpful to support some of the claims made along this article. We work with units in which $c=\hbar=1$.


\section{Gaussian states and symplectic transformation for bosonic systems of finite modes}\label{sec:prelims}

\subsection{Symplectic transformations and Gaussian states}\label{sec:gauss_formalism}
\paragraph*{Symplectic transformations:}
Our focus is on scattering processes between a finite and well-defined set of ``in" modes to a well-defined set of ``out" modes, with an interaction region in between. We presume that the interaction region is described by a Hamiltonian that is quadratic in the basic variables, with coupling coefficients that may otherwise be complicated functions of space and time. Since the interaction is quadratic, the relations between the in-modes and out-modes is linear. We then wish to provide some sort of a scattering matrix from the in-modes to the out-modes (and vice versa by unitarity). Fortunately, there exists a general formalism---the Gaussian formalism---that succinctly describes the evolution of a finite number of bosonic modes under the influence of a quadratic Hamiltonian. Gaussian states include vacua, coherent, thermal, and squeezed states and, therefore, although the formalism is restricted to Gaussian states, it is sufficiently general to describe most of the  states one can easily create and manipulate in the laboratory.  To make this paper self-contained, we provide a general discussion and some details surrounding the properties of such transformations. For explicit details and extensive derivations, see Refs.~\cite{weedbrook2012,serafini17QCV}.

To describe linear transformations, we work in the effective phase-space of the modes. Consider a finite set of in-modes and out-modes, with annihilation operators $\hat{a}^{\rm(in)}_{J}$ and $\hat{a}^{\rm(out)}_{J}$ with $J\in\{1,2,\dots,N\}$, where $N$ is the number of relevant interacting modes. (An example of a mode is an electromagnetic wave with frequency $\omega$ and wavenumber $\vec k$.)  Define the canonical  operators (or  quadrature operators, as they are often called),
\begin{align} \label{QPA}
    \hat{Q}_J&\equiv\frac{1}{\sqrt{2}}\left(\hat{a}_J+\hat{a}_J^\dagger\right),\\ \hat{P}_J&\equiv\frac{i}{\sqrt{2}}\left(\hat{a}_J^\dagger-\hat{a}_J\right),
\end{align}
which we define for both the in- and out-modes and which, from the standard commutation relations between annihilation and creation operators, obey the canonical commutation relations $[\hat{Q}_J,\hat{P}_K]=\im\delta_{JK}$. Now, define the (column) vector of canonical operators as the direct sum of the canonical pairs for each mode, i.e.
\begin{equation}
    \hat{\bm{R}}\equiv(\hat{Q}_1, \hat{P}_1,\dots,\hat{Q}_N, \hat{P}_N)^\top,
\end{equation}
where the transpose is with respect to the implicitly introduced vector space, $\mathbb{R}^{2N}$, which is the phase-space of the modes. A point that we wish to emphasize in this section is that, in the Gaussian setting, quantum dynamics reduces to the dynamics of vectors and matrices in this $2N$-dimensional phase-space, without reference to the (infinite dimensional) Hilbert space or Schr\"odinger equations. This is a significant simplification. 

In terms of the vector $\hat{\bm{R}}$, the canonical commutation relations can be succinctly written as 
\begin{equation}
    [\hat{\bm{R}}^i,\hat{\bm{R}}^j]=\im\, \bm{\Omega}^{ij},\qq{}\bm\Omega\equiv
    \bigoplus_N \begin{pmatrix}
    0 & 1\\
    -1 & 0
    \end{pmatrix} ,
\end{equation}
where $i,j,\cdots \in\{1,\dots,2N\}$ are indices running from 1 to $2N$ (contrary to capitalized latin labels $I,J,\cdots$, which we choose to run from 1 to $N$),
and we have defined the $2N\times2N$ anti-symmetric matrix, $\bm\Omega$, which is the (inverse of the) symplectic form. 

Given any unitary (i.e.\ invertible) transformation generated by a quadratic Hamiltonian, which takes in-modes to out-modes, one can  describe such a transformation by a $2N\times2N$ symplectic matrix, $\bm S$, such that,
\begin{equation}
    \hat{\bm R}^{\rm(out)}=\bm{S}\cdot \hat{\bm R}^{\rm(in)}.\label{eq:in_out_general}
\end{equation}
One can think about this transformation as the familiar Heisenberg evolution of the canonical operators---which is linear because we are assuming the Hamiltonian is quadratic.
The matrix $\bm{S}$ is symplectic in the sense that it preserves the symplectic form, i.e. $\bm{S}\bm\Omega\bm{S^\top}=\bm{\Omega}$, which is the only condition that $\bm S$ must satisfy. The set of such matrices $\bm{S}$ forms the symplectic group $Sp(\mathbb{R},2N)$; they are made of the subset of canonical transformations that are also linear. The matrix $\bm{S}$  encodes the relevant dynamical processes within the interaction region and it thus serves as a scattering matrix. Appendix \ref{sqzbs} describes two examples of symplectic transformation we use extensively in this paper: two-mode squeezing and beam-splitters.

\paragraph*{Gaussian states:} We restrict our analyses to a class of quantum states known as Gaussian quantum states (or just Gaussian states). Physically, pure Gaussian states are states which one can generate with symplectic transformations applied to the ground state of some quadratic Hamiltonian. More generally, acting with symplectic transformations on a thermal state of some quadratic Hamiltonian, one can build all mixed Gaussian states. A Gaussian state has the property that it is completely and uniquely described in terms of its first and second moments (defined below) of the basic variables (e.g.\ the $Q$'s and $P$'s); higher-order moments can be derived from the second moments, in exactly the same way that the statistical moments of a Gaussian probability distribution are all determined from the first and second moments (hence the name Gaussian). 

We define the mean vector (the vector of first moments) and covariance matrix (the matrix of second moments) for a generic quantum state as, 
\begin{align}
    \bm{\mu}^i&\equiv\ev*{\hat{\bm{R}}^i},\\
    \bm{\sigma}^{ij}&\equiv\ev{\left\{\hat{\bm{R}}^i-\bm\mu^i,\hat{\bm{R}}^j-\bm\mu^j\right\}},\label{eq:cov_general}
\end{align}
where the expectation value $\ev{\cdot}$ is taken with respect to the quantum state under consideration (either pure or mixed), and $\{\cdot,\cdot\}$ denotes the symmetric, anti-commutator. [One subtracts $ \mu_i$ to avoid having redundant information in the first and second moments. The focus on the symmetric part of the second moments is because the anti-symmetric part is fully determined by the canonical commutation relations and is thus state independent. Therefore, the pair  $(\bm{\mu},\bm{\sigma}^{ij})$ is the minimum information needed to fully characterize a Gaussian state.]  Note that these definitions apply to any quantum state whatsoever. It is only for Gaussian quantum states that such quantities are sufficient in describing the quantum state in its entirety. 

To give a few examples, the vacuum of a set of $N$ oscillators is characterized by $\bm{\mu}=\bm{0},\,   \bm{\sigma}=  \bm{\bm I}_{2N}$ ($\bm{I}_{2N}$ is the $2N\times 2N$ identity matrix); a coherent state by $\bm{\mu}\neq \bm{0},\,   \bm{\sigma}=  {\bm{I}}_{2N}$ (same covariance matrix as vacuum, but different first moments; for this reason coherent states are called displaced vacua); a squeezed state in general has $\bm{\sigma}\neq  {\bm{I}}_{2N}$, and a thermal state  $\bm{\mu}=\bm{0},\,   \bm{\sigma}=  \bigoplus_I (1+2\, n_I)\,  {\bm{I}}_{2}$, where $n_I$ is the mean number of thermal quanta in the mode $I=1,\cdots, N$. Thermal states are mixed states.

Many invariant properties of a Gaussian state can be extracted directly from $\bm{\sigma}$. For instance, Heisenberg uncertainty relations are completely and elegantly  captured in the relation $\bm{\sigma}+\im\bm{\Omega}\geq0$;\footnote{More explicitly, $(\bm{\sigma}^{ij}+\im\bm{\Omega}^{ij}) \bar{{\bm v}}_i{\bm v}_j$ is a non-negative real number for all vectors $\bm v\in \mathbb{C}^{2N}$ (the bar denotes complex conjugation).}  a Gaussian state is pure, if and only if the eigenvalues of the matrix $\bm{\sigma}^{ik}\bm\Omega_{kj}$ are all equal to $\pm \im$ \cite{weedbrook2012,serafini17QCV}; etc.

From Eq.~\eqref{eq:in_out_general}, given a generic set of in-moments $(\bm{\mu}^{\rm(in)},\bm{\sigma}^{\rm(in)})$ and a symplectic transformation $\bm S$, the out-moments are obtained by simple multiplication with  $\bm S$, as follows
\begin{align} 
    \bm{\mu}^{\rm(out)}&=\bm{S}\bm{\mu}^{\rm(in)},\label{eq:out_mu_general}\\
    \bm{\sigma}^{\rm(out)}&=\bm{S}\bm{\sigma}^{\rm(in)}\bm{S}^\top.\label{eq:out_sigma_general}
\end{align}
This is a direct consequence of the linearity of evolution. Let us assume the initial state is a Gaussian quantum state with in-moments $(\bm{\mu}^{\rm(in)},\bm{\sigma}^{\rm(in)})$. These quantities completely determine the initial quantum state. Then, since symplectic transformations map Gaussian states to Gaussian states and since the first and second moments completely determine all properties of a Gaussian quantum state, Eqs.~\eqref{eq:out_mu_general} and~\eqref{eq:out_sigma_general} completely determine all properties of the out quantum state in terms of the in quantum state. Hence, the full quantum dynamics of the system of $N$ bosonic modes can be reduced to the dynamics of vectors and matrices in the $2N$-dimensional phase-space, $\mathbb{R}^{2N}$. Since we are evolving expectation values, there is no need to differentiate between Schr\"odinger or Heisenberg pictures here.

Finally, let us state a simple and useful formula for the mean number of quanta for any Gaussian state (pure or mixed) of $N$ modes with mean $\bm\mu$ and covariance matrix $\bm\sigma$,
\begin{equation}
    \ev{\hat{n}}=\frac{1}{4}\text{Tr}\{\bm\sigma\}+\frac{1}{2}\bm{\mu}^\top\bm{\mu} -\frac{1}{2}N.\label{eq:mean_quanta}
\end{equation}
In particular, the mean $\bm\mu$ and covariance matrix $\bm\sigma$ could describe the reduced moments of an $N$-dimensional subsystem of an  $N+M$ Gaussian quantum state. (See  Appendix \ref{sqzbs} for additional details.)


\subsection{Quantum entanglement}\label{sec:logneg}
An essential feature of the Hawking process is the quantum correlations (entanglement) generated therefrom. For pure quantum states, the most popular entanglement-quantifier is the von Neumann entropy. However, for mixed quantum states (as generic sub-systems of a pure quantum state), the von Neumann entropy for a bi-partition no longer captures quantum correlations alone. There exists several measures, witnesses, and quantifiers of entanglement, with increasing levels of technicality and computational difficulty. We consider a few such quantifiers and witnesses of entanglement here. See Appendix~\ref{sqzbs} for some example applications. 

\paragraph{Logarithmic Negativity}
One easily-computable measure of entanglement for pure states and mixed states alike is the logarithmic negativity (LN) \cite{peres96, plenio05}. A non-zero value of the LN is in one-to-one correspondence with the violation of the Positivity of Partial Transpose (PPT) criterion for quantum states \cite{plenio05}---a criterion that separable quantum states faithfully obey. In simpler words, a non-zero value of LN implies the existence of entanglement (or non-separability), however, the reverse is not generically true---some entangled states have zero LN. On the other hand, when restricting to Gaussian states and when one of the subsystems is made of a single mode ($N_A=1$), regardless of the size of the other subsystem, the LN is different from zero {\em if and only if} the state is entangled. Importantly, the LN is an entanglement monotone; it is a faithful quantifier  of entanglement, in the sense that higher LN means more entanglement. Later in this paper, we will use the LN to quantify entanglement precisely.

For Gaussian quantum states (and other types of quantum states), the value of the LN has an operational meaning as the exact cost (where the currency is Bell pairs or entangled bits, ebits) that is required to prepare or simulate the quantum state under consideration \cite{wilde2020ent_cost, wilde2020alpha_ln}. Hence, if a quantum state $\rho$ has a larger LN value than a quantum state $\tau$, then one can quantitatively and confidently claim that $\rho$ possesses more entanglement than $\tau$, as it is more expensive to simulate $\rho$ than it is to simulate $\tau$. This is a useful fact. For consider that $\rho$ and $\tau$ differ by a continuous change in a parameter $\vartheta$, such that $\rho(\vartheta)$ and $\tau=\rho(\vartheta+\delta\vartheta)$. Then, one can make definitive statements about the role that $\vartheta$ plays in entanglement by monitoring the changes  it induces in the LN. For instance, if the LN monotonically increases with the parameter $\vartheta$, then this parameter quantifies a valuable resource (such as squeezing). On the other hand, if the LN monotonically decreases with $\vartheta$, then this parameter is obviously a nuisance (such as loss and thermal noise) and should be minimized as much as possible.

The LN for a Gaussian quantum state can be directly computed from its covariance matrix (i.e., the LN does not dependent on the first moments $\bm{\mu}$). Consider an $N+M$-mode Gaussian quantum state $\rho_{AB}$ with covariance matrix $\bm\sigma_{AB}$, where the sub-system $A$ consists of $N$ modes and $B$ consists of $M$ modes. The LN for the bi-partition can be computed from the symplectic eigenvalues\footnote{The symplectic eigenvalue of a covariance matrix  $\bm\sigma$, are defined as the absolute value of the eigenvalues of the matrix $\bm\sigma^{ik}\bm\Omega_{kj}$. The properties of $\bm\sigma$ guarantee that the $2N$ eigenvalues of this matrix are purely imaginary and they come in pairs, $\pm \im \nu_I$, with $I=1,\cdots, N$ and $\nu_I$ real for all $I$. Hence, by taking the absolute values, we are left with $N$ distinct symplectic eigenvalues equal to $\nu_I$.}
of the covariance matrix $\widetilde{\bm\sigma}_{AB}$ of the partially transposed quantum state $\rho_{AB}^{\top_B}$, which can be written in terms of $\bm\sigma_{AB}$ as
\begin{equation} \label{PT}
    \widetilde{\bm\sigma}_{AB} = \bm T\bm\sigma_{AB}\bm T,
\end{equation}
where $\bm T= \bm{I}_{2N}\oplus\bm\Sigma_{M}$ and $\bm\Sigma_M=\oplus_M \sigma_z$ is a direct sum of $M$ $2\times2$ Pauli-z matrices; see, for instance, Ch. 7.1 of Ref.~\cite{serafini17QCV} for details, and Appendix~\ref{sqzbs} for concrete calculations in a simple situation. Let $\{\Tilde{\nu}_j\}_{J=1}^{M+N}$ denote the symplectic eigenvalues of $\widetilde{\bm\sigma}_{AB}$. The LN for the Gaussian quantum state $\rho_{AB}$ is then given as~\cite{weedbrook2012,serafini17QCV},
\begin{equation}
    LN(\rho_{AB})=\sum_{J=1}^{M+N}\max\left[0, -\log_2(\tilde{\nu}_J)\right].
\end{equation}
Observe that a sufficient condition for quantum entanglement is $\min\{\tilde{\nu}_J\}<1$. For instance, we use this condition later on to evaluate precise, yet physically-motivated entanglement-criteria based on relationships between the Hawking temperature of an analog-gravity system and the temperature of initial thermal fluctuations.

\paragraph{A Cauchy-Schwarz inequality:}
In the analog gravity literature, there has been much focus on a particular Cauchy-Schwarz inequality (to be defined precisely below) to evaluate entanglement, which was first introduced in Refs.~\cite{Busch:2013gna,busch14} in the context of BECs; see, e.g., Refs.~\cite{steinhauer2015PRD,de2015iop,steinhauer2016} for further usage. We describe such below.

Consider two modes, labelled $\hat{a}$ and $\hat{b}$, occupying the two-mode quantum state $\rho_{AB}$. Define the mean occupation numbers $n_a=\ev*{\hat{a}^\dagger\hat{a}}$ and $n_b=\ev*{\hat{b}^\dagger\hat{b}}$ and the parameter $c_{ab}\equiv\ev*{\hat{a}\hat{b}}$, where the expectation is taken with respect to the state $\rho_{AB}$.  It was shown in Refs.~\cite{Busch:2013gna,busch14} that the following Cauchy-Schwarz (CS) inequality,
\begin{equation}
    \Delta\equiv n_an_b - \abs{c_{ab}}^2 <0,\label{eq:cs_inequality}
\end{equation}
is a sufficient condition for the two modes, $\hat{a}$ and $\hat{b}$, to be quantum mechanically entangled in the state under consideration. 

The inequality for $\Delta$ above is simple and can be evaluated by measuring only a handful of observables, leading to an approachable method for witnessing entanglement in the lab. However, it is important to be aware of its limitations. The witness $\Delta$ does not quantify entanglement. In other words, a larger violation of the inequality (a more negative $\Delta$) does not imply more entanglement is present within the system. Moreover, the inequality is only sufficient to witness entanglement---not  sufficient \textit{and} necessary; this implies that for some entangled states we may find $\Delta >0$. In Ref.~\cite{busch14}, the inequality was shown to be necessary and sufficient for a certain class of two-mode quantum states (so-called `stationary states'; cf. Refs.~\cite{Busch:2013gna,busch14}), though this does not generally hold. Hence, even when restricting to Gaussian states and to situations when one subsystem is made of a single mode (conditions under which the LN is a necessary and sufficient condition for entanglement), a non-negative value of $\Delta$ does not guarantee the absence of entanglement (see Appendix \ref{sqzbs} for an explicit example); $\Delta$ is faithful only when the additional condition of `stationarity', defined in \cite{Busch:2013gna,busch14}, is met.

As pointed out in Appendix A of Ref.~\cite{busch14}, the inequality $\Delta<0$ is actually subsumed by a more generic inequality based on the PPT criterion~\cite{simon2000criterion}. Consider the following covariance matrix for a two-mode Gaussian state, $\rho_{AB}$,
\begin{equation}
\bm\sigma_{AB}=\begin{pmatrix}
\bm A & \bm C \\
\bm C^\top & \bm B
\end{pmatrix},
\end{equation}
where $\bm A$ ($\bm B$) is the covariance matrix for the sub-system $A$ ($B$), which we assume to be made of a single mode,  and $\bm C$ describes the correlations among them. Now define the PPT parameter,\footnote{Our definition here differs from Refs.~\cite{simon2000criterion,busch14} by a numerical factor of 1/16 due to our definition of the covariance matrix, Eq.~\eqref{eq:cov_general}.}
\begin{multline}
    \mathcal{P}_-\equiv\det\bm A\det\bm B + \left(1-\abs{\det\bm C}\right)^2\\ -\Tr{\bm A\bm\Omega_1\bm C\bm\Omega_1\bm B\bm\Omega_1\bm C^\top\bm\Omega_1}-\det\bm A - \det\bm B,\label{eq:pminus}
\end{multline}
where $\bm\Omega_1=\big(\begin{smallmatrix}
  0 & 1\\
  -1 & 0
\end{smallmatrix}\big)$ is the single-mode symplectic form. Actually, $\Delta$ happens to be equal  to the first line of the last equation, namely
\be\label{Delta}  \Delta=\det\bm A\det\bm B + \left(1-\abs{\det\bm C}\right)^2 \, .\ee
A necessary and sufficient condition for a Gaussian two-mode quantum state to be entangled is $\mathcal{P}_-<0$, which is derived from the PPT criterion as applied to Gaussian states; see Ref.~\cite{simon2000criterion} for explicit details. Note also that $\mathcal{P}_-$ is  a binary indicator of entanglement, not a quantifier. The drawback to the more general quantity $\mathcal{P}_-$, as opposed to $\Delta$, is that computing the former requires access to the entire covariance matrix of the quantum state. For Gaussian states, this is equivalent to performing quantum-state tomography on the two-mode system (see, e.g., Ref.~\cite{raymer2009} for details about quantum-state tomography of optical systems). However, as shown in Ref.~\cite{busch14}, for a certain class of Gaussian quantum states (`stationary states'), ${\mathcal{P}_-<0\iff \Delta<0}$. Hence, for this sub-family of states, the simpler quantity $\Delta$ is as useful as $\mathcal{P}_-$. [Again, we insist that neither is a quantifier of entanglement; see Appendix \ref{sqzbs}.]

\paragraph{Remarks:}
The LN precisely quantifies the amount of entanglement within a system under the circumstances described above (Gaussian states and when one of the subsystems is made of a single mode). However, computing it in practice requires knowledge of the entire covariance matrix, which amounts to knowing the quantum state of said system (i.e.\ via quantum-state tomography~\cite{raymer2009}). Entanglement witnesses, such as $\Delta$, on the other hand, can relieve the burden of full quantum-state tomography, as such are typically computed by measuring only a few observables---often appearing in the form of Cauchy-Schwarz inequalities. The trade-off, however, is that entanglement witnesses have a binary outcome, telling us only if a state is entangled or not (they do not quantify entanglement), and they are not necessarily always faithful. 

In this paper, we utilize the LN to genuinely quantify entanglement and analyze how certain parameters (e.g., thermal noise, inefficiencies) affect quantum entanglement within a system of modes. We briefly consider entanglement witnesses discussed in previous paragraphs, due to their practical appeal, and point out discrepancies between such and the LN.  


\subsection{Simple yet practical noise models}\label{sec:noise_model}

Since there is currently much experimental effort to observe Hawking radiation in the lab with analog systems, it seems appropriate to include potential imperfections in a hypothetical experimental setup which may distort or degrade certain features of the Hawking effect. In this paper, we focus on two potential sources of noise:
\begin{itemize}
    \item Background temperature (i.e.\ initial thermal fluctuations), $T_{\rm env}$
    \item Attenuation or losses, $\eta$
\end{itemize}
where $T_{\rm env}$ is the temperature of the `environment' and $0\leq 1-\eta\leq1$ can represent the probability that a mode scatters into some set of inaccessible modes (attenuation); or $\eta$ can be the efficiency to which one can couple outgoing modes to a detector setup.

The former is relevant since, generally, the Hawking temperature of analog-gravity systems is meager, and thus, there will be a competition between the quantum fluctuations produced in the Hawking effect versus the initial thermal fluctuations (more on this in Section~\ref{subsec:correlations_bh}). 

We include the effects of a background temperature within our model by simply taking the initial state be in thermal equilibrium at temperature $T_{\rm env}$. In the Gaussian context, this means that the initial first and second moments of each mode, prior to any interactions or operations, are $\bm\mu_{\Theta}=\bm0$ and $\bm\sigma_{\Theta}=(1+2n_{{\rm env}})\bm{I}_2$ (and different modes are uncorrelated) where $n_{{\rm env}}$ is the mean occupation-number of a harmonic oscillator (e.g.\ a mode of a photon gas) at temperature $T_{\rm env}$. Generally, each mode can be of different frequency, as is the case for analog models with Bose-Einstein condensates~\cite{bruschi2013}, and for them 
$n_{{\rm env}}\equiv 1/(\e^{\omega/T_{\rm env}}-1)$ with $\omega$ the frequency of the mode. It is straightforward to generalize to noise which may have different spectral profiles than simple black body-like occupation numbers.

A second source of noise, which plagues all experimental platforms, is unwanted scattering of the system modes into a set of inaccessible modes (i.e., attenuation), which could arise from, e.g., inefficiencies within the setup. This leads to a loss of quanta, which in turn reduces quantum correlations between the outgoing-modes of the system and, as we show below, it may cause serious issues if the goal is to measure entanglement generated during the Hawking process. We provide a simple model for such in the form of a homogeneous, thermal-loss channel (also known as an attenuation map), $\mathcal{L}_\eta^{N}$, which preserves the Gaussianity of the state, and maps the mean $\bm\mu$ and covariance matrix $\bm\sigma$ of an $N$-mode system to
\begin{align} \label{noisemodel}
    \bm\mu\overset{\mathcal{L}_\eta^{N}}{\longrightarrow}\,&\sqrt{\eta}\, \bm\mu\\
    \bm\sigma\overset{\mathcal{L}_\eta^{N}}{\longrightarrow}\,&\eta\, \bm\sigma+N_{\rm noise} \, (1-\eta)\bm{I}_{2N},
\end{align}
where $1-\eta$ is the attenuation parameter quantifying the inefficiencies (the probability to scatter into an inaccessible channel; $\eta=1$ is no attenuation). The presence of the parameter   $N_{\rm noise} \geq 1$ allows for the possibility that the the inaccessible scattering channel is not merely a pure  losses channel, but it also injects noise to the system. This model assumes that the inaccessible modes, to which the system modes scatter into, act homogeneously on the system and are uncorrelated. These assumptions allow us to discuss generic features of unwanted scattering processes in simple terms and to evaluate the impact of losses on quantum aspects of the Hawking process. As an aside, our analyses can be extended by introducing a more elaborate model for the open-system dynamics, perhaps leveraging the input-output formalism for open systems of bosonic modes (see for instance Ch.\ 6 of Ref.~\cite{serafini17QCV}).


\section{Example 1: Black hole}\label{sec:primer}

In this section, we apply the formalism presented in Section~\ref{sec:prelims} to the (usual) Hawking process triggered by the formation of a black hole by the gravitational collapse of a spherically symmetric body, and introduce a simple symplectic circuit associated with this process. The motivation is to illustrate the use of these tools in a simple context. 

\begin{figure}
    \centering
    \includegraphics[width=\linewidth]{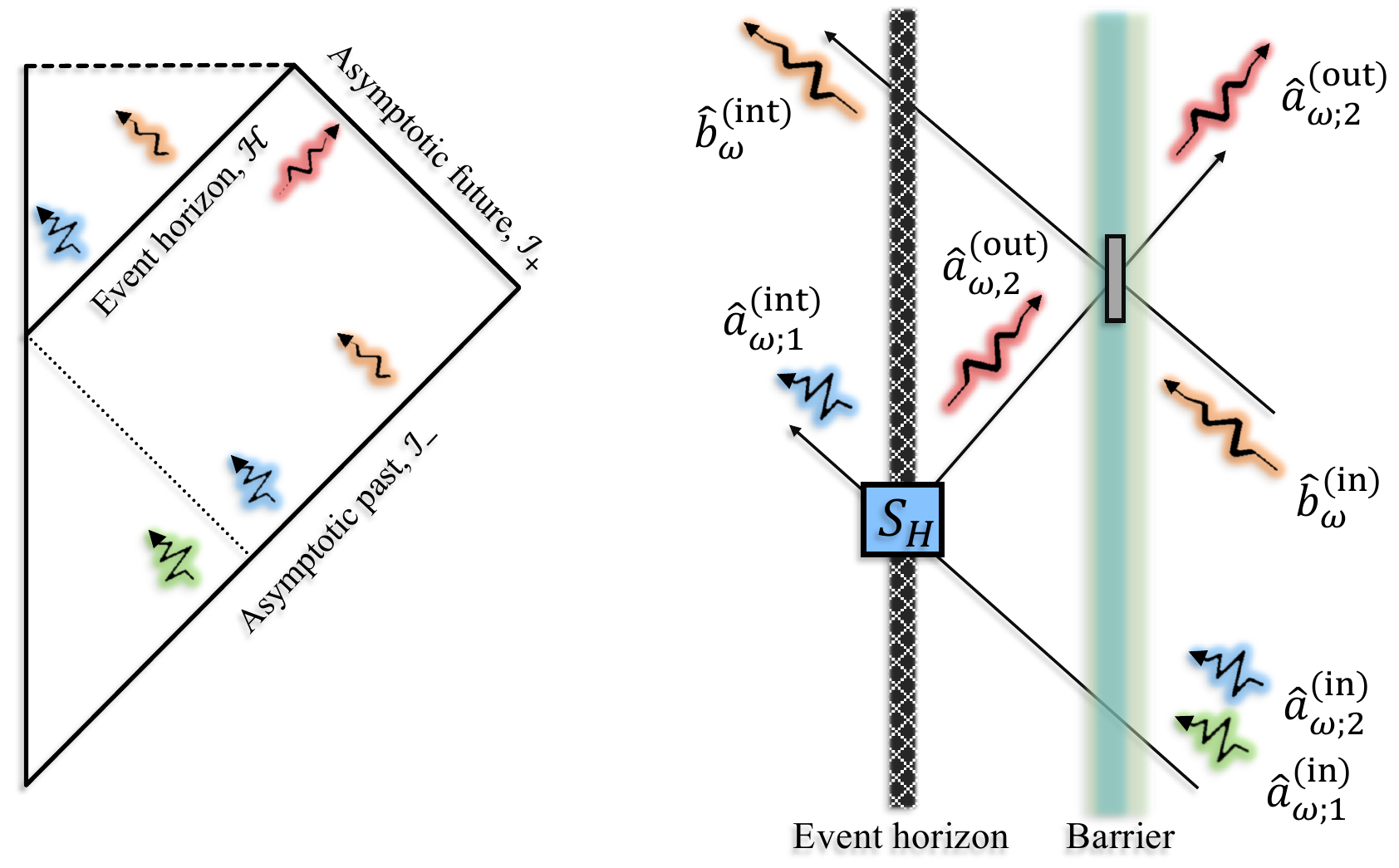}
    \caption{Conventional illustrations of the Hawking process. (Left) Space-time diagram of a black hole formed by gravitational collapse depicting ingoing modes, interior modes, and outgoing Hawking radiation. (Right) A close-up of the different scattering events. The origin-story of the long-wavelength Hawking radiation in the asymptotic future typically follows by tracing the evolution of the outgoing Hawking radiation `back in time' along geometric rays to the asymptotic past. Doing so, one finds that the Hawking radiation has contributions from particle-pair creation at the event horizon (Hawking effect) as well as contributions from a classical scattering process (back-scattering) at a gravitational potential barrier. In absence of back-scattering, the outgoing Hawking radiation has a blackbody spectrum due to the Hawking effect. However back-scattering modulates the blackbody spectrum by a grey-body factor, which is just the probability to transmit through the barrier.}
    \label{fig:spacetime_diagram}
\end{figure}

\subsection{Physical picture, symplectic-circuit, and example calculations}

\paragraph*{Physical picture:} Imagine a scattering experiment consisting of a well-defined and finite set of in-modes and out-modes, with an interaction region in between that is well-described by a quadratic interaction Hamiltonian,  with coupling coefficients that are dependent on space and time. The interaction then leads to a linear transformation between the in- and out-modes, as in Section~\ref{sec:prelims}. 

One usually derives the astrophysical Hawking effect in precisely this context, in which case there is a free (bosonic) quantum field, described by a quadratic Hamiltonian, evolving on the spacetime-dependent (classical) metric of a collapsing body.  The relevant in-modes are the modes which define the standard vacuum in the asymptotic in-region, at past infinity (past null infinity for massless fields, which are the type of fields that dominate the Hawking effect \cite{Page:1976df}), while the out-modes consist of the Hawking radiation emitted by the black hole, which make it to the asymptotic out-region at future infinity,  asymptotically far from the black hole, together with the `Hawking-partner' modes, as well as back-scattered modes, which fall through the event horizon and into the black hole. The conventional story is illustrated in Fig.~\ref{fig:spacetime_diagram}.

\begin{figure}
    \centering
    \includegraphics[width=\linewidth]{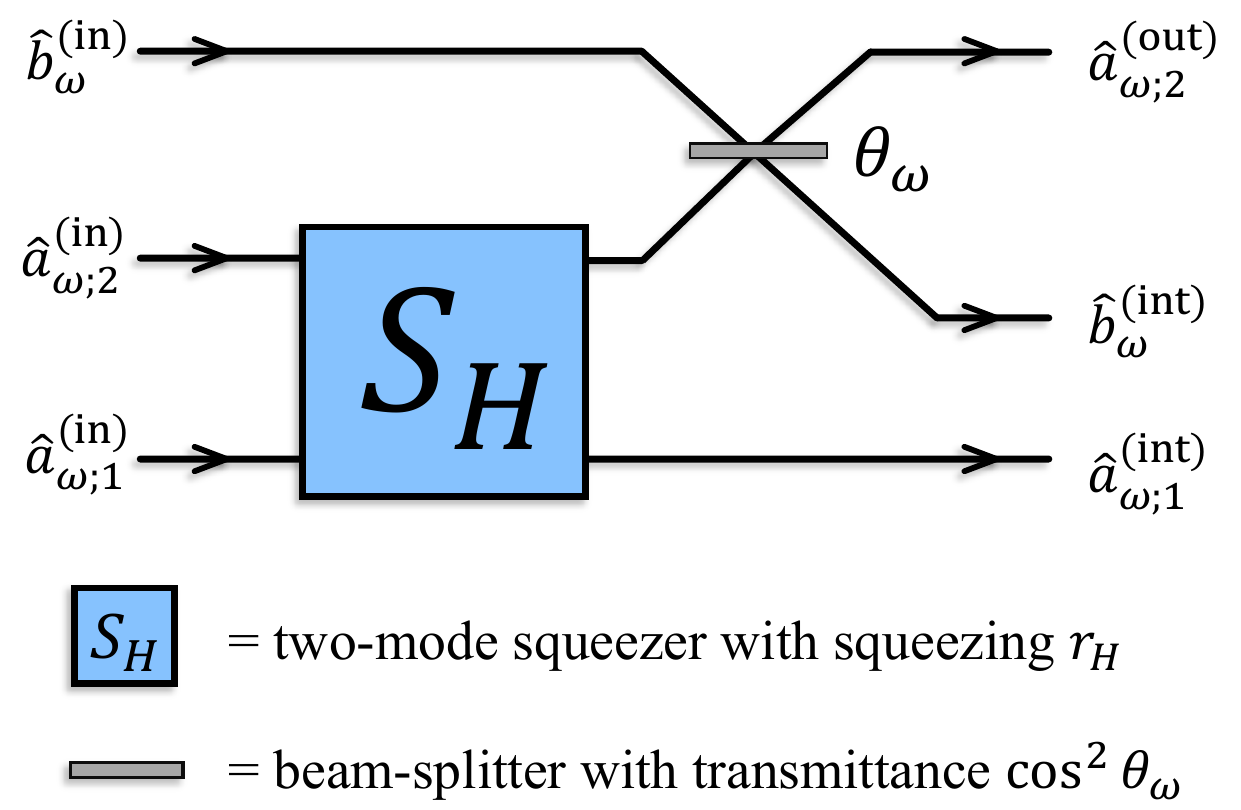}
    \caption{Symplectic-circuit of the Hawking process. Three in-modes ---$\hat a_{\omega;1}^{\rm(in)}$, $\hat a_{\omega;2}^{\rm(in)}$, and a back-scattering mode $\hat b_{\omega}^{\rm(in)}$---scatter to three out-modes. The outgoing Hawking radiation occupies the mode $\hat a_{\omega;2}^{\rm(in)}$. The other two modes, $\hat a_{\omega}^{\rm(int)}$ and $\hat b_{\omega}^{\rm(int)}$, propagate into the interior of the black hole. The former is the Hawking-partner mode while the latter, the back-scattered mode, carrying Hawking radiation that was reflected back into the black hole by a potential barrier.}
    \label{fig:bh_circuit}
\end{figure}

Though there is generically an infinite number of modes to deal with in a quantum field theory in curved space-time, the Hawking process can be reduced to interactions between only a handful of modes, as heuristically alluded to above and illustrated in Figs.~\ref{fig:spacetime_diagram} and \ref{fig:bh_circuit}, such that one can directly apply the Gaussian formalism of Section~\ref{sec:gauss_formalism} to the process. This can be done by using the ``Wald's basis'' \cite{Wald:1975kc} (see also \cite{Frolov:1998wf}). 
We sketch the essential ingredients here and provide a reconstructive argument starting with Hawking's final result~\cite{Hawking75} and working backwards (see \cite{Wald:1995yp,Frolov:1998wf,fabbri05} for detailed derivations). We start by ignoring back-scattering, which we introduce later. The arguments below apply to any field that is emitted in the Hawking process, but for concreteness, we focus on massless fields (e.g., electromagnetic radiation). 

Let $\mathscr{H}_{\rm out}$ be the Hilbert space of the field at future infinity. The Hawking process leads to a quantum state in this Hilbert space corresponding to an outgoing  flux of uncorrelated blackbody-radiation, beginning with only vacuum fluctuations in the asymptotic past~\cite{Hawking74,Hawking75}. Such state has a density matrix of the form
\begin{equation}
    \hat{\Theta}_{\rm out}=\bigotimes_{\omega}\hat{\Theta}_{\rm out,{\omega}},\label{eq:hawking_out}
\end{equation}
where
\begin{equation}
    \hat{\Theta}_{\rm out,{\omega}} = \frac{1}{1+n_\omega}\sum_{N=1}^\infty \left(\frac{n_\omega}{1+n_\omega}\right)^N\dyad{N}_{\rm out,{\omega}},\label{eq:hawking_out_omega}
\end{equation}
and $n_\omega\equiv1/(e^{\omega/T_H}-1)$ is the Bose-Einstein distribution characterized by the Hawking temperature $T_H$, which is frequency independent. Here, $\omega$ labels positive frequency modes---of which there are infinitely many---and, hence, $\ket{N}_{\rm out,{\omega}}$ is a Fock state of $N$ quanta in the out frequency-mode $\omega$ reaching  future null infinity. The quantity $n_\omega$ can be precisely interpreted as the (average) number of quanta emitted by the black hole per frequency per time.

In words, Eq.~\eqref{eq:hawking_out} asserts that the frequency modes in the outgoing Hawking-flux are completely uncorrelated among each other, while Eq.~\eqref{eq:hawking_out_omega} asserts that, for an individual frequency, the quantum state $\hat{\Theta}_{\rm out, \omega}$ is completely mixed, in the sense that its photon-statistics\footnote{By photon-statistics, we are referring to the statistical fact that $N$ is a discrete random-variable, governed by the probability distribution $\frac{1}{1+n_\omega} \left(\frac{n_\omega}{1+n_\omega}\right)^N$, which is often referred to as a completely mixed (or `thermal') distribution in the quantum optics literature for any form of $n_\omega$.} is dictated by the probability distribution given in Eq.~\eqref{eq:hawking_out_omega}. Moreover, since $n_\omega$ follows a Bose-Einstein distribution, the entire ``out" state, $\hat{\Theta}_{\rm out}$, is precisely that of an uncorrelated quantum-gas of black-body radiation, characterized by a single number---the Hawking temperature, $T_H$.

The state $\hat{\Theta}_{\rm out}$ is obviously a mixed state, but this is only because we are restricting ourselves to the ``out" Hilbert space $\mathscr{H}_{\rm out}$, whereas the entire Hilbert space is $\mathscr{H}_{\rm int}\otimes\mathscr{H}_{\rm out}$, with $\mathscr{H}_{\rm int}$ the Hilbert space of modes falling into the balck hole horizon. By introducing the ``int" degrees of freedom, we can \textit{purify} the out-state, such that $\hat{\Theta}_{\rm out}\rightarrow\hat{\Psi}_{\rm int,out}$ and $\hat{\Psi}_{\rm int,out}\in\mathscr{H}_{\rm int}\otimes\mathscr{H}_{\rm out}$ is the pure quantum state which arises from evolving the in-vacuum. The correct purifying modes where discovered in Ref.~\cite{Wald:1975kc}. Their explicit form is not important for the arguments written below, and they can be found in Refs.~\cite{Wald:1975kc,Wald:1995yp,Frolov:1998wf,fabbri05}. We will denote the creation operators defined from them (or, as usual, by wave-packets built from them \cite{Hawking75,Wald:1975kc}) by $\hat a_\omega^{({\rm int})}$, labeled by the out-frequency $\omega$ (note that the int-modes in Wald's basis do not have well defined frequency with respect to any natural notion of time translation at the horizon). The total state in $\mathscr{H}_{\rm int}\otimes\mathscr{H}_{\rm out}$ is the simplest purification $\hat{\Psi}_{\rm int,out}$ of the out-state $\hat{\Theta}_{\rm out}$, namely
\begin{equation}
    \hat{\Psi}_{\rm int,out}=\bigotimes_\omega \hat{\Psi}_{\rm int,out; \omega},\label{eq:purif_out}
\end{equation}
where
\begin{equation}
    \hat{\Psi}_{\rm int,out; \omega}=\frac{1}{\sqrt{1+n_\omega}}\sum_{N=1}^\infty \left(\frac{n_\omega}{1+n_\omega}\right)^{N/2}\ket{N, N}_{\rm int, out; \omega},\label{eq:purif_out_omega}
\end{equation}
and $\{\ket{N}_{\rm int}\}$ is the number state associated with the number operator defined from $\hat a_\omega^{({\rm int})}$. This is a `thermo-field double' state~\cite{israel1976thermo} (although we need to keep in mind that at the horizon there is no natural notion of time which allows us to physically interpret the interior state as black body radiation). Using Eqs.~\eqref{eq:purif_out} and~\eqref{eq:purif_out_omega}, it is easy to check that $\hat{\Theta}_{\rm out}=\Tr_{\rm int}(\hat{\Psi}_{\rm int,out})$, justifying that $\hat{\Psi}_{\rm int,out}$ is indeed a purification of $\hat{\Theta}_{\rm out}$. In words, Eqs.~\eqref{eq:hawking_out}-\eqref{eq:purif_out_omega} tell us that there exists two quantum-gases of blackbody radiation: the Hawking radiation escaping to spatial infinity and another gas falling into the black hole. Moreover, these quantum gases are entangled, per Eqs.~\eqref{eq:purif_out} and~\eqref{eq:purif_out_omega}.

The state $\hat{\Psi}_{\rm int, out}$ is dynamically related to the ``in" vacuum, $\ket{\rm vac}_{\rm in}\in\mathscr{H}_{\rm in}$. There exists a basis of in-modes~\cite{Wald:1975kc} such that this dynamical relation is just a direct-product of two-mode transformations (one for each $\omega$). We denote Wald's in-modes by $\hat{a}_{\omega;1}^{\rm(in)}$ and $\hat{a}_{\omega;2}^{\rm(in)}$. They are related with the int- and out-modes via~\cite{Wald:1975kc} 
\begin{align}\label{eq:tms1_hawking}
    \hat{a}_\omega^{\rm(out)}&=\frac{1}{\sqrt{1-\e^{-\omega/T_H}}}\left(\hat{a}_{\omega;1}^{\rm(in)} +\e^{-\omega/2T_H}\hat{a}_{\omega;2}^{\rm(in)\dagger}\right)\\
    \hat{a}_\omega^{\rm(int)}&=\frac{1}{\sqrt{1-\e^{-\omega/T_H}}}\left(\hat{a}_{\omega;2}^{\rm(in)} +\e^{-\omega/2T_H}\hat{a}_{\omega;1}^{\rm(in)\dagger}\right). \label{eq:tms2_hawking}
\end{align}
Wald's in modes are labeled by the out-frequency $\omega$ that they are the progenitors of. They do not have well defined in-frequency but, rather, are made by combining positive frequency in-modes, without contribution from negative-frequency in-modes. This guarantees that Wald's basis is an explicit basis to define the unique in-vacuum---i.e., ${\hat{a}_{\omega;k}^{(\rm in)}\ket{\rm vac}_{\rm in}=0}$, $k=1,2$. The advantage of Wald's basis is that it makes the dynamical relation between in- and int-out-modes a direct-product of two-mode transformations (one for each $\omega$). We can therefore focus on one $\omega$ at a time and think of the Hawking-process as just a two-mode interaction mapping two ``in" modes to the ``out" and ``int" modes. The full interaction is then a direct-product over all the modes $\omega$. 

Back-scattering can then be introduced by noticing that the out-modes do not actually propagate to infinity, since they are partially scattered back by the gravitational potential barrier surrounding the black hole. Only a portion $\Gamma_{\ell, \omega}$ of the initial out-mode makes it to infinity, while the rest falls back into the black hole.

\paragraph*{Symplectic circuit:} We now show how the previous results can be described in an exact and efficient manner using the Gaussian formalism summarized in Section~\ref{sec:prelims}. The reformulation is simple, and allows us to extend the calculation to any initial Gaussian state, and to compute all aspects of the final state in an efficient manner, including entanglement between any subsystem of modes. The reformulation comes together with useful diagrams, which we refer to as symplectic circuits, which help to derive non-trivial qualitative information about the Hawking process. The name `symplectic circuit' is motivated by the fact that the diagrams depict the dynamical evolution by a concatenation of elementary symplectic transformations (squeezers, beam-splitters and phase shifters). They describe both the classical and the quantum evolution. 

For the Hawking effect, we notice that Eqs.~\eqref{eq:tms1_hawking} above, which constitute the core of the Hawking effect, correspond precisely to a process of {\em two-mode squeezing} (see Appendix \ref{sqzbs}), while the process of back scattering is described by a beam-splitter (see below). The physics of the Hawking process can be described diagrammatically by the symplectic circuit depicted in Fig.~\ref{fig:bh_circuit}, which consists of a discrete concatenation of physical processes in a concrete manner. The circuit is simple: First, the in-modes $a_{\omega;1}^{\rm(in)}$ and $a_{\omega;2}^{\rm(in)}$ pass through a two-mode squeezer, which we label as $\bm{S}_{\rm H}$ (H stands for Hawking), leading to particle-pair creation. Formally, this process corresponds to the two-mode symplectic squeezing transformation, written in $2\times2$ blocks as 
\begin{equation}
    \bm{S}_H=
    \begin{pmatrix}
    \cosh r_H\, \bm{I}_2 & \sinh r_H\, \bm\sigma_z\\
    \sinh r_H\, \bm\sigma_z & \cosh r_H\, \bm{I}_2
    \end{pmatrix},
\end{equation}
which acts on the canonical operators as described in Section~\ref{sec:gauss_formalism} (note that this matrix is the particularization of Eq.~\eqref{squeezer} to $r=r_H$ and $\phi=0$). Here, $\bm\sigma_z$ is the standard Pauli-z matrix, and $r_H$ are frequency dependent real numbers given by $\tanh^2 r_H\equiv\e^{-\omega/T_H}$, where $T_H$ is the Hawking temperature of the system. The expression for $r_H$ accounts for the black-body spectrum at the Hawking temperature $T_H$ of the outgoing radiation.

Following pair-creation in the near horizon region, one mode of the pair, the one described by the int-mode $a_{\omega;1}^{\rm(int)}$ (the Hawking-partner mode), falls into the black hole, while the other outgoing mode $a_{\omega;2}^{\rm(out)}$ (the long wavelength, outgoing Hawking radiation) must climb the potential barrier on its way to infinity. This process of back-scattering is described completely by a simple process in which two-modes scatter to two-modes. The incoming modes are the mode $a_{\omega;2}^{\rm(out)}$ produced in the near horizon region (this mode, is in many references, called the ``up'' mode; see e.g.~\cite{Frolov:1998wf,fabbri05}) and one long-wavelength incoming mode from past infinity, which we will denote by the annihilation operator $\hat b^{\rm(in)}_\omega$. The result of the scattering process are the out-modes which make it to the asymptotic out-region as Hawking radiation, together with long-wavelength modes back-scattered into the black hole, which we will denote as $\hat b_{\omega}^{\rm(int)}$ (see Fig.~\ref{fig:spacetime_diagram} for an illustration). Formally, the back-scattering process corresponds to a beam-splitter, described by an orthogonal-symplectic transformation $\bm O_\theta$, written in $2\times2$ blocks as,
\begin{equation}
    \bm O_{\theta}=
    \begin{pmatrix}
    \cos\theta_{\ell, \omega}\, \bm{I}_2 & \sin\theta_{\ell, \omega}\, \bm{I}_2 \\
    -\sin\theta_{\ell, \omega}\, \bm{I}_2 & \cos\theta_{\ell, \omega}\, \bm{I}_2
    \end{pmatrix},
\end{equation}
where the angle $\theta_{\ell, \omega}$ encodes the grey-body factor $\Gamma_{\ell,\omega}$ through the transmission probability $\cos^2\theta_{\ell, \omega}=\Gamma_{\ell, \omega}$. We represent this classical scattering in the circuit by a beam-splitter element [to simplify our notation we will omit the labels $\ell$ (the angular quantum number) and $\omega$ in $\theta$ and $\Gamma$]. At the level of annihilation operators, the beam-splitter imposes the following transformation (see Appendix \ref{sqzbs} for further details)
\begin{align}
    \hat{a}_{\omega;2}^{\rm (out)} &\longrightarrow  \cos \theta_\omega\, \hat{a}_{\omega;2}^{\rm (out)} + \sin \theta_\omega\,  \hat{b}_{\omega}^{\rm (in)}\, ,\\
     \hat{b}_{\omega}^{\rm (in)} &\longrightarrow    \cos \theta_\omega\, \hat{b}_{\omega}^{\rm (in)} - \sin \theta_\omega\,  \hat{a}_{\omega;2}^{\rm (out)} \, .
\end{align}
Observe that this process does not create particles, easily seen by the fact that there are no creation operators present in the output modes. 

We now combine these elements to write a scattering matrix corresponding to the circuit of Fig.~\ref{fig:bh_circuit}. Let us choose the following order for the vector of canonical operators
\begin{align}
    \hat{\bm{R}}^{\rm(in)}&\equiv&\left(\hat{Q}_{\omega;1}^{\rm(in)},\hat{P}_{\omega;1}^{\rm(in)},\hat{Q}_{\omega;b}^{\rm(in)},\hat{P}_{\omega;b}^{\rm(in)},\hat{Q}_{\omega;2}^{\rm(in)},\hat{P}_{\omega;2}^{\rm(in)}\right)^\top, \nonumber \\
     \hat{\bm{R}}^{\rm(out)}&\equiv&\left(\hat{Q}_{\omega;1}^{\rm(int)},\hat{P}_{\omega;1}^{\rm(int)},\hat{Q}_{\omega;b}^{\rm(int)},\hat{P}_{\omega;b}^{\rm(int)},\hat{Q}_{\omega;2}^{\rm(out)},\hat{P}_{\omega;2}^{\rm(out)}\right)^\top, \nonumber
\end{align}
where, e.g., $\hat{Q}_{\omega;1}^{\rm(in)}=(\hat{a}_{\omega;1}^{\rm(in)}+\hat{a}_{\omega;1}^{{\rm(in)}\dagger})/\sqrt{2}$ and $\hat{P}_{\omega;1}^{\rm(in)}=-i\, (\hat{a}_{\omega;1}^{\rm(in)}-\hat{a}_{\omega;1}^{{\rm(in)}\dagger})/\sqrt{2}$, and similarly for the other pairs of canonical variables. Under this ordering, the full symplectic-transformation taking the in-modes to the out/int-modes is easily found as
\begin{widetext}
\begin{equation}
    \bm S_{\rm BH} =  
    \begin{pmatrix}\bm{I}_2 & 0 & 0\\ 
  0 &  \cos\theta_\omega\bm{I}_2 & -\sin\theta_\omega\bm{I}_2 \\
    0 & \sin\theta_\omega\bm{I}_2 & \cos\theta_\omega\bm{I}_2
    \end{pmatrix}\cdot  \begin{pmatrix}
    \cosh  r_H\bm{I}_2 &0 &  \sinh  r_H\bm\sigma_z\\
    0 & \bm{I}_2& 0\\ 
    \sinh  r_H\bm\sigma_z &0 &  \cosh  r_H\bm{I}_2
    \end{pmatrix}=  
    \begin{pmatrix}
    \cosh{ r_H}\bm{I}_2 & 0 & \sinh{ r_H}\, \bm\sigma_z \\
    \sin{\theta_\omega}\sinh{ r_H}\, \bm\sigma_z & \cos\theta_\omega\bm{I}_2 & \sin\theta_\omega\cosh{ r_H}\, \bm{I}_2\\
    \cos\theta_\omega\sinh{ r_H}\, \bm\sigma_z & -\sin\theta_\omega\, \bm{I}_2 &\cos\theta_\omega\cosh{ r_H}\, \bm{I}_2 
    \end{pmatrix}. 
    \label{eq:s_matrix_bh}
\end{equation}
\end{widetext}
We note that a similar parameterization was given in Ref.~\cite{de2015iop,nambu21} in the context of an analog black hole formed in BECs, but no diagrammatic decomposition (nor generalizations thereof to multi-mode scattering scenarios) was discussed.

From the $S$-matrix, and given the first and second moments of a Gaussian initial state, one can derive all aspects of the final state in a simple manner. We emphasize that the circuit simply provides a re-formulation of Hawking's original derivation~\cite{Hawking75} and, in particular, the circuit does not introduce any extra assumption or approximation.

Let us compute, for instance, the particle density for the outgoing Hawking radiation, assuming a vacuum input. The vacuum state is a three-mode Gaussian state with moments $\bm\mu_{\rm vac}^{\rm(in)}=\bm0$ and $\bm \sigma_{\rm vac}^{\rm(in)}=\bm{I}_6$. The out/int moments can then be found by acting with the scattering matrix of Eq.~\eqref{eq:s_matrix_bh} on the in-moments, using the general transformations of Eqs.~\eqref{eq:out_mu_general} and~\eqref{eq:out_sigma_general}, namely $\bm\mu_{\rm vac}^{\rm(out)}=   \bm S_{\rm BH} \, \bm\mu_{\rm vac}^{\rm(in)}$ and $ \bm \sigma_{\rm vac}^{\rm(out)}=\bm S_{\rm BH}\, \bm  \sigma_{\rm vac}^{\rm(in)}\, \bm S_{\rm BH}^{\top}$. One then finds the quantum state of the outgoing Hawking radiation from these results by simply taking the reduced moments of the single-mode phase-space corresponding to the mode $a_{w;2}^{\rm(out)}$ (just extract the relevant matrix coefficients of the final out/int moments). After a little algebra, the mean vector and covariance matrix for the outgoing Hawking mode are
\begin{align}
    \bm\mu_{w;2}^{\rm(out)}&=\bm0,\\ \bm\sigma_{w;2}^{\rm(out)}&=\left(1+2\cos^2\theta_\omega\sinh^2 r_H\right)\bm{I}_2.
\end{align}
This is the covariance matrix of a thermal state with mean number of quanta,
\begin{equation}
    \ev{\hat{n}_H}=\frac{1}{4}\text{Tr} \bm\sigma_{w;2}^{\rm(out)}-\frac{1}{2}=\cos^2\theta_{\ell, \omega}\sinh^2 r_H=\frac{\Gamma_{\ell,\omega}}{\e^{\omega/T_H}-1},\label{eq:hawking_flux} 
\end{equation}
where we have used Eq.~\eqref{eq:mean_quanta}, the correspondence $\cos^2\theta_{\ell,\omega}=\Gamma_{\ell,\omega}$ (and restored the labels $\ell$ and $\omega$ in this expression) and the Hawking relation $\tanh^2{ r_H}=\e^{-\omega/T_H}$. Eq.~\eqref{eq:hawking_flux} is the standard result for the outgoing Hawking-flux in the asymptotic out-region, physically describing a quantum gas of blackbody radiation modulated by a grey-body coefficient, $\Gamma_{\ell,\omega}$.

\begin{figure*}
    \centering
    \includegraphics[width=.42\linewidth]{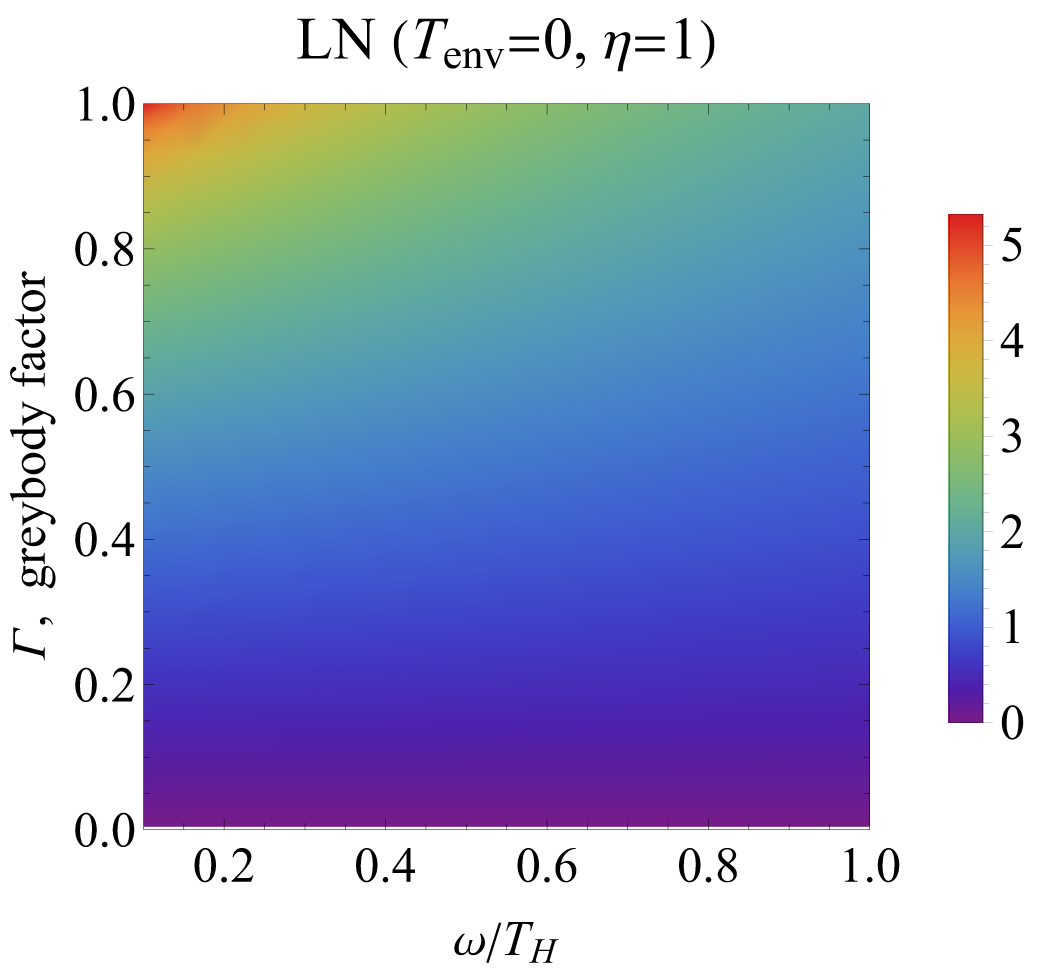}
    \includegraphics[width=.42\linewidth]{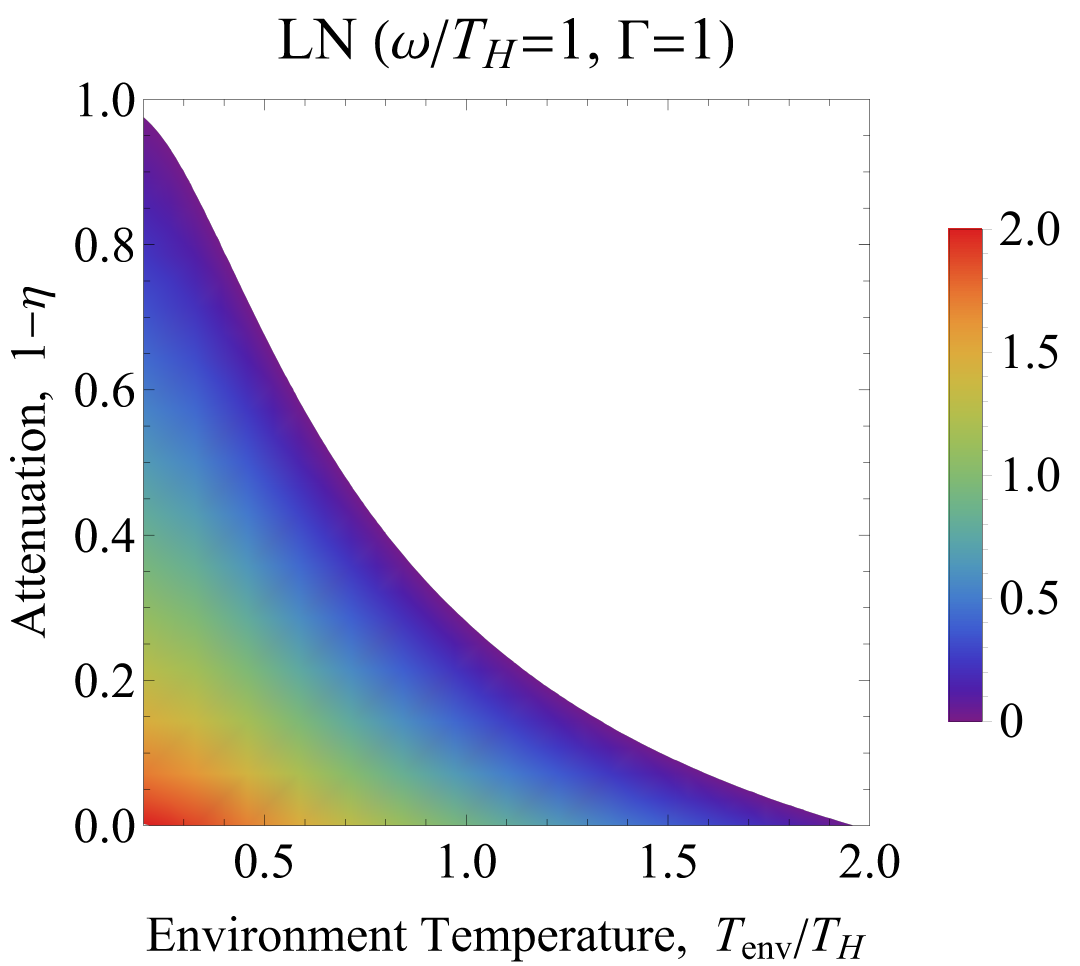}
    \caption{Entanglement (quantified by LN) between the `Hawking-pair' of modes $\hat a_{\omega;2}^{(\rm int)}$ and $\hat a_{\omega;1}^{(\rm out)}$ in the Hawking process. Left panel: Variation of entanglement across system-parameters, $(\Gamma, \omega/T_H)$, assuming ideal conditions of zero background temperature, $T_{\rm env}=0$, and perfect efficiency, $\eta=1$ (zero attenuation). Right panel: Variation of entanglement in noisy parameter space $(1-\eta,T_{\rm env}/T_H)$ at $\omega/T_H=1$ and $\Gamma=1$. This plot assumes isotropic thermal noise (an unphysical assumption)---i.e., the same number of thermal quanta in the three in modes. 
    Null-space indicates zero entanglement. Observe the boundary curve where entanglement vanishes identically. Behavior does not change with choice of $\omega/T_H$.}
    \label{fig:logneg_bh}
\end{figure*}

\subsection{Quantum correlations and entanglement degradation}\label{subsec:correlations_bh}

Next, we use the black hole circuit to analyze the entanglement between the modes in the final state. From the symplectic circuit, we can obtain analytical expressions for the entanglement in any bipartition or any pair of modes in the 3-mode network, using techniques described in Sections~\ref{sec:gauss_formalism} and~\ref{sec:logneg}. In an analog-gravity experiment, it is likely that entanglement will only be measured pair-wise for the outgoing radiation, as measuring entanglement between any-or-all bipartitions requires multi-mode measurements (or multi-mode quantum-state tomography)---an experimentally challenging feat. In this section, we therefore restrict our attention to the modes, $a_{\omega;2}^{(\rm int)}$ and $a_{\omega;1}^{(\rm out)}$ (the ``original Hawking pairs"), and quantify the entanglement by the LN. (We discuss multi-mode entanglement in Section~\ref{quantcorr}.) 

We also incorporate the effects of initial thermal fluctuations and attenuation. Defining efficiency $\eta$ ($1-\eta$ measures attenuation), the covariance matrix of the process is,
\begin{equation}
    \bm\sigma^{(\rm out)}=\eta\, \bm S_{\rm BH}\bm\sigma^{(\rm in)} \bm S_{\rm BH}^\top+ (1-\eta)\, \bm{I}_6,\label{eq:sigma_bh_open}
\end{equation}
where $\bm S_{\rm BH}$ is the symplectic matrix of Eq.~\eqref{eq:s_matrix_bh} and $ \bm\sigma^{(\rm in)}=(1+2n_{{\rm env}})\, \bm{I}_6$ is an initial thermal state at e.g. temperature $T_{\rm env}$. Notice that we have assumed isotropic initial thermal fluctuations in the inputs for simplicity (i.e., $n_{{\rm env}}$ is the same for the three input modes). In fact, isotropic thermal noise is highly unphysical, since the in-modes $\hat a_{\omega;1}^{(\rm in)}$ and $\hat a_{\omega;2}^{(\rm in)}$ have support only on ultra-high frequencies as measured by an inertial observer at past infinity, while the back-scattering mode $\hat b_{\omega}^{(\rm in)}$ has frequency of the same order as the Hawking quanta (of the order of the Hawking temperature $T_H$). Therefore, for any realistic thermal environment for astrophysical black holes (e.g. the cosmic microwave background radiation), the in-modes $\hat a_{\omega;1}^{(\rm in)}$ and $\hat a_{\omega;2}^{(\rm in)}$ would be meekly populated due to their ultra-high frequencies. However, since our main goal in this section is to illustrate the effects that thermal noise have in the generation of entanglement and to extract conclusion for analog black holes for which all in-modes can be realistically thermally populated, we consider the situation of isotropic thermal noise. Though, it is straightforward to relax this assumption.\footnote{We can generalize this to non-isotropic noise and also include noise added from the loss channel, $\bm\sigma^{(\rm out)}=\eta\, \bm S_{\rm BH}(\oplus_{i=1}^3N_{k_i}\bm{I}_2)\bm S_{\rm BH}^\top + (1-\eta)\,(\oplus_{i=1}^3N_{k_i}\bm{I}_2)$, where $N_{k_i}=1+2n_{k_i}$ and $n_{k_i}$ is the number of noisy quanta in the $i$th mode. Here, we have assumed the added quanta from the loss channel is equal to the initial noise population.} Similarly, the physical meaning of  the attenuation parameter $\eta$ is more clear for analog black holes---where $\eta$ is related to the peculiarities of the experimental set up or to the efficiency of detectors---than for astrophysical black holes. Anyway, we study the effects that varying $\eta$ has on the  entanglement of the final state as a pedagogical preparation for the section \ref{sec:imperfect_wb}.

The covariance matrix $ \bm\sigma^{(\rm out)}$ is a $6\times6$ matrix for the three modes of the black hole circuit (corresponding to a mixed 3-mode Gaussian quantum state), from which we take the reduced covariance matrix for the modes $a_{\omega;2}^{(\rm int)}$ and $a_{\omega;1}^{(\rm out)}$.

Given the reduced covariance matrix for Hawking pairs in the final state, we compute the LN as a function of all the open parameters and plot the result; see Fig.~\ref{fig:logneg_bh}. The left panel is the entanglement in the system-parameter space, $(\omega/T_H, \Gamma)$, for ideal operating conditions, $n_{\rm env}=0$ and $\eta=1$. We show this to simply illustrate that there is entanglement in this pair of modes, over all system parameters (vanishing only asymptotically as $\Gamma\rightarrow 0$ or $\omega/T_H\rightarrow\infty$). Note that the entanglement decreases with the greybody factor, $\Gamma$, as some quanta are lost to the back-scattering channel, $\hat b_{\omega}^{(\rm int)}$. However, in the absence of environmental thermal fluctuations (characterized by $n_{\rm env}$), there is always residual entanglement between the modes $\hat a_{\omega}^{(\rm int)}$ and $\hat a_{\omega}^{(\rm out)}$, no matter the amount of back-scattering $\Gamma>0$. The entanglement within the entire system of three modes, however, does not change with the grey-body factor, as back-scattering simply shifts quanta around. On the other hand, we note that, if initial thermal fluctuations are present, non-negligible back-scattering will generally degrade (or even destroy) entanglement between Hawking-pair modes if the thermal fluctuations are strong enough (not shown here).

The right panel of Fig.~\ref{fig:logneg_bh} illustrates the effects of noise due to thermal fluctuations and attenuation, assuming $\Gamma=1$ and $\omega/T_H=1$. Both of these effects are harmful to the entanglement generated in the Hawking process. For instance, ignoring attenuation for the moment (i.e., restricting to the bottom line at $\eta=1$), once we ``turn on" the Hawking effect, the Hawking process will cause initially independent thermal fluctuations to spread amongst the modes, leading to (classically) correlated noise within the system, but these correlations have nothing to do with entanglement. Even worse, such classical-correlations can dominate over the genuine quantum-correlations produced during the Hawking process---rendering any entanglement null, as depicted by the white-space in the right panel of Fig.~\ref{fig:logneg_bh}. 

For near-unit efficiency, a critical condition is found $T_{\rm env}/T_H<2$, above which entanglement is zero (under the assumption of isotropic thermal noise). This is an equivalent condition as that discussed in the BEC context of Ref.~\cite{bruschi2013}. We extend these prior results to include attenuation---thus obtaining a family of critical-conditions, $T_{\rm env}/T_H<f(\eta)$, where $0\leq f(\eta)\leq2$, as seen by the boundary curve in Fig.~\ref{fig:logneg_bh}. For instance, there is entanglement for all $T_{\rm env}/T_H<1$ for $1-\eta\approx.3$. We remark that this family of conditions are independent of the ratio $\omega/T_H$ in this simple setup.

The LN explicitly shows the behavior of entanglement in the noise parameter space $(1-\eta,T_{\rm env}/T_H)$ (see Fig.~\ref{fig:logneg_bh}): the entanglement is monotonically decreasing in both the attenuation, $1-\eta$, and in the environment temperature, $T_{\rm env}$, as intuitively expected. Hence, the take home message is that the entanglement generated in the Hawking process is fragile to both loss and ambient noise; this is an important lesson to keep in mind in analog scenarios discussed below. 

Contrariwise, one cannot faithfully deduce such behavior from the entanglement witnesses $P_-$ or $\Delta$ of Eq.~\eqref{eq:cs_inequality} for the same set of parameters, as they are witnesses of entanglement and not measures of it. Indeed, there are regions in the parameter space $(1-\eta,T_{\rm env}/T_H)$ which exhibit a larger violation of the CS inequality, $\Delta<0$, when the noise is \textit{increased} (see Fig.~\ref{fig:Delta_eta_T} in Appendix~\ref{app:ent_compare}). Although $\Delta$ cannot be used as a quantifier of entanglement, we point out that, in this simple set up, $\Delta=0$ genuinely highlights the critical line in the $(1-\eta,T_{\rm env}/T_H)$ plane where entanglement vanishes, consistent with the vanishing of the LN (the boundary line in Fig.~\ref{fig:logneg_bh}), as should be the case for a good entanglement witness. [On the other hand, Appendix \ref{app:ent_compare} shows other situations where $\Delta=0$ does {\em not} faithfully capture the boundary in the $(1-\eta,T_{\rm env}/T_H)$ plane where entanglement vanishes.]

\section{Simple recipe for circuit construction}\label{sec:recipe_lim}

From the simple circuit decomposition detailed in the previous section, we ascertain a few rules-of-thumb that we can follow to assess more generic settings (e.g., multi-mode settings with multiple event horizons) in order to find effective circuit descriptions therein. The rules are: (1) Insert a two-mode squeezer, with squeezing intensity $r_H$ characterized by the Hawking temperature $T_H$ via $\tanh^2 r_H=\e^{-\omega/T_H}$, for each Hawking-pair creation mechanism; (2) Insert a two-mode beam-splitter for each classical scattering-event between two modes; and finally, (3) Arrange circuit elements appropriately to highlight the physics at play. If white hole and black hole horizons are both present, then one must take into consideration the inverse nature of the white hole scattering process and the order of scattering events (see next section).   

In a more mathematical language, a pair creation occurs when a mode with positive symplectic norm\footnote{Given a complex solution $\varphi(\vec x,t)$ to the equations of motion describing one mode of the system, its symplectic norm is defined from the symplectic product $(\varphi_1,\varphi_2)\equiv i\, \int_{\Sigma} d^3\Sigma\, n^{\mu}\,  (\bar \varphi_1  \nabla_{\mu}\varphi_2- \nabla_{\mu}  \bar \varphi_1\,\varphi_2)$, where $d^3\Sigma$ is the volume element of  the Cauchy hypersurface $\Sigma$ and $n^{\mu}$ its fututre-oriented unit normal (see, e.g.~\cite{fabbri05,Wald:1995yp}). The name `symplectic' is motivated from the fact that such product is defined from the symplectic structure of the classical phase space.}
mixes with a mode with negative symplectic norm (the mixing is, of course, dictated by the equations of motion). Such interaction is described, therefore, by a two-mode squeezer. On the contrary, when two modes with the same sign of their symplectic norm interact, the process is described by a beam-splitter. Free propagation corresponds to phase shifters. Therefore, if one knows the dynamics, it is simple to translate it to a simpler and more intuitive circuit.

When the details of the dynamics are unknown, the rules-of-thumb (1)-(3) enumerated above are heuristic and not necessarily algorithmic,\footnote{Though, using physical arguments, one can be slightly more rigorous and, e.g., place bounds on the number of free parameters required to specify/construct the desired circuit; see Appendix~\ref{app:counting} for further comments on this matter.} nor are such rules entirely general. For instance, the role of phases are not mentioned in the rules above, which can lead to important consequences in certain scenarios where resonance effects matter (such as lasing in a white-black hole~\cite{corley1999lasers,katayama2021circuit}). Moreover, our circuit diagrams are not a substitute for the micro-physics, since a micro-physical description is required in order to specify, e.g., the functional form of the greybody factor, $\Gamma$, as well as the Hawking temperature, $T_H$. Nevertheless, with some ingenuity, such a circuit diagram proves to be immensely useful as an explanatory tool and supplementary guide, especially when an analytical description is not straightforward, as we demonstrate in the next section for an optical-analog, white-black hole generated by a strong electromagnetic pulse.

\section{Example 2: White-black hole pair}\label{sec:imperfect_wb}

In this section we put the techniques described above in action for the more complicated scenario where a black hole event-horizon and a white hole event-horizon are both present and interacting. We construct a symplectic circuit using the recipe given above and derive analytical expressions for different aspects of the out-state therefrom (regarding both particle creation and multi-mode entanglement). Finally, in order to quantitatively assess the accuracy of the circuit in describing the underlying physics, we  compare the results derived from such with numerical solutions to the equations of motion for the particular case of optical systems.

\subsection{Physical picture, symplectic circuit, and example calculations}

\paragraph*{Physical picture:} Consider the circumstance where a black hole event-horizon and a white hole event-horizon share a mutual interior region. This is common in optical setups where the analog space-time is generated from a strong pulse with a rising tail (corresponding to the white hole) and a descending front (corresponding to the black hole) \cite{philbin08, drori19}. For concreteness, the analysis of this section will have this system in mind, but we expect a similar scenario for any background flow with a rise-then-fall profile and a (approximate) trapped/anti-trapped region in between. 

\begin{figure}[t]
    {\centering     
\includegraphics[width = 0.48\textwidth]{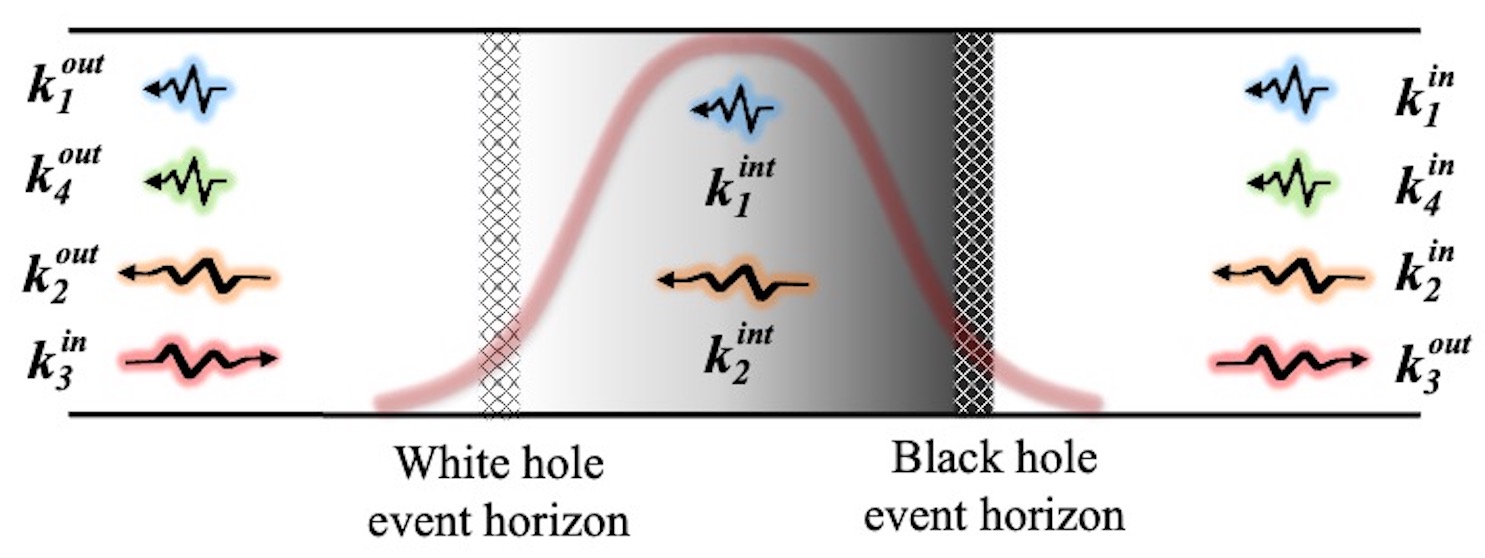}
}
\caption{Illustration of the structure of $in$, $int$, and $out$ modes for an optical analog white-black hole in the comoving frame~\cite{agullo2022prl}. The analog white-black hole is generated by a strong electromagnetic pulse via the Kerr effect. There are four in-modes (three arriving at the black hole horizon and one at the white hole horizon), and four outgoing-modes. There are two real, propagating interior modes (int-modes) between horizons; the other two modes are evanescent (and thus exponentially suppressed) within this region.}\label{modestructure}
\end{figure}

The ``white-black hole", as we shall call it, partitions two asymptotically flat space-time regions. The exterior region of the white hole (ideally) has no access whatsoever to the exterior region of the black hole, due to the causal-impossibility of traversing the white hole event horizon. However, the converse is not true. Anything that falls into the black hole emerges from the white hole, though slightly perturbed by the Hawking pair-creation mechanisms occurring near each event horizon.  Figure~\ref{modestructure} illustrates this configuration, together with the structure of modes involved in the process, separated in three groups: incoming, outgoing, and interior. In optical systems, these are the modes associated with a single frequency $\omega$ but different wavenumber $k$ (as defined in the frame comoving with the white-black hole pair), and modes with different frequencies do not interact (more details below, and in Ref.~\cite{linder16}). The discussion in this section applies to any system with a similar mode structure, regardless of its physical origin.

The Hawking process of the white hole is the time-reverse process of the black hole. Instead of emitting bland, thermal Hawking-radiation, the white hole emits quantum-correlated Hawking pairs of modes $k^{(\rm out)}_1$ and $k^{(\rm out)}_4$. However, for the tangled white-black hole system colloquially described above, the white hole pair-creation process is not spontaneous, as the infalling Hawking-partner generated at the black hole event horizon stimulates the white hole. Thus, the Hawking-pairs generated by the white hole are slightly tainted and in a mixed state---purified only by the outgoing Hawking radiation in the exterior region of the black hole.

\paragraph*{Symplectic circuit:}

For simplicity, we assume that the white hole and black hole event-horizons are identical. In other words, the pulse generating the white-black hole is symmetric about its center. One can straightforwardly generalize this to include asymmetries in the flow profile induced by the pulse. 

To describe the modes involved in the scattering process, it is simpler to work in the frame comoving with the pulse, in which the white-black hole pair is at rest and stationary. The frequency $\omega$ of waves defined in this frame is conserved because of stationarity, and modes with different frequencies do not mix with each other. Therefore, the evolution factorizes into uncoupled $\omega$-sectors, reducing the problem to a finite set of interacting modes for each $\omega$. The analysis of the dispersion relation for this system (see e.g.~\cite{linder16} for an analytical description), shows that, far away from the strong pulse producing the white-black hole system (in the asymptotic regions), there exist four different modes for each frequency $\omega$, which we label as $k_1,k_2,k_3,k_4$. Three of them move to the left, and the fourth moves to the right (see Fig.~\ref{modestructure} to see our conventions). Hence, considering both the white and black hole sides of the pulse, we have four modes coming in, and four modes going out, which is a slightly more complicated setup than that of astrophysical black holes, where we had only three in-modes and three out-modes and a single event horizon. In spite of such complications, the analysis is conceptually similar. 

We choose the following order for the in-vector of canonical operators,
\begin{multline}
    \hat{\bm{R}}^{\rm(in)}\equiv\\\left(\hat{Q}_{k_1}^{\rm(in)},\hat{P}_{k_1}^{\rm(in)},\hat{Q}_{k_2}^{\rm(in)},\hat{P}_{k_2}^{\rm(in)},\hat{Q}_{k_3}^{\rm(in)},\hat{P}_{k_3}^{\rm(in)}, \hat{Q}_{k_4}^{\rm(in)},\hat{P}_{k_4}^{\rm(in)}\right)^\top,
\end{multline}
where, e.g., $\hat{Q}_{k_3}^{\rm(in)}$ is the only in-mode to the white hole and is the time-reverse of the outgoing Hawking mode emitted by the black hole (thus, $\hat{Q}_{k_3}^{\rm(in)}$ is  the ``ingoing Hawking mode"). We choose a similar order for the out-modes.

\begin{figure}
    \centering
    \includegraphics[width=\linewidth]{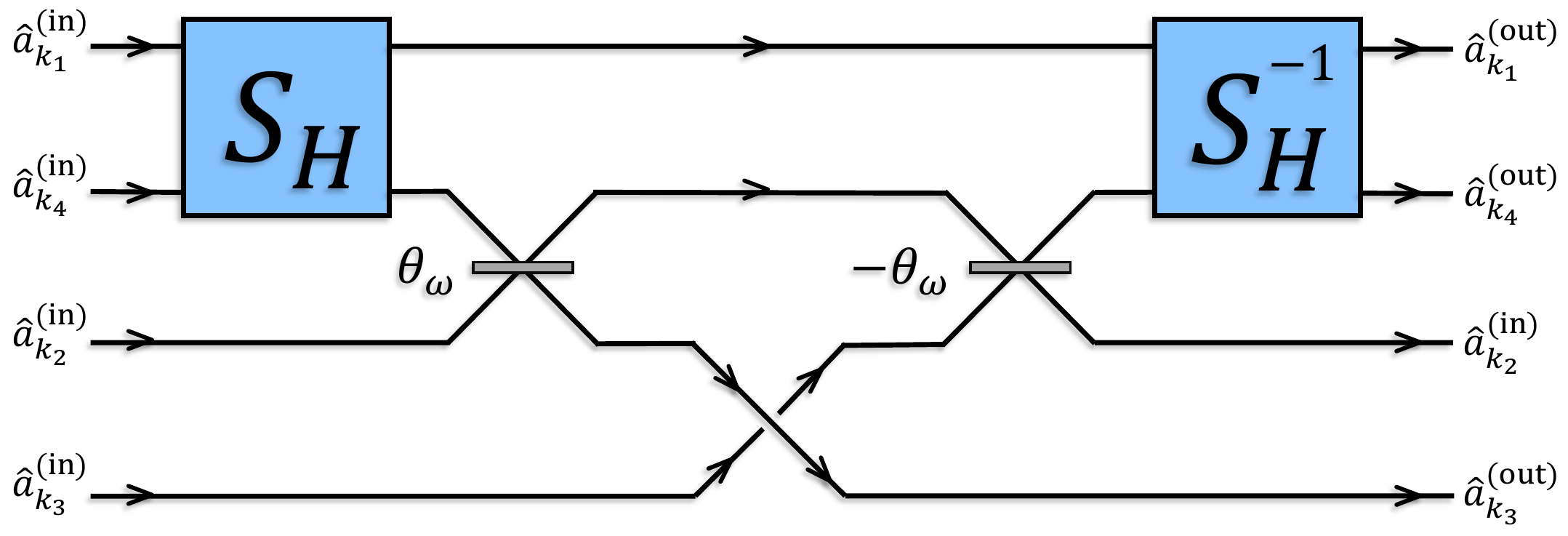}
    \caption{Minimal ansatz for the symplectic-circuit of the Hawking process for the white-black hole. The squeezer and beam-splitter on the left corresponds to the black hole, while the elements to the right correspond to the white hole horizon. The white hole scattering process is the time reverse of the black hole scattering process; hence the in-modes of the black hole are the out-modes of the white hole and vice versa. $\bm S_H$ is a two-mode squeezer with squeezing intensity $r_{H}$, while beam-splitters are labeled by the parameter $\theta_{\omega}$.}
    \label{fig:wbh_circuit}
\end{figure}

In Ref.~\cite{agullo2022prl}, we introduced a symplectic-circuit to describe the Hawking process for a white-black hole (shown here in Fig.~\ref{fig:wbh_circuit}), which serves as a powerful explanatory tool to describe the physics behind the distinct scattering processes of a white-black hole. Here, we provide a more detailed analysis of the circuit description. This circuit is built following the recipe provided in the previous section; namely, each event horizon introduces a two-mode squeezer accounting for the pair-creation and a beam-splitter to implement a possible process of back-scattering. As discussed in Refs.~\cite{Corley:1996ar,linder16}, back-scattering primarily affects long-wavelength modes ($k_2$ and $k_3$ in this scenario), while it can be neglected for the short-wavelength modes $k_1$ and $k_4$. Observe that the symplectic circuit of Fig.~\ref{fig:wbh_circuit} arises by simply gluing together the circuit for the black hole alone (Fig.~\ref{fig:bh_circuit}) with its mirror image (because a white hole is the time reversal of a black hole). In the next section, we show that this simple circuit captures the underling physics with great accuracy in the regime where the analogy with the Hawking effect occurs---i.e., in the low frequency regime where dispersive effects do not dominate.

The next step is to convert the symplectic circuit into a scattering matrix, from which we can derive analytical expressions depending on the parameters of the circuit. Let us label each symplectic operation in the circuit by $\bm S_{i}$, where $i\in\{1, 2, 3, 4\}$ labels the order of operations in time (from left to right in Fig.~\ref{fig:wbh_circuit}). For instance, $\bm S_1$ is the black hole two-mode squeezer which, with this mode ordering, has a matrix description
\begin{equation}
\bm S_1= \begin{pmatrix}
\cosh{ r_H}\, \bm{I}_2& 0 & 0 & \sinh{ r_H}\, \bm\sigma_z \\
0 & \bm{I}_2 & 0 & 0 \\
0 & 0 & \bm{I}_2 & 0 \\
\sinh{ r_H}\, \bm\sigma_z & 0 & 0 & \cosh{ r_H}\, \bm{I}_2
\end{pmatrix}.
\end{equation}
Next, we have a scattering process in the exterior of the black hole horizon, described by 
\begin{align}
    \bm S_2 &= \begin{pmatrix} \bm{I}_2 & 0 & 0 & 0 \\ 
    0 & \cos{\theta_\omega}\bm{I}_2 & 0 &-\sin\theta_\omega\bm{I}_2 \\
    0 & 0 & \bm{I}_2 & 0 \\ 
    0 & \sin\theta_\omega\bm{I}_2 & 0 & \cos{\theta_\omega}\bm{I}_2 \end{pmatrix}, 
    \end{align}
 followed by the counterpart of these two operations for the white-hole,
     \begin{align}  \bm S_3 &= \begin{pmatrix} \bm{I}_2 & 0 & 0 & 0 \\
    0 & \cos{\theta_\omega}\bm{I}_2 & \sin\theta_\omega\bm{I}_2 & 0 \\
    0 & -\sin\theta_\omega\bm{I}_2 & \cos{\theta_\omega}\bm{I}_2 & 0 \\
    0 & 0 & 0 & \bm{I}_2 \end{pmatrix}, \\
    \bm S_4 &= \begin{pmatrix}
    \cosh{ r_H}\bm{I}_2 & 0  & -\sinh{ r_H}\bm\sigma_z & 0 \\
    0 & \bm{I}_2 & 0 & 0 \\
    0 & 0  & 0 & \bm{I}_2 \\
    -\sinh{ r_H}\bm\sigma_z & 0 & \cosh{ r_H}\bm{I}_2 & 0 
\end{pmatrix}.
\end{align}
The relative minus signs between $S_1$ and $S_4$, and $S_2$ and $S_3$ are due to the inverse character of the white hole relative to the black hole. 
The full scattering matrix describing the transformation from the four in-modes to the four out-modes is then found by matrix multiplication, which we write formally as $\bm S_{\rm WB} = S_4S_3S_2S_1$
where the subscript WB refers to ``white-black" hole. Written out fully,

\begin{widetext}

\begin{equation}
\bm S_{\rm WB}=
\begin{pmatrix}
 (1+\cos ^2\theta \sinh ^2r_H)\bm{I}_2 &  \cos\theta\sin\theta\sinh r_H\bm{\sigma}_z & -\cos\theta\sinh r_H\bm{\sigma}_z & \cos ^2\theta \cosh{r_H}\sinh{r_H}\bm{\sigma}_z\\
 -\cos\theta\sin\theta\sinh{r_H}\bm{\sigma}_z & \cos^2\theta\bm{I}_2 & \sin\theta\bm{I}_2 & -\cos\theta\sin\theta\cosh{r_H}\bm{I}_2 \\
\cos\theta\sinh{r_H}\bm{\sigma}_z  & \sin\theta\bm{I}_2  & 0 & \cos\theta\cosh{r_H}\bm{I}_2 \\
 -\cos^2\theta\cosh{r_H}\sinh{r_H}\bm{\sigma}_z & - \cos\theta\sin\theta\cosh{r_H}\bm{I}_2 & \cos\theta\cosh{r_H}\bm{I}_2  & (\sin^2\theta-\cos ^2\theta\sinh^2r_H)\bm{I}_2
\end{pmatrix}\label{eq:SWH}
\end{equation}
\end{widetext}
Due to the multi-mode structure of the setup and the many physically distinct operations in play, it is instructive to perform some simple calculations to gain some insight into the dynamics of the network. We will thus begin by restricting ourselves to the ideal scenario where all modes initially contain only vacuum fluctuations. Then, the output covariance matrix for the 4-mode pure state is given by $\bm\sigma^{(\rm out)}_{\omega}=\bm S_{\rm WB}\cdot \bm\sigma_{\omega}^{(\rm in)} \cdot\bm S_{\rm WB}^\top$, with $\bm\sigma^{(\rm in)}_{\omega}=\bm{I}_8$; the mean of the state is zero. All properties of the quantum state of the modes can then be written in terms of the elements of $\bm\sigma_{\omega}^{(\rm out)}$.

We first calculate the outgoing spectrum of Hawking particles emitted by the black hole. This can be computed by taking the reduced covariance matrix of $\bm\sigma_{\omega}^{(\rm out)}$ of the single-mode sub-space corresponding to the mode $a_{k_3}^{(\rm out)}$, $\bm\sigma_{\omega;k_3}^{(\rm out)}$. It is straightforward to see that this covariance matrix describes a thermal state, $\bm\sigma_H^{(\rm out)}=(1+2\ev*{\hat{n}_H^{(\rm out)}})\, \bm{I}_2$, as expected. That is, the outgoing Hawking radiation from the black hole is completely oblivious to the presence of the white hole, as required by causality (or an approximate form thereof for analog horizons). 

Utilizing the general formula~\eqref{eq:mean_quanta} for the mean number of quanta, the mean number of quanta in the mode $k^{\rm (out)}_3$ is
\begin{equation}
    \ev{\hat{n}_H^{(\rm out)}}=\cos^2\theta_\omega\, \sinh^2 r_H=\frac{\Gamma_\omega}{\e^{\omega/T_H}-1},
\end{equation}
where we have used the definition $\Gamma_\omega\equiv\cos^2\theta_\omega$, and we have assumed the Hawking relation $\tanh^2 r_H = \e^{-\omega/T_H}$ holds (see \cite{linder16} for a justification using an analytical model, valid in the low frequency regime where dispersive effects do not dominate; see below for a numerical confirmation). We see that the outgoing Hawking spectrum is a modulated blackbody spectrum, analogous to the original Hawking effect, with modulation coming from back-scattering via the greybody factor $\Gamma_\omega$.

Beyond the Hawking spectrum alone, another interesting quantity is the total number of quanta generated in the full process, i.e., the quanta emitted by both the black and white hole (per unit frequency and unit time). Using the circuit and the relations above, we find
\begin{multline}
   \sum_{\rm all\, modes}\ev{\hat{n}_{i}^{(\rm out)}}=\\2\cos^2\theta_\omega\sinh^2r_H\left(2+\cos^2\theta_\omega\sinh^2r_H\right). \label{eq:quanta_gamma}
\end{multline}
Interestingly, this is a decreasing function of the greybody factor, $\Gamma_\omega=\cos^2\theta_\omega$. Hence, back-scattering, if present, acts as a self-stabilising mechanism for the white-black hole. We further note that this result is agnostic to individual phases which might accumulate between the black hole and white hole pair creation processes in Fig.~\ref{fig:wbh_circuit}. [We do not include such phases in our circuit decomposition, but it is straightforward to check this claim.] In other words, adding a phase to each mode in the middle of the circuit, which preserves the space-time symmetry of the diagram (a symmetry which is induced by the symmetry of the pulse), does not change the total particle number at the output.\footnote{This is in contrast to, e.g., white-black hole lasers which support a resonant enhancement of particle creation~\cite{corley1999lasers,katayama2021circuit}. Though, we note that the laser setup is quite different than the white-black hole here, since, in the laser system, the output of the white hole seeds the black hole and vice versa, leading to a cascaded succession of pair-creation events.}

The reduction in the total number of particles by the grey-body factor is unique to the white-black hole case (it does not arise for a single horizon) and has an interesting story behind which can be easily understood by simple inspection of the circuit. In short, if $\Gamma_{\omega}\to 0$ (completely reflecting potential barrier for the long-wavelength modes $k_2$ and $k_3$) the white hole undoes exactly the pair-production at the black hole horizon. In more detail, consider the $a_{k_4}^{(\rm in)}$ mode approaching the black hole in Fig.~\ref{fig:wbh_circuit}). This mode converts into outgoing Hawking radiation via the Hawking process, while its entangled partner falls into the black hole. On the way out, the Hawking radiation meets a barrier and  scatters back into the black hole as a $k^{\rm (int)}_2$ mode. This back-scattered Hawking radiation then traverses through the interior region of the white-black hole, emerging from the white hole unperturbed (the back-scattered mode mixes negligibly with the mode $k_1$ in the interior). Following its escape from the white hole horizon, the back-scattered radiation meets another barrier in the exterior region of the white hole; at which point, it reflects back towards the white hole---now as an ingoing Hawking-mode, $k^{\rm (in)}_3$, seeding the white hole pair-creation mechanism. This ``reflective-seeding" process undoes the Hawking process initiated by the black hole, since the white hole and black hole unitary dynamics are time-reversals of one another---thus resulting in an overall reduction in the total number of particles. One can see this directly by treating each of the beam-splitter elements in Fig.~\ref{fig:wbh_circuit} as perfect mirrors (such that ingoing quanta bounce off the beam-splitters and cannot traverse through them) and tracing the paths of the modes through the circuit. Of course, the exact cancellation we just described only occurs under the assumption of symmetric pulse, which amounts to saying that the white hole is the exact time-reversal of the black hole. 

This example illustrates the usefulness of symplectic circuits to understand complex aspects of system by mere visual inspection.

\begin{figure}
    \centering
    \includegraphics[width=\linewidth]{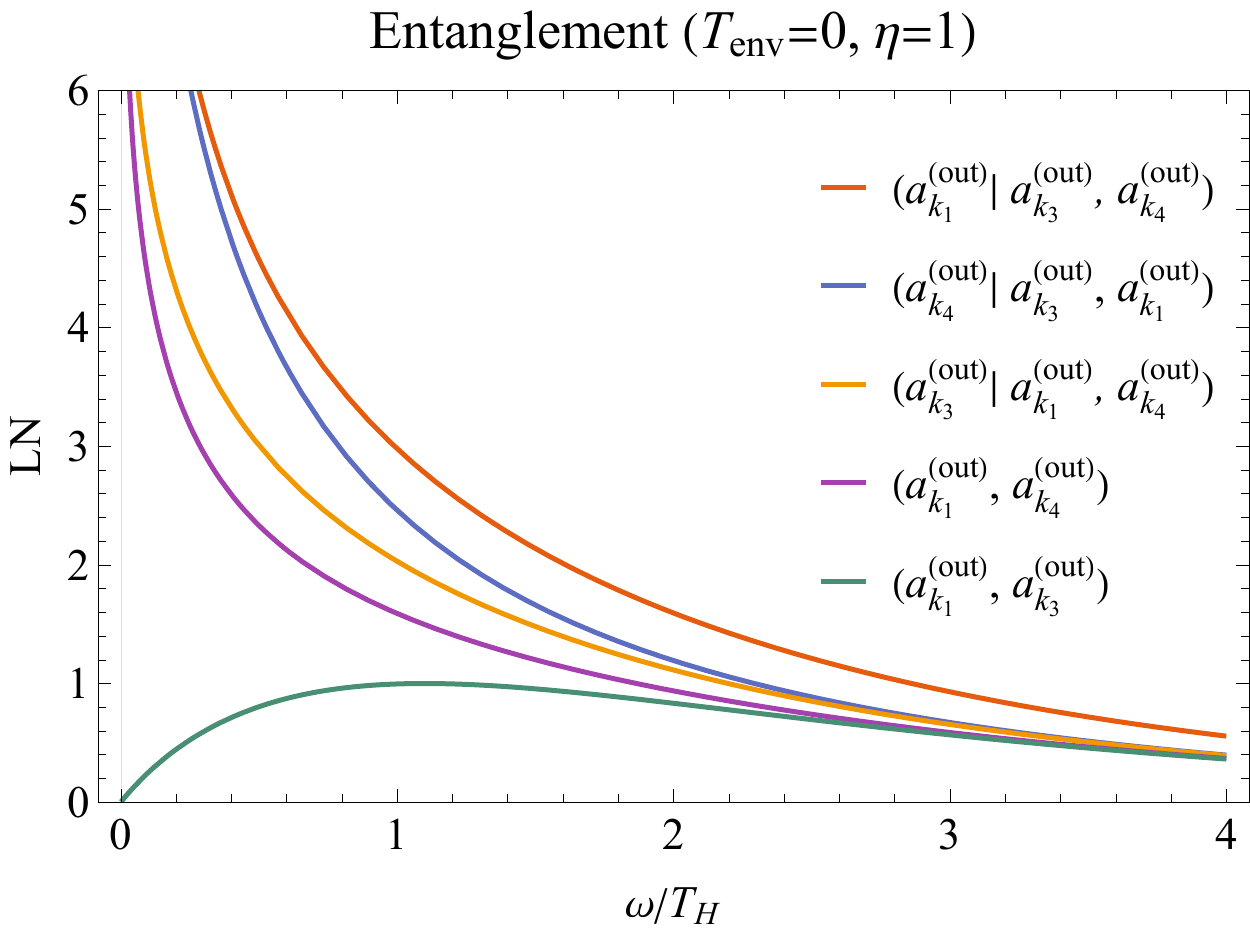}
    \caption{Entanglement predicted by the circuit of Fig.~\ref{fig:wbh_circuit} in all possible bi-partitions and mode-pairs for the out-modes of the analog white-black hole, as measured by the logarithmic negativity (LN); assuming only initial vacuum fluctuations are present and taking $\Gamma_{\omega}=1\,\forall\,\omega/T_H$.}
    \label{fig:modes_logneg}
\end{figure}

\subsection{Quantum correlations}\label{quantcorr}

We now investigate the quantum correlations generated by the circuit depicted in Fig.~\ref{fig:wbh_circuit}. We will show results for $\Gamma_{\omega}=1$ (negligible back-scattering), not only for pedagogical reasons, but also because in the numerical simulations of this systems shown below (see also \cite{agullo2022prl}), we find that $\Gamma_{\omega}$ is indeed very close to one for all frequencies of interest. For the squeezing intensity, as mentioned above, we use $\tanh^2{r_H}=\e^{-\omega/T_H}$, with $T_H$ a single number (i.e., independent of frequency). The output covariance matrix can be found from the symplectic elements introduced in the previous section, namely $\bm S_{\rm WB}\cdot \bm\sigma^{(\rm in)}_{\omega}\cdot \bm S_{\rm WB}^\top$, with $\bm S_{\rm WB}$ given in Eq.~\eqref{eq:SWH}. We start by discussing vacuum input, for which $\bm\sigma^{(\rm in)}_{\omega}=\bm{I}_8$.

Due to the entangling effects of the two-mode squeezers, there is genuine multi-mode entanglement within the network. To assess the entanglement, we consider pairwise entanglement for a reduced set of modes as well as bi-partite entanglement for the entire system of modes. Since the scattering intensity is $\Gamma_{\omega}\approx 1$, the mode $k_2$ weakly mixes with the rest and does not get entangled with the other modes. We are left with $k_1$, $k_3$ and $k_4$, out of which there are 2 entangled mode-pairs corresponding to\footnote{Entanglement is contained only in mode pairs of opposite symplectic norm.} $(\hat a^{\rm (out)}_{k_1},\hat a^{\rm (out)}_{k_4})$ and $(\hat a^{\rm (out)}_{k_1},\hat a^{\rm (out)}_{k_3})$. There are also 3 multi-mode bi-partitions corresponding to $(\hat a^{\rm (out)}_{k_1}|\hat a^{\rm (out)}_{k_4},\hat a^{\rm (out)}_{k_3})$, $(\hat a^{\rm (out)}_{k_4}|\hat a^{\rm (out)}_{k_3},\hat a^{\rm (out)}_{k_1})$, and $(\hat a^{\rm (out)}_{k_3}|\hat a^{\rm (out)}_{k_1},\hat a^{\rm (out)}_{k_4})$. Figure~\ref{fig:modes_logneg} depicts the entanglement between all such mode-pairs and bi-partitions; the entanglement in the pair $(\hat a^{\rm (out)}_{k_4},\hat a^{\rm (out)}_{k_3})$ is identically zero (see below for an explanation) and is thus not shown. [The circuit produces analytical expressions for all the curves shown in Fig.~\ref{fig:modes_logneg}, but they are lengthy and provide no real insight.] 

\begin{figure}[t]
    \centering
    \includegraphics[width=\linewidth]{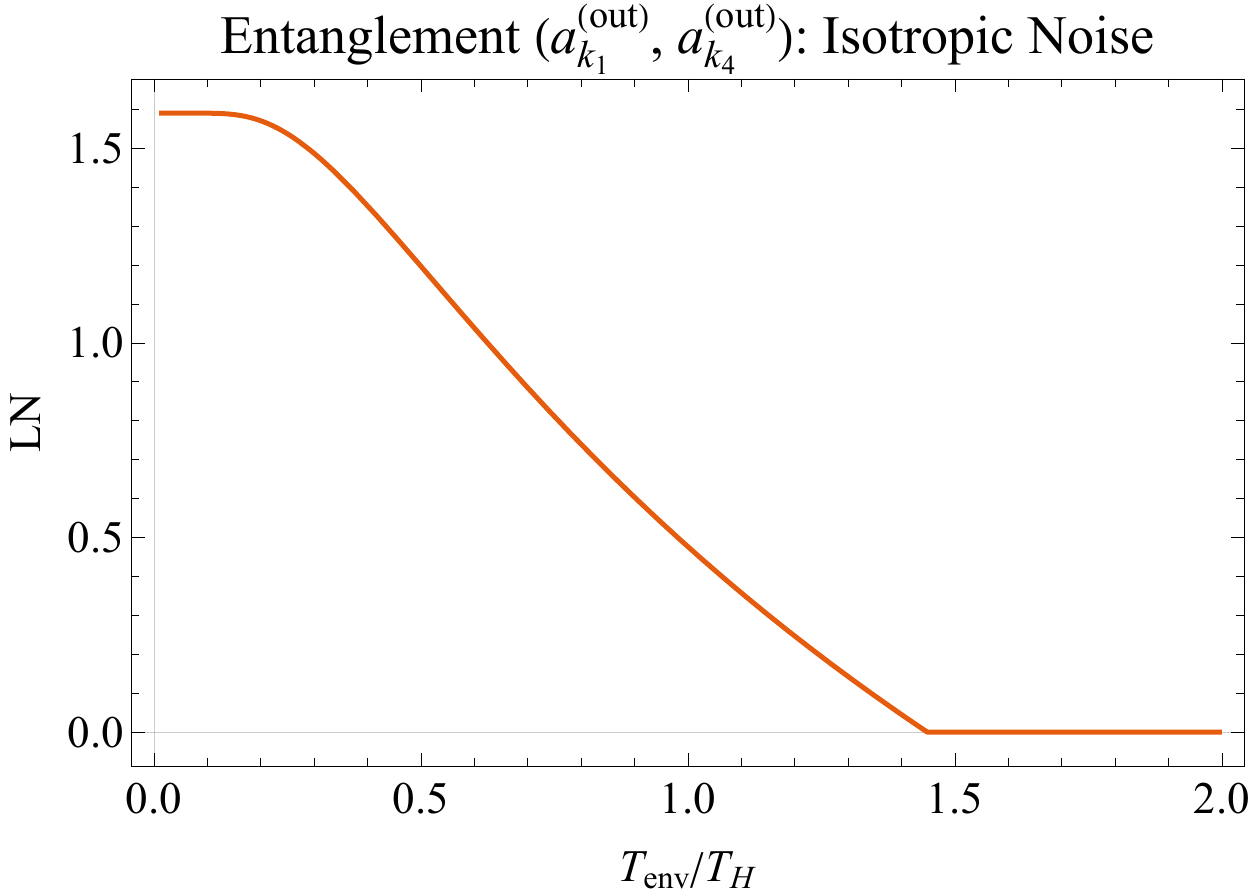}
    \caption{Entanglement between ``Hawking pairs'' emitted by the white hole horizon, $(\hat{a}_{k_1}^{(\rm out)},\hat{a}_{k_4}^{(\rm out)})$, at $\omega/T_H=1$ for varying environmental temperatures $T_{\rm env}/T_H$; assumes initial, isotropic thermal fluctuations of temperature $T_{\rm env}$ in the comoving frame for all modes.}
    \label{fig:ln14_therm}
\end{figure}

Note that the bi-partition $(\hat a^{\rm (out)}_{k_3}|\hat a^{\rm (out)}_{k_1},\hat a^{\rm (out)}_{k_4})$ physically represents the partition between the black hole and white hole exterior regions. By looking at the circuit, it is easy to understand that the entanglement across the black hole/white hole bi-partition is formally equivalent to that of a single two-mode squeezed vacuum state, with squeezing governed by the Hawking relation $\tanh^2{r_H}=\e^{-\omega/T_H}$. This is because the white hole squeezer entangles the modes $\hat a^{\rm (out)}_{k_1}$ and $\hat a^{\rm (out)}_{k_4}$, but this is a local transformation within one subsystem in the bi-partition $(\hat a^{\rm (out)}_{k_3}|\hat a^{\rm (out)}_{k_1},\hat a^{\rm (out)}_{k_4})$ and as such it does not alter the entanglement in the bi-partition (recall that entanglement is invariant under local transformations---where `local' here means within a subsystem of modes). Furthermore, Fig.~\ref{fig:modes_logneg} shows that the black hole/white hole bi-partition hosts the least amount of entanglement out of all 3-mode bi-partitions, which is due to the fact that the white hole squeezer does not contribute.

It is interesting to point out that, for vacuum input, the black hole horizon generically produces more entanglement than the white hole one. This is because, as it is manifest by simple inspection of the circuit, the white hole horizon receives thermal radiation emitted by the black hole in a flux of Hawking partners. Hence, the white hole emission is always stimulated by thermal radiation, and we know from the analysis of the previous section and also from Appendix \ref{sqzbs} that such thermal input degrades the production of entanglement. It is important to emphasize that this, as well as all the features regarding entanglement discussed above, can be easily understood by simple inspection of the circuit. More than that, the circuit provides analytical expressions for all these quantities. 

Generally, the entanglement between bi-partitions of the 3 modes is greater than or equal to the entanglement between the mode-pairs because the LN is an entanglement monotone~\cite{plenio05}. This is quantitatively manifest in Fig.~\ref{fig:modes_logneg}. Also, observe that the bi-partition with the greatest amount of entanglement is $(\hat a^{\rm (out)}_{k_1}|\hat a^{\rm (out)}_{k_4},\hat a^{\rm (out)}_{k_3})$, since the mode $\hat a^{\rm (out)}_{k_1}$ participates in each pair-creation mechanism this system experiences. The mode-pair with the largest amount of entanglement are the Hawking-pairs emitted from the white hole, $(\hat a^{\rm (out)}_{k_1},\hat a^{\rm (out)}_{k_4})$. Oddly, the maximal amount of entanglement between the original Hawking pair $(\hat a^{\rm (out)}_{k_1},\hat a^{\rm (out)}_{k_3})$ (the pair generated by the black hole) carries, at most, precisely 1 bit of entanglement,\footnote{The LN is in units of entangled bits or ebits.} which is maximal at about one Hawking wavelength. 

\begin{figure}[t]
    \centering
    \includegraphics[width=\linewidth]{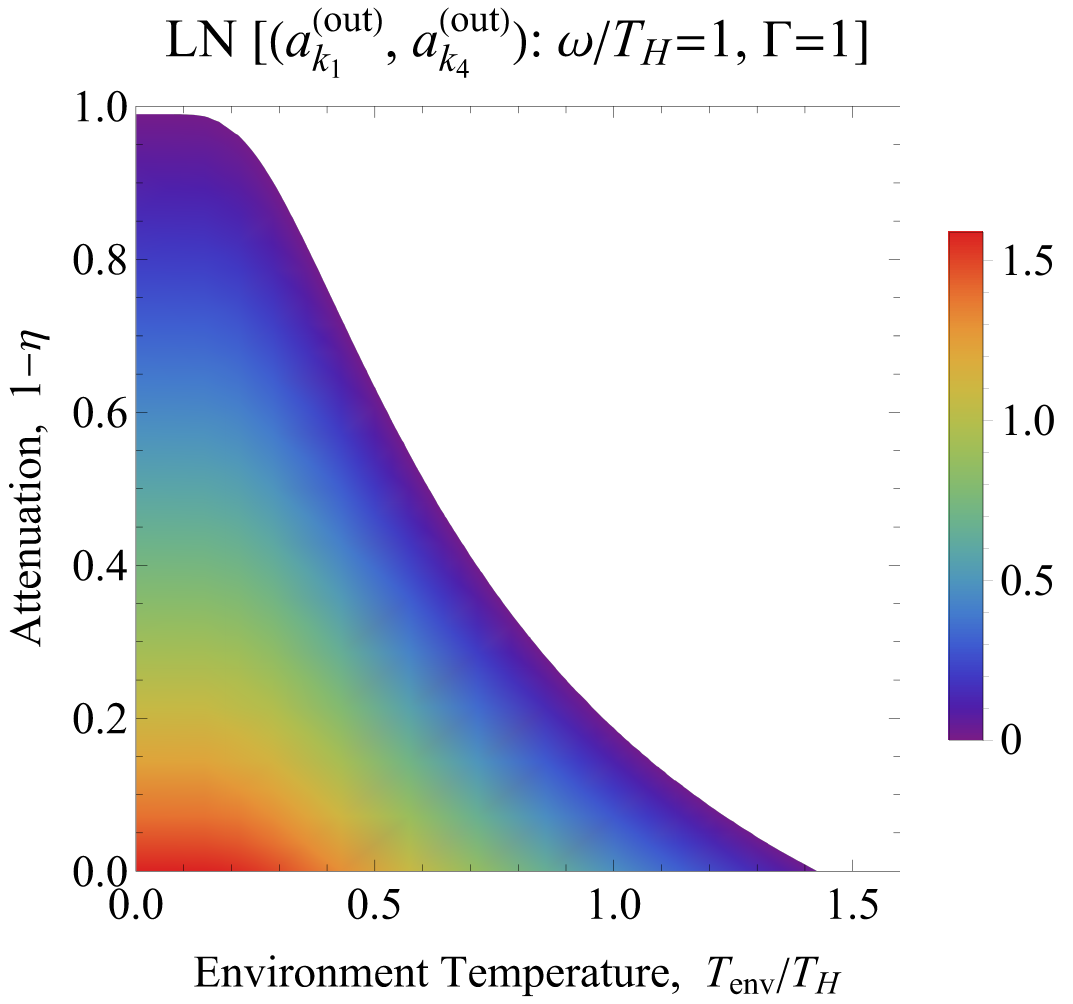}
    \caption{Entanglement between Hawking pairs emitted by the white hole horizon, $(\hat{a}_{k_1}^{(\rm out)},\hat{a}_{k_4}^{(\rm out)})$, at $\omega/T_H=1$ and $\Gamma=1$ for varying environmental temperatures $T_{\rm env}/T_H$ and attenuation, $1-\eta$; assumes initial, isotropic thermal fluctuations of temperature $T_{\rm env}$ in the comoving frame.}
    \label{fig:ln14_eta}
\end{figure}


We now include initial thermal fluctuations assuming attenuation is negligible. We consider (the unphysical but simplistic assumption of) isotropic thermal noise in the comoving frame---i.e., the same amount of noisy quanta in all input modes---and parameterize the thermal noise by a single temperature $T_{\rm env}$, which determines the number of noisy quanta by means of $n_{\rm env}=[{\rm exp}(\omega/T_{\rm env})-1]^{-1}$. We show in Fig.~\ref{fig:ln14_therm} the LN for the sub-system $(\hat a^{\rm (out)}_{k_1}|\hat a^{\rm (out)}_{k_4})$ (again, this is the Hawking-pair emitted by the white hole) versus $T_{\rm env}/T_H$ for $\omega/T_H=1$. In order to discuss one problem at a time, we consider no losses for the moment ($\eta=1$). The plot shows the fragility of entanglement in noisy environments. This fragility is actually frequency dependent, and we report in Table~\ref{table:table} the values of $T_{\rm env}$ for which the entanglement in different bi-partitions is lost, for low ($\omega/T_H\rightarrow\infty$) and high ($\omega/T_H\rightarrow0$) particle emission regimes.
In the low particle regime (large frequency), almost all the entanglement conditions collapse to $T_{\rm env}<2T_H$ [with the exception of the $(\hat a^{\rm (out)}_{k_1}|\hat a^{\rm (out)}_{k_4},\hat a^{\rm (out)}_{k_3})$ bi-partition], which is the entanglement-condition for a single horizon in BECs found by Bruschi et al ~\cite{bruschi2013}, and can also be deduced from the convergence of curves in Fig.~\ref{fig:modes_logneg} as $\omega/T_H\rightarrow\infty$. This simple condition is thus a good heuristic proxy for entanglement in bosonic analogue-gravity systems, at least in the low-particle limit. The situation changes, however, as frequencies approach the typical Hawking frequency (high-particle limit), which one can observe by scanning the far-right column of Table~\ref{table:table} or the curves in Fig.~\ref{fig:modes_logneg} as $\omega/T_H\rightarrow0$. We stress, however, that these results correspond to an isotropic temperature \textit{in the comoving frame} of the white-black hole. For optical analogs, the conclusions drawn henceforth can drastically change due to large boosts between the lab frame and the comoving frame, as we discuss briefly below. 

Note that these results represent the best-case scenario, when attenuation is negligible ($\eta=1$), but such can be extended to include attenuation, leading to a family of critical conditions parameterized by $(1-\eta,T_H, \omega)$, similar to the black hole case discussed in Section \ref{subsec:correlations_bh}. This information is shown in Fig.~\ref{fig:ln14_eta} for the entanglement in the Hawking pair emitted by the white hole, $(\hat a^{\rm (out)}_{k_1},\hat a^{\rm (out)}_{k_4})$, for $\omega/T_H=1$ and $\Gamma=1$.

\begin{table*}[t]
		\renewcommand{\arraystretch}{2}
		\centering
		\begin{tabular}{c c c}
			\hline\hline Sub-systems &$(\omega/T_H\rightarrow \infty)$ & $(\omega/T_H\rightarrow 0)$\\ \hline
			$(\hat a^{\rm (out)}_{k_1}|\hat a^{\rm (out)}_{k_3})$  & $T_{\rm env}<2T_H$ & 0
			\\ 
			$(\hat a^{\rm (out)}_{k_1}|\hat a^{\rm (out)}_{k_4})$  & $T_{\rm env}<2T_H$ & $T_{\rm env}<T_H$
			\\ 
			$(\hat a^{\rm (out)}_{k_3}|\hat a^{\rm (out)}_{k_1},\hat a^{\rm (out)}_{k_4})$ & $T_{\rm env}<2T_H$ & $T_{\rm env}<2T_H$ \\
			$(\hat a^{\rm (out)}_{k_4}|\hat a^{\rm (out)}_{k_3},\hat a^{\rm (out)}_{k_1}) \ \ $   & $T_{\rm env}<2T_H$ &\hspace{.5em} $T_{\rm env}<4T_H(\frac{T_H}{\omega})$ \\
			$(\hat a^{\rm (out)}_{k_1}|\hat a^{\rm (out)}_{k_4},\hat a^{\rm (out)}_{k_3}) \ \ $   & $T_{\rm env}<\frac{2T_H}{1-\ln(2)T_H/\omega}$ &\hspace{.5em} $T_{\rm env}<4T_H(\frac{T_H}{\omega})$
			\\
			\hline\hline
		\end{tabular}
\caption{Necessary criteria for entanglement between different sub-systems of out-modes for the white-black hole, in different particle-number regimes, ($\omega/T_H\rightarrow\infty$) and ($\omega/T_H\rightarrow0$). Assuming initial, homogenous thermal fluctuations at temperature $T_{\rm env}$ in the comoving frame and no loss ($\eta=1$).}\label{table:table}
\end{table*}

A situation of practical interest is when the thermal bath is at rest in the laboratory frame. This case is more subtle since, on the one hand, the modes $\{k_i\}$ have different populations (due to having different lab frequencies; their comoving frequencies are the same) and, on the other hand, the large Lorentz boost between the lab and comoving frames implies that higher temperatures are needed for the thermal noise to appreciably affect  entanglement. We find that the threshold lab temperature $T^{\rm (lab)\,\star}_{\rm env}$ above which entanglement is extinguished is $T^{\rm (lab)\,\star}_{\rm env}/T_{\rm H}\sim \order{10^3}$---significantly higher than the Hawking temperature, as qualitatively alluded to in previous studies of optical analogs~\cite{philbin08}. This is a subtle and distinctly interesting point that warrants focused attention, which we leave for future study.

\subsection{Accurateness of the white-black hole circuit}

The circuit of Fig.~\ref{fig:wbh_circuit} has been proposed using physical arguments resting on the analogy with the Hawking effect in gravitational black holes. A legitimate question is whether this circuit actually captures correctly the physics of optical white-black hole pairs. The goal of this section is to quantitatively address this query. Optical white-black hole pairs rest on complex physics; even in situation where the effect of higher order non-linearities can be neglected, the dispersive character of these systems may introduce additional complications, which can break the analogy with the (gravitational) Hawking effect. Our circuit has been built assuming the analogy works well. Therefore, any deviation from the predictions of the circuit are actually signaling a break down of the analogy between the physics of these optical systems and the Hawking effect. We show in this subsection that there is a region in the parameter space where the analogy with the Hawking effect is on solid ground and our circuit describes the system with great accuracy. We also find regions in the parameter space where dispersive effects lead to a breakdown of the analogy with the Hawking effect and identify physical origins of such.

The analysis of this subsection rests on a numerical simulation of the optical systems described above (see \cite{rubino2012soliton,petev2013blackbody,Gaona-Reyes:2017mks,Moreno-Ruiz:2019lgn,macher2009,Busch:2013gna,Finazzi:2010yq,Finazzi:2011jd,Finazzi:2012iu,Michel:2014zsa,Michel:2016tog} for previous numerical efforts in similar systems). Our simulations are based on the microscopic analytical model proposed in Ref.~\cite{linder16}, building on previous work \cite{belgiorno15}, and based on the Hopfield model in which the dielectric material is modeled by a collection of oscillators~\cite{hopfield1958}. The interaction with a strong electromagnetic pulse modifies the natural frequency of the oscillators, and this modification, in turn, is experienced by weak electromagnetic probes propagating thereon as a change in the effective index of refraction. As described in Ref.~\cite{linder16}, the dynamics of weak probes in this model is governed by a 4th order ordinary differential equation in the comoving coordinates (Eq.~(11) in Ref.~\cite{linder16}). For each (comoving) frequency $\omega$, we numerically solve this equation and compute the evolution of in-wavepackets which, asymptotically far from the horizons, are peaked on each of the four wavenumbers $k_1,k_2,k_3,k_4$ (these are the four solutions of the dispersion relations for a fixed $\omega$). We propagate these wavepackets and track the evolution to outgoing wavepackets propagating asymptotically far away from the strong pulse. From this, we obtain the $S$-matrix relating in- and out-modes. [A companion paper~\cite{paper3}, which goes over the details of this microphysical model and of the numerical simulations, is in preparation.]

In our numerical simulations, we model the perturbation of the refractive index as ${\delta n(x,t)=\delta n_0\,\sech^2\left(\frac{t-x/u}{D}\right)}$, a profile which is a common and convenient choice~\cite{philbin08, drori19}, where $u$ is the group velocity of this perturbation, and $x$ and $t$ are space-time coordinates in the lab frame. This profile is parameterized by two real positive numbers, $\delta n_{0}$ and $D$; they determine the amplitude and width of the perturbation, respectively. 
We have performed simulations for $\delta n_{0}$ in the interval $0.01$ to $0.1$, and $D$ ranging from $2$ fs to $10$ fs, respectively. These ranges are chosen by demanding that analog white and black hole event horizons are present even for the fastest mode (since the dispersion relation of this system is sub-luminal, low frequency modes are faster). This is also the range of parameters for which the analogy with the Hawking effect is closer, as we discuss below.

Our numerical calculations are independent of the circuit of Fig.~\ref{fig:wbh_circuit}. The inputs of our numerics are the parameters $\delta n_{0}$ and $D$, and the output is an $8\times8$ matrix of numbers representing the $S$-matrix (there is an $S$-matrix for each value of the frequency $\omega$). We say that the circuit describes well the physics of the system when there exist values of the free parameters of the circuit, $r_H(\omega)$ and  $\theta_{\omega}$ 
for which the analytical $S$-matrix associated with the circuit predicts values of physical quantities of interest that are equal to those obtained from the numerical $S$-matrix. This is a non-trivial demand. A symplectic evolution of four modes generically requires 36 free parameters (this is the dimension of the symplectic group in four dimensions), though this can be reduced to 16 using information from the dispersion relation (see Appendix~\ref{app:counting} for more discussion on this)---whereas our circuit ansatz contains only three free parameters. 

We do not expect, however, our circuit to describe every single aspect of the dynamics since, as discussed above, we have not included in the circuit, e.g., ``single-mode phase shifters'' to account for possible phases that different modes acquire during the propagation. In other words, we expect our circuit to describe the system ``up to phases''. It is not difficult to include these phases, but we consider unnecessary to add this complication for the following reason. We have found that the quantities we are interested in---namely, the mean number of quanta in each out-mode and the entanglement between modes---are insensitive to phases. In particular, none of the quantities we plot below depends on, for instance, the squeezing phase $\phi$ in our circuit. Hence, we will leave this phase unspecified. 

Keeping this in mind, to compare both calculations we will focus attention on their physical predictions, rather than merely comparing the $S$-matrices. We proceed as follows. For each frequency $\omega$, we determine the two circuit-parameters---$r_H(\omega)$ and $\theta_{\omega}$---by demanding that two components of the circuit $S$-matrix matches the same components of the numerical $S$-matrix. More concretely, we obtain $\theta_{\omega}$ from the component (3,3) of the numerical $S$-matrix, which, according to the circuit, is equal to $\cos^2\theta_{\omega}$ [see Eq.~\eqref{eq:SWH}]; 
$r_H(\omega)$ is then obtained from the component (7,5), which corresponds to $\cos\theta_{\omega}\, \cosh r_H(\omega)$, according to the circuit. In this way, we determine $r_H(\omega)$ and $\theta_{\omega}$ (we repeat the the calculation for each value of $\omega$ in the regime of interest). Next, we substitute these values into the $S$-matrix written in Eq.~\eqref{eq:SWH}, and use it to evaluate quantities of interest. The values for physical quantities obtained in this way are then compared with the results derived with the numerically computed $S$-matrix. If both calculations agree, we say the circuit captures the correct physics. We insist that this is a non-trivial check, since the circuit-calculation uses inputs from just two-components of the numerical $S$-matrix, while we demand the two calculations to agree in quantities which depend on complicated combinations of all components. This summarizes the strategy we follow. We now show the results. 

\begin{figure}[t]
    \centering
    \includegraphics[width=\linewidth]{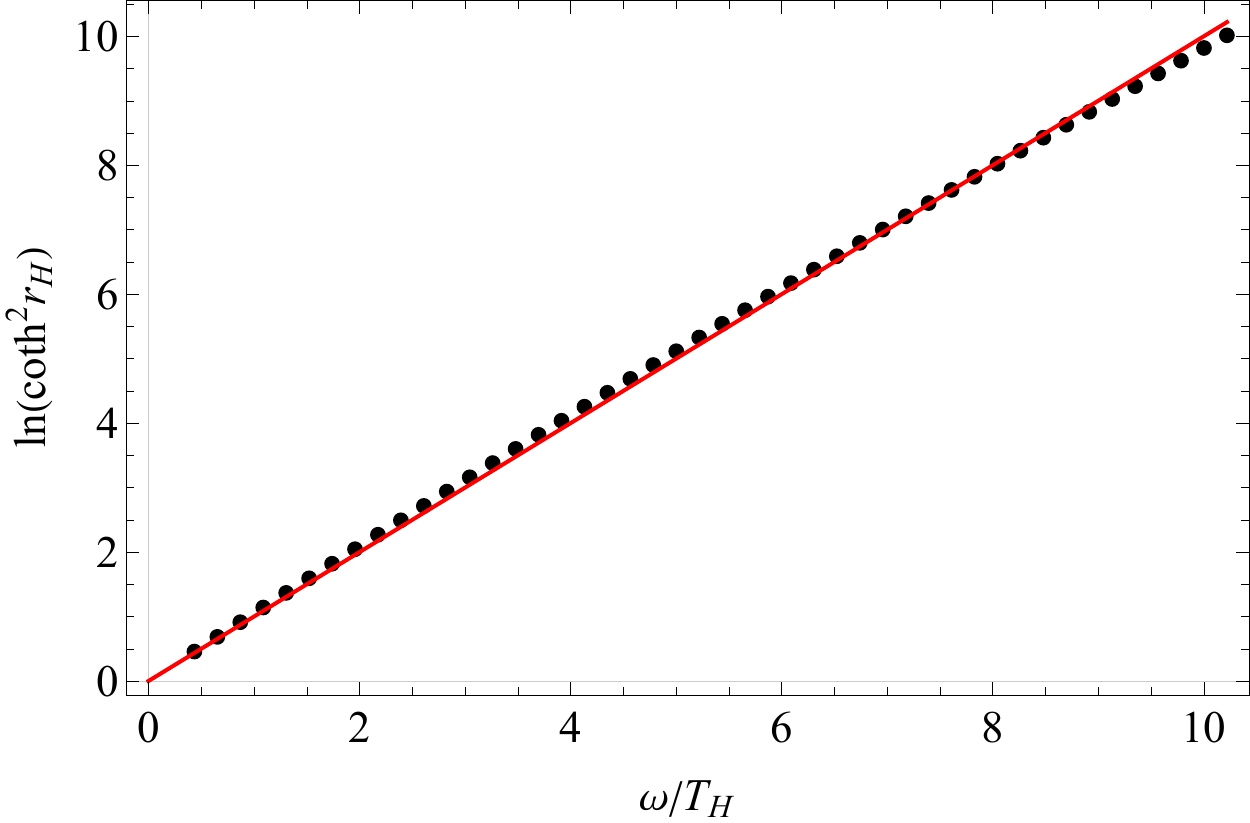}
    \caption{The dots correspond to the numerically determined values of $\ln(\coth^2 r_H)$ versus $\omega$, for a strong pulse with $D=6\, \rm{fs}$ and $\delta n_0=0.05$. We also show a straight line, $\omega/T_H$, which is a good fit to the data for $T_H=3.51\, {\rm K}$. This plot shows that the squeezers representing each horizon produce radiation with a frequency spectrum in good agreement with a black-body distribution at temperature $T_H=3.51\, {\rm K}$. Stronger deviations from thermality emerge at high frequencies.}
    \label{fig:cothrvsw}
\end{figure}

\begin{figure}[t]
    \centering
    \includegraphics[width=\linewidth]{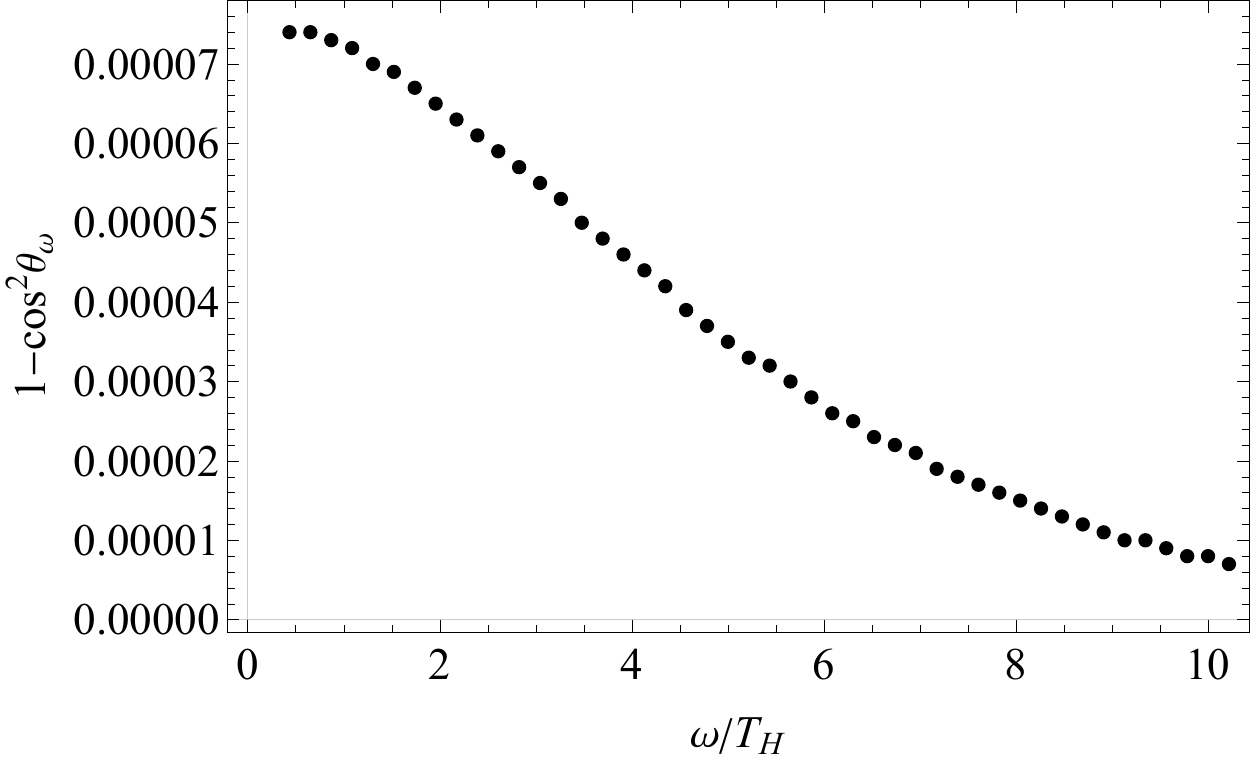}
    \caption{Numerically determined value of $1-\cos^2\theta$ versus $\omega/T_H$, for a strong pulse with $D=6\, \rm{fs}$ and $\delta n_0=0.05$. The plot shows that the transmission probability of the beam-splitters ($\cos^2\theta$) of the circuit is very close to one. This implies the mode $k_2$ weakly interacts with the other modes in the system.}
    \label{fig:costheta}
\end{figure}

First, we plot the two-circuit parameters $r_H(\omega)$ and $\theta_{\omega}$ versus frequency $\omega$ (in this simulation we use $D=6 \, {\rm fs}$ and $\delta n_0=0.05$). More concretely, Figs.~\ref{fig:cothrvsw} and \ref{fig:costheta} show $\ln(\coth^2 r_H)$ and $1-\cos^2\theta$, respectively. On the one hand, $\ln(\coth^2 r_H)$ helps us to evaluate whether the squeezers associated with both white and black hole horizons emit radiation with a black-body frequency spectrum (as expected from the analogy with the Hawking effect) since, in such a situation, $r_H$ must depend on frequency as $r_H= {\rm arcoth}\, (e^{\omega/2 T_H})$, with $T_H$ a frequency independent real number. This in turn implies that $\ln(\coth^2 r_H)=\omega/T_H$ is a linear function of $\omega$. Figure~\ref{fig:cothrvsw} confirms that this is in fact the case to a great approximation. From this plot, we can read the value of the Hawking temperature, from which we obtain $T_H=3.51 \, {\rm K}$. This value agrees with analytical approximations derived in Ref.~\cite{linder16} within a few percent.

The quantity $\cos^2\theta$ informs us about the transmission probability of the potential barrier (recall, $\Gamma_{\omega}=\cos^2\theta_{\omega}$). Fig.~\ref{fig:costheta} shows that $\cos^2\theta$ is very close to one for all frequencies in the range of interest, and it is closer to one for large $\omega$. Since the beam splitters in the circuit are the means by which the mode $k_2$ interacts with the rest, these results tell us that $k_2$ evolves almost entirely in solitude, with little interaction with the rest of modes. [Physically, this is because the wavelength of this mode in the lab frame is much larger than the rest, and therefore this mode probes the system at different scales.] 

\begin{figure*}[t]
    \centering
    \hfill\includegraphics[width=.485\linewidth]{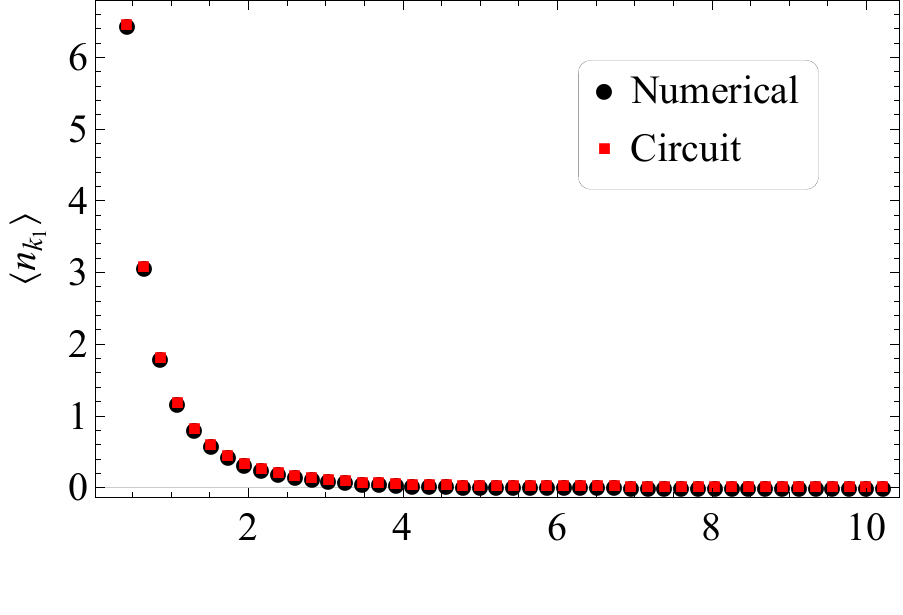}\hfill
     \includegraphics[width=.495\linewidth]{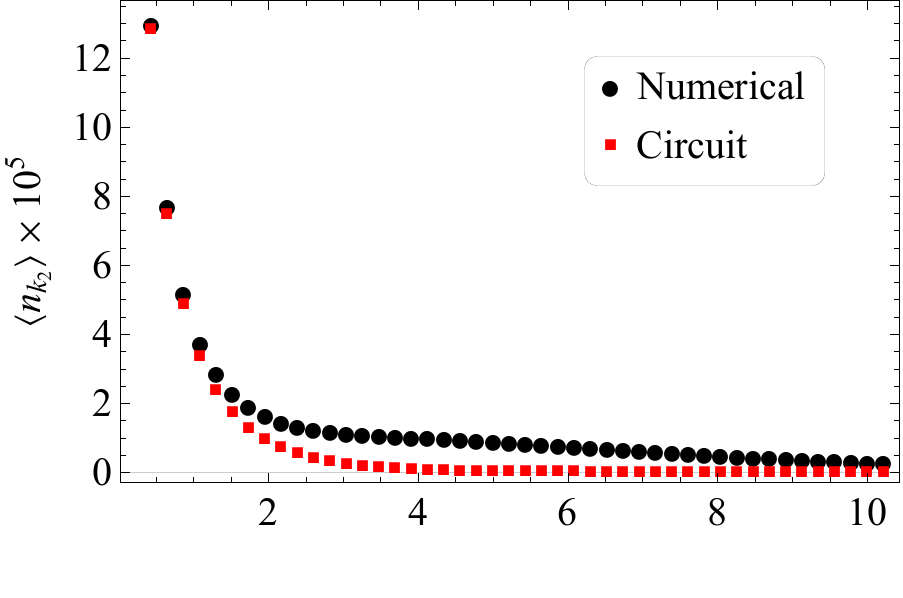}
      \includegraphics[width=.495\linewidth]{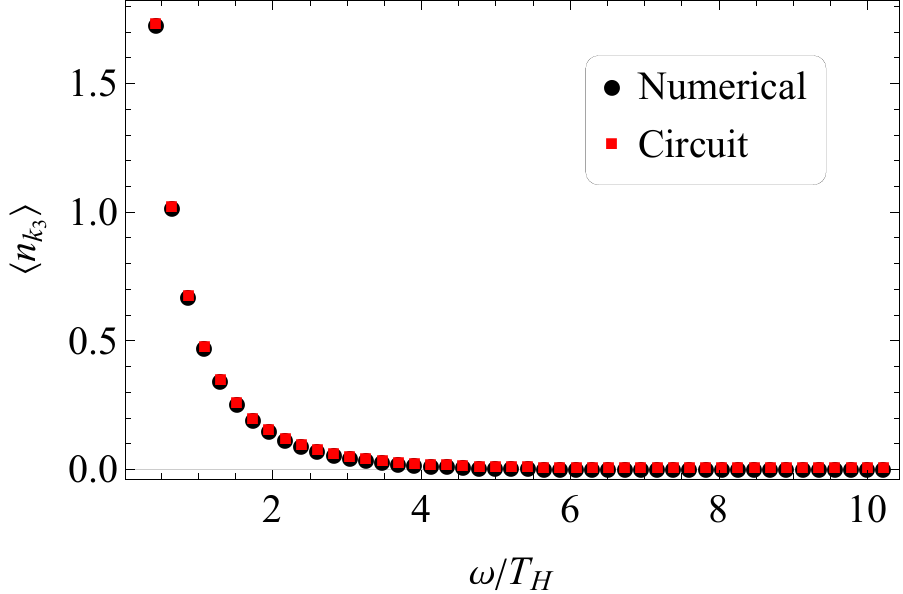}\hfill
       \includegraphics[width=.485\linewidth]{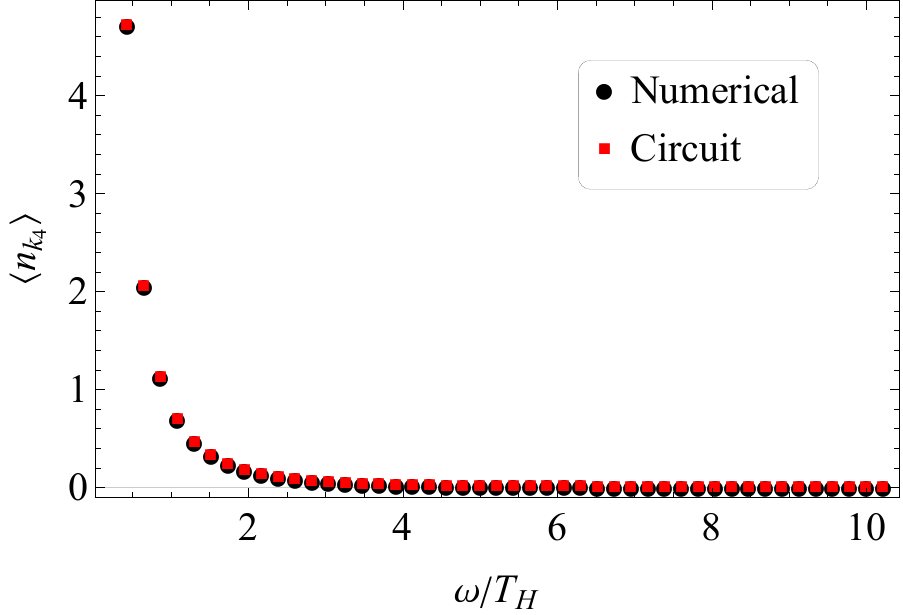}\hfill
    \caption{Expectation values of the number operator of the out-modes $k^{\rm(out)}_1$, $k^{\rm(out)}_2$, $k^{\rm(out)}_3$, and $k^{\rm(out)}_4$ resulting from evolving the vacuum state (spontaneous creation of particles), for a strong pulse with $D=6\, \rm{fs}$ and $\delta n_0=0.05$. The red squares represent the predictions of our circuit, while the black dots correspond to the results of our numerical code. The number of quanta produced in the mode $k^{\rm(out)}_2$ is several order of magnitude smaller than in the rest of modes. We observe great agreement between the predictions of the circuit and our numerical code for the modes $k^{\rm(out)}_1$, $k^{\rm(out)}_3$ and $k^{\rm(out)}_4$, which are the ones involved in the Hawking process. Discrepancies appear in the mode $k^{\rm(out)}_2$; the circuit underestimates the quanta produced in this mode. This indicates a (weak) coupling between the modes $k^{\rm(out)}_1$ and $k^{\rm(out)}_2$ not captured by the circuit.}\label{fig:number}
\end{figure*} 

Next, we compare the predictions for the mean number of quanta in the out-modes $\hat{a}^{\rm(out)}_{k_1}$, $\hat{a}^{\rm(out)}_{k_2}$, $\hat{a}^{\rm(out)}_{k_3}$, and $\hat{a}^{\rm(out)}_{k_4}$, starting from vacuum input. This is shown in different panels of Fig.~\ref{fig:number}. A few interesting messages emerge from these plots. The number of quanta produced in the mode $\hat{a}^{\rm(out)}_{k_2}$ is much smaller than the rest, by about four orders of magnitude. This confirms that the mode $k_2$ is a spectator in this process. The symplectic circuit also produces results in great agreement with the numerical code for the number of quanta in the modes $\hat{a}^{\rm(out)}_{k_1}$, $\hat{a}^{\rm(out)}_{k_3}$ and $\hat{a}^{\rm(out)}_{k_4}$ (differences are of order of one part in thousand, or smaller). However discrepancies do appear for the mode $\hat{a}^{\rm(out)}_{k_2}$. 

\begin{figure*}[t]
    \centering
    \includegraphics[width=.48\linewidth]{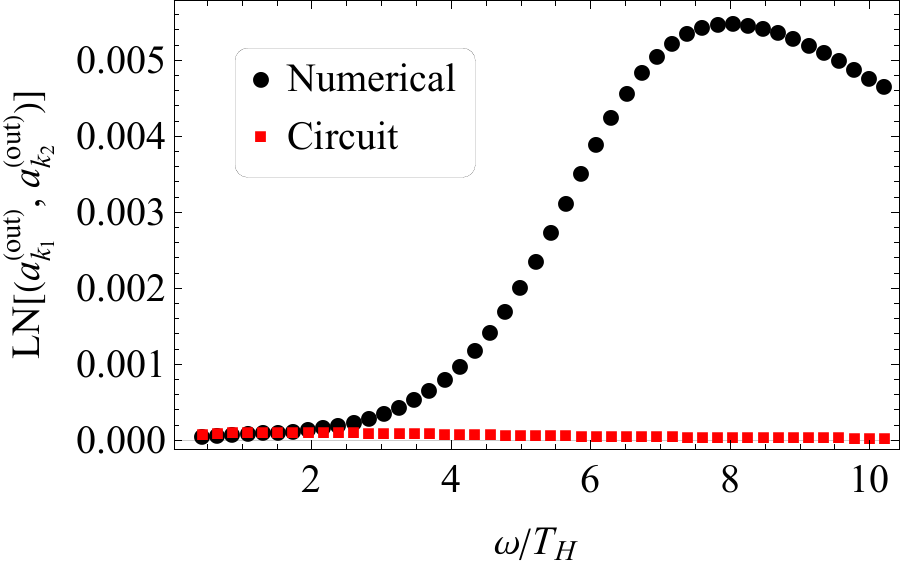}
    \hfill
     \includegraphics[width=.48\linewidth]{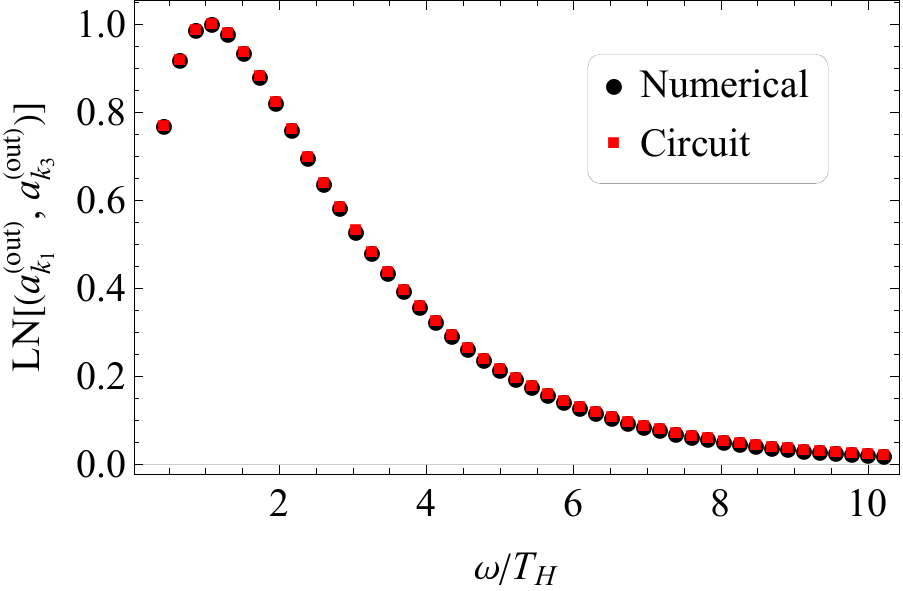}
     {\includegraphics[width=.48\linewidth]{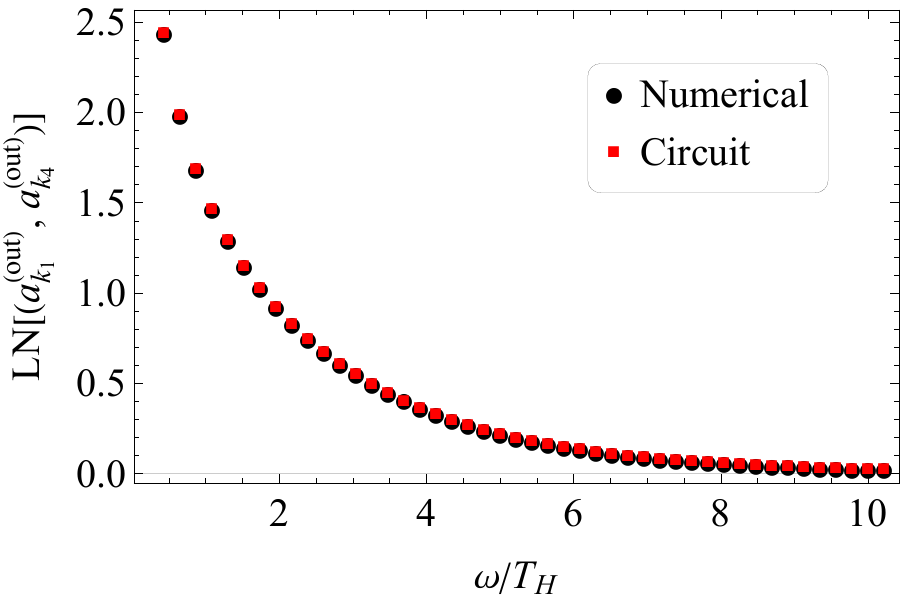}}
    \caption{Logarithmic negativity between pairs of modes in the out-state corresponding to vacuum input, for a strong pulse with $D=6\,\rm{fs}$ and $\delta n_0=0.05$. The red squares represent the predictions of our circuit, while the black dots correspond to the results of our numerical code. The plot shows great agreement for the pairs ($k^{\rm(out)}_1$, $k^{\rm(out)}_3$) and ($k^{\rm(out)}_1$, $k^{\rm(out)}_4$), which are the pairs that dominate the entanglement in the final state. The value of LN for the pair ($k^{\rm(out)}_1$, $k^{\rm(out)}_2$) is several orders of magnitude smaller and is not well captured by our circuit.}\label{fig:LN}
\end{figure*}

We show in Figs.~\ref{fig:LN} the entanglement between pair-wise out-modes. First of all, such entanglement is exactly zero unless the mode $\hat{a}^{\rm(out)}_{k_1}$ is a member of the pair. This is true both in our circuit as well as in our numerical calculations, and it has a simple explanation: entanglement can only be generated if modes with different symplectic norms interact. The mode $\hat{a}_{k_1}$ is the only one having negative symplectic norm~\cite{linder16}. The remaining modes have positive norm (which is a consequence of the the dispersion relation of the system \cite{linder16}). Our circuit respects this fact, since both squeezers involve a positive- and a negative-norm mode. Hence, in Figs.~\ref{fig:LN}, we only show the non-zero pair-wise entanglements. We see that the entanglement in the pairs $(\hat{a}^{\rm(out)}_{k_1},\hat{a}^{\rm(out)}_{k_3})$ and $(\hat{a}^{\rm(out)}_{k_1},\hat{a}^{\rm(out)}_{k_4})$ shows again a great agreement between the outcomes of the circuit and the numerical calculations. The entanglement in the pair $(\hat{a}^{\rm(out)}_{k_1},\hat{a}^{\rm(out)}_{k_2})$ predicted by our circuit is several orders of magnitude too small.

The discrepancy in the predictions involving the mode $\hat{a}^{\rm(out)}_{k_2}$ for both the number of quanta and entanglement, have a common origin---namely, the absence of a direct coupling between the modes $\hat{a}^{\rm(out)}_{k_1}$ and $\hat{a}^{\rm(out)}_{k_2}$ in our circuit. Since these modes have symplectic norms of different signs, such coupling would imply a source of particle-pair production, which would account for the extra quanta in the mode $\hat{a}^{\rm(out)}_{k_2}$ shown in the numerical simulation, as well as the extra entanglement between the modes $\hat{a}^{\rm(out)}_{k_1}$ and $\hat{a}^{\rm(out)}_{k_2}$. However, the fact that our circuit captures very well the behavior of the mode $\hat{a}^{\rm(out)}_{k_1}$ implies that such coupling is very weak and does not significantly affect the rest of the modes in the system. [For very large frequencies, $\omega \gg T_H$, this coupling has a larger relative effect, since the number of quanta and entanglement in the rest of modes fall of exponentially with $\omega$, due to its black-body character, while the coupling of the mode $k_2$ with the rest, falls at a slower rate.] It is not difficult to modify our circuit to account for the missing coupling between modes $\hat{a}^{\rm(out)}_{k_1}$ and $\hat{a}^{\rm(out)}_{k_2}$, by simply adding an extra squeezer. Though, we find this additional complication unnecessary, since such coupling is unrelated to the Hawking effect and, more importantly, it produces negligible effects for the modes actually involved in the Hawking process for frequencies in the range $\omega\in (0, 10 \,T_H)$. A similar dispersive effect was previously identified in \cite{Corley:1996ar}, in a different system.

Our numerical simulations reveal another dispersive effect, which indicates a second departure from the analogy with the Hawking effect and the predictions of our circuit. We identify the physical origin of this effect as being due to the tunneling of the mode $k_3$ (the ingoing Hawking mode) from outside the white hole to the exterior of the black hole horizon---hence, indicating that these are not perfect event horizons. This limitation is intrinsic to this (and other) analog models, and originates from the fact that, in between horizons, there actually exist modes propagating in both directions, but the modes propagating from the white hole to the black hole have complex wavenumbers \cite{Corley:1996ar,linder16}. [These modes are not drawn in Fig.~\ref{modestructure}]. They are evanescent modes (not propagating modes), in the sense that their amplitude falls off exponentially in the interior region. If the pulse generating the horizons is narrow enough, a significant portion of the mode can emerge on the black hole side. This is not very different from the familiar tunneling effect in wave mechanics. From general arguments, this effect is expected to be more important for narrow (small $D$) and weaker pulses (small $\delta n_0$), as well as for long wavelength modes. In agreement with this expectations, we observe this effect only for the mode $k_3$, since this is the lone, longest wavelength mode propagating to the right (with the conventions of Fig.~\ref{modestructure}) and for low frequencies. This effect is sizable already for $D=4$ fs, and $\delta n_0=0.01$ (although only for very low frequencies) and quickly disappears when these two parameters are increased. 

\begin{figure}[t]
    \centering
    \includegraphics[width=\linewidth]{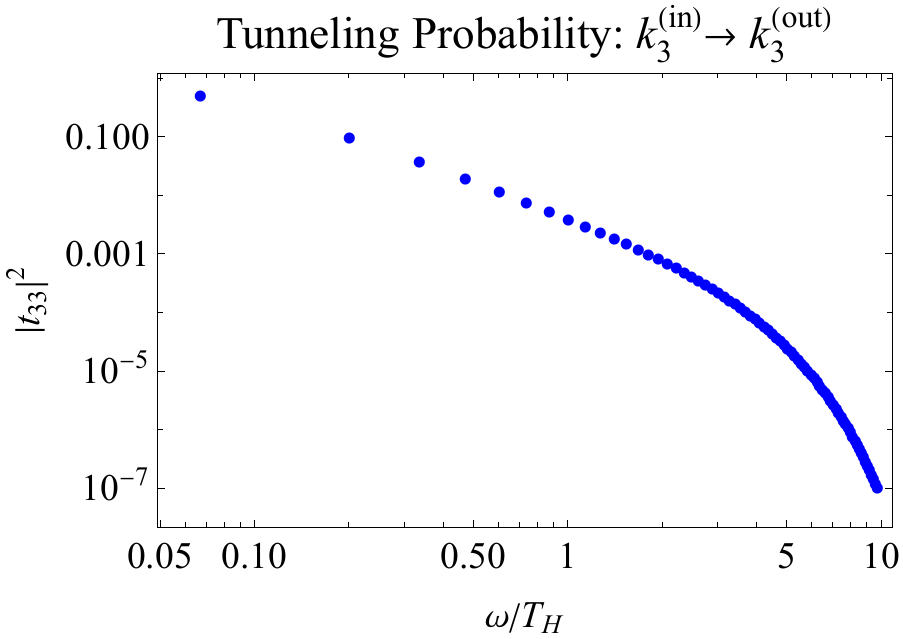}
    \caption{Tunneling probability, $\abs{t_{33}}^2$, across the white-black hole. In the ideal case, when the analogy to gravity is firm, the probability to tunnel through the white-black hole is zero and the mode $k_3^{(\rm out)}$ (the outgoing Hawking radiation from the black hole) has no contribution from the mode $k_3^{(\rm in)}$ (the ingoing mode to the white hole). However, as the height of the strong pulse, which generates the effective horizons, decreases and the width narrows, transmission through the white hole horizon increases. The parameters of the strong pulse used in this figure are $\delta n_0=.01$ and $D=4$ fs. An (approximate) Hawking temperature can still be found for these parameter values, $T_H\approx 1.14$ K.}
    \label{fig:tunnel}
\end{figure}

Tunneling by the $k_3$ mode (from the white hole exterior to the black hole exterior) obviously breaks the analogy with the Hawking effect, since nothing should emerge from the interior of an actual black hole. Not surprisingly, this effect also produces significant differences between the predictions of our circuit and the numerical simulations. These differences manifest primarily in a non-zero tunneling probability, $\abs{t_{33}}^2$, from $k_3^{(\rm in)}$ to $k_3^{(\rm out)}$ (which is identically zero in the perfect analog case; see Fig.~\ref{fig:wbh_circuit}). In Fig.~\ref{fig:tunnel}, we plot the $\abs{t_{33}}^2$ for a pulse characterized by $D=4\, \rm{fs}$ and $\delta n_0=0.01$. The tunneling probability is higher for low-frequency modes ($\omega/T_H\lesssim 1$), as expected on physical grounds\footnote{Low frequency modes have longer wavelengths, and thus the pulse is relatively narrower for such modes.}, and is non-negligible ($\abs{t_{33}}^2\sim.001-.1$) for this choice of pulse parameters. As a comparison, for $D=6\, \rm{fs}$ and $\delta n_0=0.05$, where the analogy to the Hawking effect holds well, we find $\abs{t_{33}}^2\lesssim10^{-6}$ in the frequency range of interest. We are currently developing a more detailed analysis of tunneling, which will appear in a later paper~\cite{paper3}, as such is prominent in realistic optical platforms~\cite{philbin08,rubino2012soliton,petev2013blackbody,drori19}.

In summary, for a strong pulse with $D=6\, \rm{fs}$ and $\delta n_0=0.05$ (or larger), we find that our circuit accounts very well (with differences of order of one part in thousand, or smaller) for the evolution of the modes actually involved in the Hawking effect, namely modes $k^{\rm(out)}_1$, $k^{\rm(out)}_3$ and $k^{\rm(out)}_4$. The analogy with the Hawking effect is on firm ground. The mode $k^{\rm(out)}_2$ interacts very weakly with the rest. We have identified two dispersive effects which break the analogy with the Hawking effect and induce deviations from the predictions of our circuit. On the one hand, dispersive effects produce a coupling between the mode $k^{\rm(out)}_1$ and $k^{\rm(out)}_2$ which induce production of pairs of quanta in the modes $k^{\rm(out)}_1$ and $k^{\rm(out)}_2$. Such production is not thermal and is unrelated to the Hawking effect, and it can be neglected for low frequencies $\omega <10\, T_H $. On the other hand, if the strong pulse producing the horizon is too weak or too narrow (approximately under $D=6\, \rm{fs}$ and $\delta n_0=0.05$), the white- and black hole horizons are imperfect and modes can propagate in both directions, breaking the analogy. These are intrinsic limitations caused by the dispersion-relation underlying this---and other---analog systems. It is worth mentioning that these threshold values for $D$ and $\delta n_0$, at which tunneling becomes relevant, depend on the properties of the material.



\section{Entanglement enhancement via the stimulated effect}\label{sec:nonclassical_assist}
\subsection{General remarks}
As discussed above, practical constraints---such as initial thermal fluctuations and attenuation---are a detriment to the quantum correlations generated during the Hawking process. These are serious hurdles for experimental platforms to overcome, in order to genuinely observe the quantumness of the Hawking process. Of course, one can always seek to improve the operating conditions (low $T_{\rm env}$ and $\eta\approx1$). However, this requires operating at temperatures near the Hawking temperature of the system, which is generally quite low, as well as having control of unwanted scattering channels (i.e., increasing efficiency of the setup). An alternative way to overcome noise is to utilize quantum resources at the inputs. We discuss this possibility for two specific strategies:

\begin{enumerate}[label=(\roman*)]

\item \textit{Entanglement resonance:} One can, in some sense, revitalize entanglement in the face of noise by injecting a controllable two-mode squeezed-vacuum state into the analog system, which is phase-matched (in resonance) with the Hawking process. The resonance condition allows for the entanglement in each source (the input and the Hawking process, respectively) to add constructively, thus boosting the total amount of entanglement in the output. This can be used to bypass the temperature conditions discussed previously. The proviso is that one must be able to phase-match the controllable two-mode squeezer to the Hawking process of the analog system in order to observe constructive effects; otherwise, one could induce destructive interference, potentially disentangling the particles. Another aspect to bear in mind here is the ability to distinguish between the entanglement generated by the input resource and the Hawking process; although this may be possible with sufficient control over the input resource and sufficient knowledge of the analog system. We do not explore this idea further, as such has already been discussed in the context of BECs in Ref.~\cite{bruschi2013}.  

\item \textit{Leverage a single-mode resource:} Likewise, injecting a single-mode squeezed vacuum into one of the modes can enhance the entanglement generated during the Hawking process and help to subdue background thermal noise~\cite{agullo2022prl}. The benefit here (contrasted with the entanglement-resonance scheme above) is that the single-mode squeezing approach does not rely on a phase-matching condition (the entanglement varies only with the amount of squeezing), and thus no fine-tuning is necessary. The drawback, however, is that this approach is more sensitive to losses. Nevertheless, we provide a quantitative assessment of this strategy in the context of an analog white-black hole in the next section, and discuss where it may be beneficial.
\end{enumerate}

The amplification of the entanglement in the stimulated Hawking process may come as a surprise at first, since the stimulated process is usually regarded as a purely classical effect. We insist that this intuition is true if the stimulation is done with a `classical' state, namely a coherent state or mixtures thereof. The covariance matrix of a coherent state is the identity---the same as for vacuum---and since entanglement is derived exclusively from the covariance matrix (with no reference to the first moments), entanglement in the output state cannot distinguish between initial vacuum or coherent input. In other words, by illuminating the system with a coherent state, we can amplify the output intensities, but there is no amplification of entanglement in the final state. This conclusion, however, ceases to be true if non-classical inputs are used to stimulate the system, even if the input is separable (not entangled). Appendix~\ref{sqzbs} shows an example of this mechanism for a simple two-mode squeezing interaction. 

Entanglement-enhancement strategies are promising, however we must also be careful of the conclusions we draw about the Hawking process itself when non-classical---even non-entangled---resources are in play. The reason being that there are passive operations that can be done on non-classical, \textit{separable} inputs which generate quantum entanglement. For instance, it is known that a network of passive elements (consisting only of orthogonal symplectic transformations, such as the beam-splitter discussed in Appendix \ref{sqzbs}) generally generates multi-mode entanglement starting from single-mode squeezed vacuum at the input \cite{kim02,jiang13}. A very simple example is the scattering of a single-mode squeezed vacuum by a potential barrier: the transmitted and the reflected beams are quantum mechanically entangled, even when the input is separable; the barrier is able to transform initial single-mode squeezing into two-mode entanglement. If we now consider the network as a blackbox, with elements unknown to us, would we deduce that the box is intrinsically quantum by nature, or is the quantumness of an output a feature solely of the input resource? 

Hence, is there a way to extract the quantum features of the output that are products of the Hawking process and not necessarily due to our input resources? One indication would be that the analog system works as an amplifier (creates particles) for the entangled modes under question and that the amount of entanglement increases with the number of particles generated by the system (holding everything else fixed). Thus, concurrently observing amplification of the modes as well as quantum correlations would provide support for the quantumness of the stimulated Hawking process in the presence of extrinsic non-classical resources. Though, more sophisticated methods for distinguishing such features may be desired in practice. Sufficient control of the inputs is thus a necessity.

\subsection{Example: Seeding with single-mode squeezed vacuum}\label{sec:sq_wbh}

We illustrate the strategy of stimulating the system with a single-mode squeezed vacuum with the example of the optical white-black hole discussed above. These results have been reported in \cite{agullo2022prl}; here we add further details omitted there and generalize the strategy to non-isotropic noise. Our strategy is to illuminate the white hole (i.e., populate the mode $a_{k_3}^{\rm(in)}$; see Fig.~\ref{modestructure}) with a single-mode squeezed-vacuum state and monitor the entanglement in the Hawking-pairs emitted by the white hole, i.e.\ the modes $a_{k_1}^{\rm(out)}$ and $a_{k_4}^{\rm(out)}$, as well as the output intensities. As we shall show, illuminating the white hole with a single-mode squeezed vacuum allows one to tune the output entanglement as a function of the input squeezing, and extract the symplectic-circuit parameters (squeezing amplitudes and beam splitter angles) of the white-black hole from the intensities and entanglement, even in the presence of thermal fluctuations and (a mild amount of) attenuation. Before moving forward, we emphasize again that such {\em entanglement enhancement} is not possible by stimulating the process with classical states, such as coherent states (or more generally, a convex combination of coherent states, e.g. a thermal state) for the reasons discussed in the previous subsection and in Appendix \ref{sqzbs}. [All results described below can be understood in the simpler context of a two-mode squeezer described in detail in Appendix~\ref{sqzbs}.]

The single-mode covariance matrix for the initially squeezed mode $a_{k_3}^{\rm(in)}$, written in terms of the squeezing strength $s$, is given by $N_{\rm env}\, \e^{2\, s\, \bm\sigma_z}$, with $\bm\sigma_z$ the familiar $z$-Pauli matrix, where we have assumed squeezing along the quadrature $P$ (and, consequently, anti-squeezing along $Q$ quadrature)---although the direction of squeezing does not alter the quantities that we are interested in. Assuming all input modes are populated by thermal fluctuations, each with individual noise factor $N_{ k_i}$, $i=1,2,3, 4$, the input covariance matrix is given by the direct sum of single-mode covariance matrices
\begin{equation}
    \bm\sigma^{\rm(in)}=N_{k_1}     \bm{I}_2\oplus N_{k_2}     \bm{I}_2\oplus     N_{k_3} \, e^{2s\bm\sigma_z}   \oplus
    N_{k_4}\,      \bm{I}_2\, .
\end{equation}
Given this input to the white-black hole and including attenuation effects, the output covariance matrix is formally given by,
\begin{equation}
    \bm\sigma_{\rm WB}^{\rm(out)}= \eta\, \bm S_{\rm WB}\, \bm\sigma^{\rm(in)}\, \bm S_{\rm WB}^\top + (1-\eta)\, \bm{I}_{6},\label{eq:wb_out_sq}
\end{equation}
(the first moments ${\bm \mu}$ of the initial state---and hence of the final state too---are chosen to be zero). From this covariance matrix all quantities of interest can be calculated.

\begin{figure}[t]
    \centering
    \includegraphics[width=\linewidth]{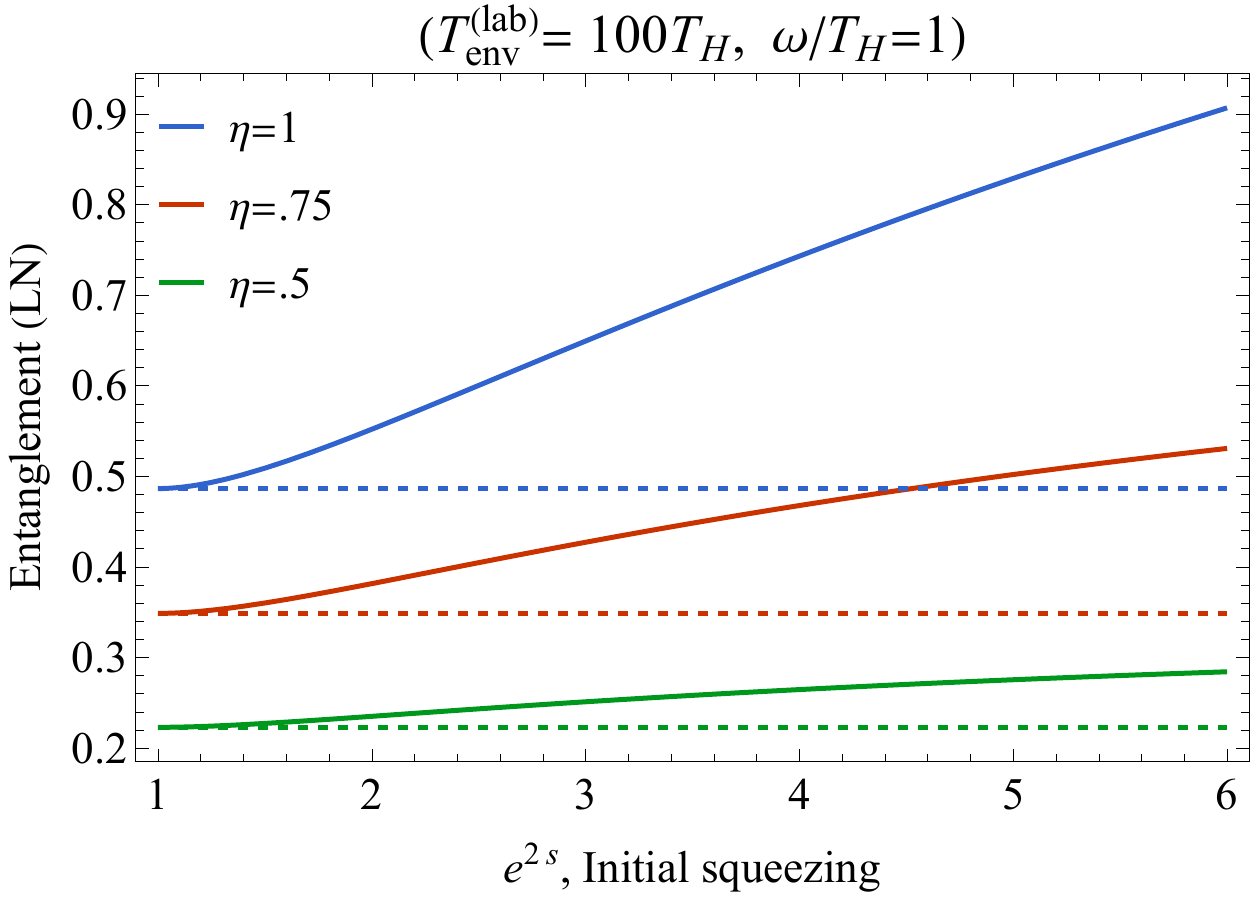}
    \caption{Entanglement between the photon-pairs emitted by the white hole, $(a_{k_1}^{\rm(out)},a_{k_4}^{\rm(out)})$, versus initial squeezing (solid curves), for various efficiency values $\eta=1,.75,.5$ (from top to bottom) and the value ${T_{\rm env}^{(\rm lab)}=100T_H}$ has been taken for the initial thermal fluctuations in the lab frame. Dashed lines represent zero initial squeezing, similarly with $\eta=1,.75,.5$ (from top to bottom).}
    \label{fig:ln_sq}
\end{figure}

\paragraph{Entanglement enhancement with squeezing:}
In the near-ideal scenario of perfect efficiency, $\eta=1$, we have found that single-mode squeezing can always be used to overcome the degrading effects of initial thermal fluctuations. In other words, for a given frequency $\omega/T_H$ and arbitrary thermal noise, there exists a value of the squeezing strength $s^\star(\omega)$ such that the LN is non-zero. Moreover, the entanglement is shown to increase with more squeezing, $s$ (see the curve corresponding to $\eta=1$ in Fig.~\ref{fig:ln_sq}). Importantly, the entanglement is not generated by the squeezing resource itself per se; in the sense that, for a set squeezing strength $s$, the entanglement between the $(a_{k_1}^{\rm(out)},a_{k_4}^{\rm(out)})$ modes increases with the Hawking-squeezing parameter, $r_H$, and {\em vanishes entirely} as $r_H$ goes to zero. In this sense, the origin of the entanglement can be attributed to Hawking pair production.

We now discuss the effects of attenuation ($\eta<1$). As shown in Fig.~\ref{fig:ln_sq}, attenuation limits the entanglement-enhancement gained by initially squeezing the white hole input, which also depends on the initial thermal fluctuations in the lab frame. However, enhancements are to be found even for very inefficient setups; for instance, initial squeezing can help even for $\eta=.5$ and $T_{\rm env}^{(\rm lab)}=100T_H$. The thermal fluctuations (in the lab frame) do not cause much harm to the entanglement generated in the system, even when efficiency effects are accounted for, thanks to the large Lorentz boost between the lab frame and the comoving frame. Our squeezing protocol is thus more robust to efficiency effects than we 
anticipated in Ref.~\cite{agullo2022prl}.


It is worth mentioning that, upon stimulating the white hole with a single-mode squeezed vacuum, the entanglement criterion imposed by the parameter $\Delta$ [$\Delta<0$; Eq.~\eqref{eq:cs_inequality}] is not sufficient to witness entanglement in our setup across all regions of parameter space; see Appendix~\ref{app:ent_compare} and Fig.~\ref{fig:delta_sq} for more details.


\begin{figure}
\centering   
    \includegraphics[width=\linewidth]{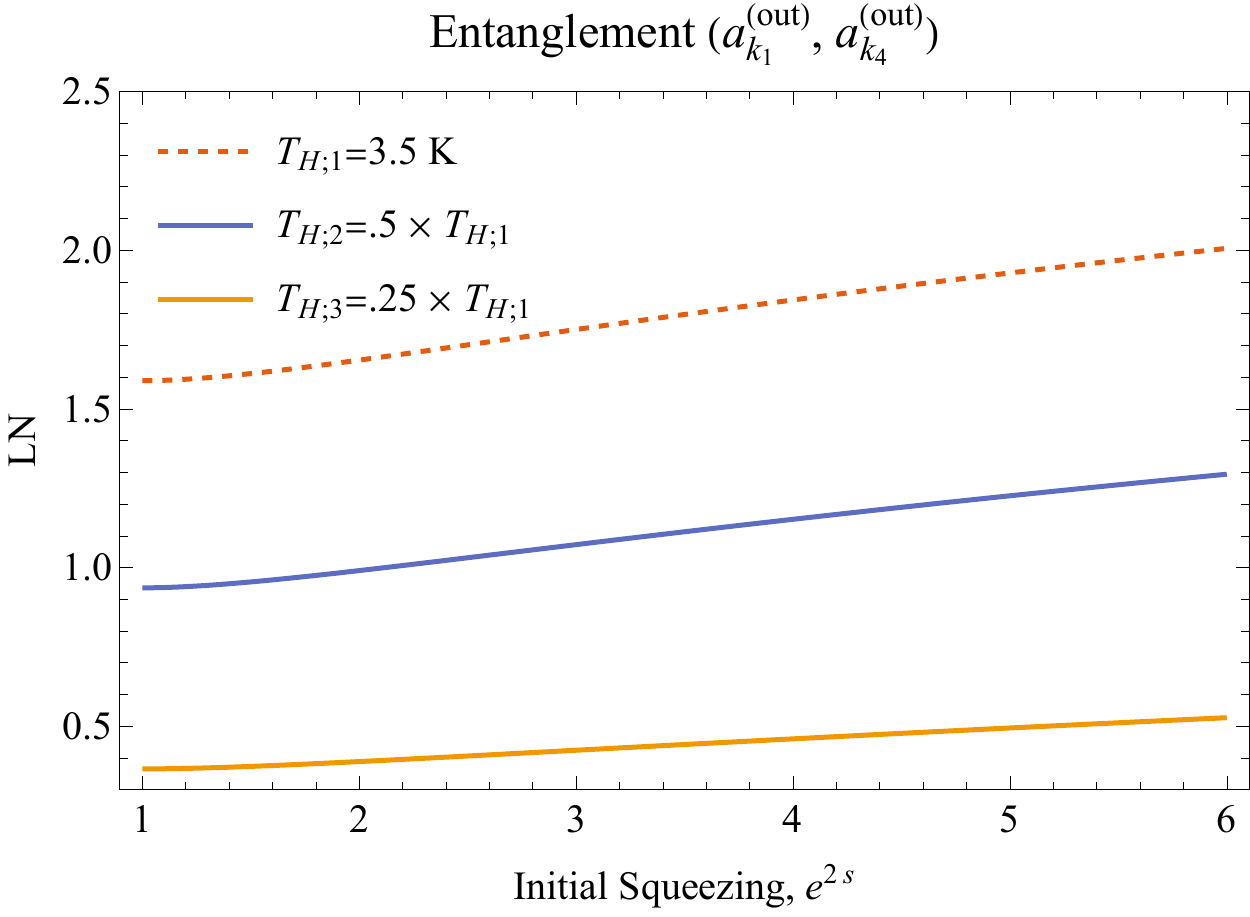}
    \caption{Logarithmic negativity between modes $\hat{a}_{k_1}^{(\rm out)}$ and $\hat{a}_{k_4}^{(\rm out)}$ forming the Hawking pairs emitted by the white hole horizon, versus the gain of the initial single-mode squeezing, $e^{2s}$. Curves correspond to different possible values of the Hawking temperature $T_{H;i}$, at a given comoving frequency $\omega$. This plot is obtained for a transmission coefficient of the potential barrier $\Gamma=0.999929$.}\label{fig:ln_example}
\end{figure}

\paragraph{Circuit parameters from intensities and LN:} 
An interesting by-product of squeezing is that it also allows us to extract the functional form of the symplectic-circuit parameters from the output intensities of the white hole only, assuming our squeezing source is characterized and tunable. Using Eq.~\eqref{eq:wb_out_sq}, we obtain an analytical expression for 
 the mean-occupation number for each mode,  and find that they all have the simple form, $\ev*{\hat{n}^{\rm (out)}_{k_i}}=m_i\sinh^2s + b_i$, where $\sinh^2s$ is the number of initially squeezed photons and the intercepts $b_i$ are independent of the squeezing parameter $s$. The slopes, $m_i$, encode the circuit parameters via the relations,
\begin{align}\label{eq:slopes}
    m_{1}&=N_{k_{3}} \, \eta\, \cos^2\theta_\omega\sinh^2r_H, \nonumber \\
    m_{2}&=N_{k_{3}}\, \eta\, \sin^2\theta_\omega , \nonumber\\
    m_{3}&=0,\nonumber \\
    m_{4}&=N_{k_{3}}\, \eta\, \cos^2\theta_\omega\cosh^2r_H,
\end{align}
where, recall, $N_{k_{3}}$ is the noise factor in the initial state of the mode $k^{\rm (in)}_3$. The expressions for $b_i$ are lengthy, and will not be used in this section (for completeness, we report them in expressions~\eqref{bs} in Appendix \ref{app:formulae}). Note that all slopes $m_{{i}}$ are proportional to the thermal noise in the input mode $k_{3}$ and to the attenuation parameter $\eta$. For a given frequency, we can monitor the output intensities of each mode as we vary the squeezing parameter $s$. If we then plot the output intensities versus the intensity of  the initial  squeezing $\sinh^2s$, we will find straight lines with slopes given as above. Taking ratios of these slopes we cancel out the factors $N_{k_{3}}$ and $\eta$, 
allowing us to map out the frequency dependent forms of the Hawking intesity $r_H(\omega)$ from the ratio,
\begin{equation}\label{rHintensity}
    \frac{m_1}{m_4}= \tanh^2r_H = \e^{-\omega/T_H}, 
\end{equation}
from which the Hawking temperature may also be derived. 

The value of the Hawking temperature, $T_H$ (or equivalently, Hawking's squeezing intensity, $r_H$) can be independently checked against the size of the entanglement between, say, the Hawking pair $k_1$ and $k_4$ emitted by the white hole, as described in \cite{agullo2022prl}. As mentioned previously, using the circuit, we can obtain analytical expressions for the LN between these two modes (and any other bi-partition), in terms of the parameters of the circuit and the initial state. These expressions are lengthy, but they can be easily obtained using any software for symbolic calculus. Rather than writing the lengthy expressions for LN, we plot its value in terms of the squeezing intensity of the initial state and for different potential values of the Hawking temperature $T_{H;i}$ in Fig.~\ref{fig:ln_example}. If the LN can be measured, for instance, by reduced-state tomography using homodyne measurements \cite{raymer2009} (a challenging task), by comparing with the theoretical curves in Fig.~\ref{fig:ln_example} it would be possible to determine the actual Hawking temperature ($T_{H;1}=3.5\,\rm{K}$ here) from a quantity of purely quantum origin. The value of $T_H$ obtained in this way must be in agreement with the one independently obtained from intensities [Eq.~(\ref{rHintensity})]; this is a non-trivial consistency test \cite{agullo2022prl}.


\section{Conclusion}\label{sec:conclude}

One of the goals of this paper is to promote a set of analytical and visual techniques to study Hawking-like phenomena in analog gravity systems. The analytical techniques are derived from the so-called Gaussian formalism for continuous variables systems, and the visual techniques correspond to symplectic circuits. The Gaussian formalism provides a powerful toolbox to describe Gaussian states of bosonic systems and their evolution under quadratic Hamiltonians. 
Importantly, the formalism comes together with a set of compact and simple expressions to extract physical quantities from the first and second moments of the state, such as particle number, purity, entropy, entanglement, etc. (see e.g.\ \cite{serafini17QCV} for an extensive treatment).

Particle creation phenomena in quantum field theory in curved spacetimes, such as the Hawking effect, precisely fall within the applicability of the Gaussian formalism. In this paper we have shown how to apply the formalism to reformulate the Hawking effect in very simple terms, from which one can derive all interesting aspects of the underling physics (see, e.g., Refs. ~\cite{bruschi2013,nambu21,Isoard:2021peb} for similar treatments in the context analog gravity). Our analysis is particularly useful to study Hawking-like effects in analog gravity, where multi-mode scattering adds complications which are otherwise more difficult to deal with. We have shown how effects of thermal noise, losses, and detector inefficiencies---ubiquitous in experimental setups---can be incorporated within the description and have provided a thorough study of the impact they have on quantities of physical interest. In particular, we have studied multi-mode entanglement generated in the Hawking process and how such is affected by noise. One important aspect we want to emphasize about our analysis is its simplicity: the {\em a priori} complex task of quantifying entanglement between any subset of modes becomes a simple exercise within the Gaussian formalism.

We have complemented the analysis with the introduction of symplectic circuits  to describe  Hawking like phenomena. The goal of these circuits is twofold: they provide an analytical approximation for the dynamics and, moreover, are powerful visual tools for understanding how particles and entanglement are created and distributed among the different output modes. By mere inspection of such circuits, one can extract valuable information about the physics of the system and write down an approximation for the evolution (scattering matrix). The name `symplectic circuit' is motivated by the fact that, in the classical theory, the circuit describes a symplectic transformation---i.e., a canonical transformation that is linear in the basic variables. These circuits are generally built by concatenating three elementary operations: a phase shifter describing free evolution, beam splitters accounting for scattering phenomena, and two-mode squeezers describing  creation of entangled particles. It is important to keep in mind that symplectic circuits are meant to approximate the dynamics of the system, and in concrete applications, one must check the accuracy of the proposed circuit against the actual dynamics.

We have applied these tools to study the Hawking effect in optical analog systems, where an analog white hole horizon is generated together with a black hole horizon, extending our previous results summarized in Ref.~\cite{agullo2022prl}. The use of symplectic circuits proves to be very convenient in this context. We have provided a detailed comparison of the predictions of our circuit ansatz against the numerical solutions of the equations describing the system and have found that the circuit ansatz indeed provides an excellent description of the system in the regime where the analogy with the Hawking effect is on firm ground. With this, we have studied the concrete way in which ambient noise and losses affect the entanglement structure of the output state, confirming that these effects are dangerous enemies of entanglement. Since  entanglement generation is the quantum signature of the Hawking effect, ambient noise and losses must be under control in any experiment aiming to observe quantum aspects of the Hawking process. 

We point out that, for an analog white-black hole, our circuit fails to accurately describe the system when the analogy with the Hawking effect breaks down due to dispersive effects. This has allowed us, in turn, to identify and characterize the physical origin of such dispersive effects in a precise manner. Nevertheless, one can extend the circuits presented here and add new elements to accommodate these additional effects.
 
Finally, we have proposed a mechanism to amplify the quantum aspects of the Hawking process via stimulating the analog system  with (single-mode) squeezed light. Such stimulation triggers additional particle creation. Stimulated Hawking radiation is commonly regarded as being purely classical and of little value to inform us about the quantum aspects of the Hawking effect; we argue otherwise. We show that the induced radiation is indeed classical when the system is stimulated with classical light, namely with a coherent or a thermal state. On the contrary, we show that quantum inputs have the ability to stimulate the generation of additional entanglement (relative to the spontaneous Hawking effect). This not only solidifies that the stimulated Hawking effect is genuinely quantum but also provides a mechanism to increase the visibility of quantum aspects of the Hawking effect in the lab. We have outlined a protocol to implement this idea in an experimental set up.  

It is our view that the tools leveraged here will be beneficial to study other analog systems beyond the ones that we have already explored, such as laser-like effects in multi-horizon scenarios \cite{corley1999lasers,Coutant:2009cu,Finazzi:2010nc,Gaona-Reyes:2017mks, katayama2021circuit}, tunneling effects \cite{philbin08,rubino2012soliton,petev2013blackbody,drori19}, super-radiant effects \cite{Jacquet:2022vak,Chelpanova:2021gnm}, etc. Indeed, one of the motivations for the pedagogical character of this paper is to make these tools readily available to the community.    

\begin{acknowledgements}
A.J.B. acknowledges Quntao Zhuang and support from the Defense Advanced Research Projects Agency (DARPA) under Young Faculty Award (YFA) Grant No. N660012014029. I.A.\ and D.K.\ are supported by the NSF grant PHY-2110273, and by the Hearne Institute for Theoretical Physics. We thank Adria Delhom for comments and discussions. 
\end{acknowledgements}


\begin{widetext}
\appendix
\section{Two examples of Gaussian evolution: two-mode squeezers and beam-spliters}\label{sqzbs}

This appendix summarizes some details of two representative examples of linear evolution of Gaussian states. Our goal is twofold: (i) to increase the pedagogical content of Section \ref{sec:prelims} by putting the techniques summarized there in action in two concrete and simple situations, and (ii) to introduce and describe the properties of the two symplectic transformations that play an important role in the bulk of this paper. In fact, the main results of this paper can be reduced, in a sense, to the properties of two-mode squeezers and beam-splitters spelled out here. 

\subsection{Two-mode squeezer}

A two-mode squeezer is a transformation between two modes 
\be   \hat{\bm R}^{\rm(in)}=(\hat Q^{\rm(in)}_1,\hat P^{\rm(in)}_1, \hat Q^{\rm(in)}_2,\hat P^{\rm(in)}_2) \nonumber \\ \nonumber \longrightarrow  \hat{\bm R}^{\rm(out)}=(\hat Q^{\rm(out)}_1,\hat P^{\rm(out)}_1, \hat Q^{\rm(out)}_2,\hat P^{\rm(out)}_2) \, \ee
defined by a scattering matrix that depends on two parameters, $r$ and $\phi$, as
\be \label{squeezer} {\bm S}_{\rm Sqz}(r,\phi) =\left(
 \begin{array}{cccc}
 \cosh r & 0 & \sinh r  \cos \phi  & \sinh r \, \sin \phi  \\
 0 & \cosh r & \sinh r \, \sin \phi & - \sinh r \, \cos \phi  \\
 \sinh r \,  \cos \phi  & \sinh r \,  \sin \phi  & \cosh r & 0 \\
 \sinh r\,  \sin \phi  & - \sinh r \, \cos \phi  & 0 & \cosh r \\
\end{array}
\right).
\ee
The parameters $r,\phi$ are real numbers called the squeezing intensity and squeezing phase, respectively. It is straightforward to show that this matrix belongs to the symplectic group ${\rm Sp}(4,\mathbb{R})$, by checking ${\bm S}_{\rm Sqz}\cdot  \Omega\cdot {\bm S}_{\rm Sqz}^\top=\Omega$ for all $r$ and $\phi$. Using \eqref{QPA}, it is easy to obtain the transformation of annihilation operators by the two-mode squeezer,
\bea \hat a^{\rm(in)}_1&\to& \hat a^{\rm(out)}_1=\cosh r\,  \hat a^{\rm(in)}_1+\e^{i\phi}\, \sinh r\,  \hat a^{\rm(in)\, \dagger}_2, \\ 
 \hat a^{\rm(in)}_2&\to& \hat a^{\rm(out)}_2=\cosh r\,  \hat a^{\rm(in)}_2+\e^{i\phi}\, \sinh r\,  \hat a^{\rm(in)\, \dagger}_1. 
\eea 
[The way creation operators transform is obtained from these equations by simple Hermitian conjugation.] Observe that the two-mode squeezer mixes creation and annihilation operators, but does so in a very concrete manner: $\hat a^{\rm(out)}_1$ is made of a combination of $\hat a^{\rm(in)}_1$ and $\hat a^{\rm(in)\, \dagger}_2$, but it does not get contributions from either $\hat a^{\rm(in)\, \dagger }_1$ nor $\hat a^{\rm(in)}_2$ (and similarly for $\hat a^{\rm(in)}_2$). We depict a two-mode squeezer by the symbol showed in Fig.~\ref{fig:squeezer}.

\begin{figure}
    \centering
    \includegraphics[width=.5\linewidth]{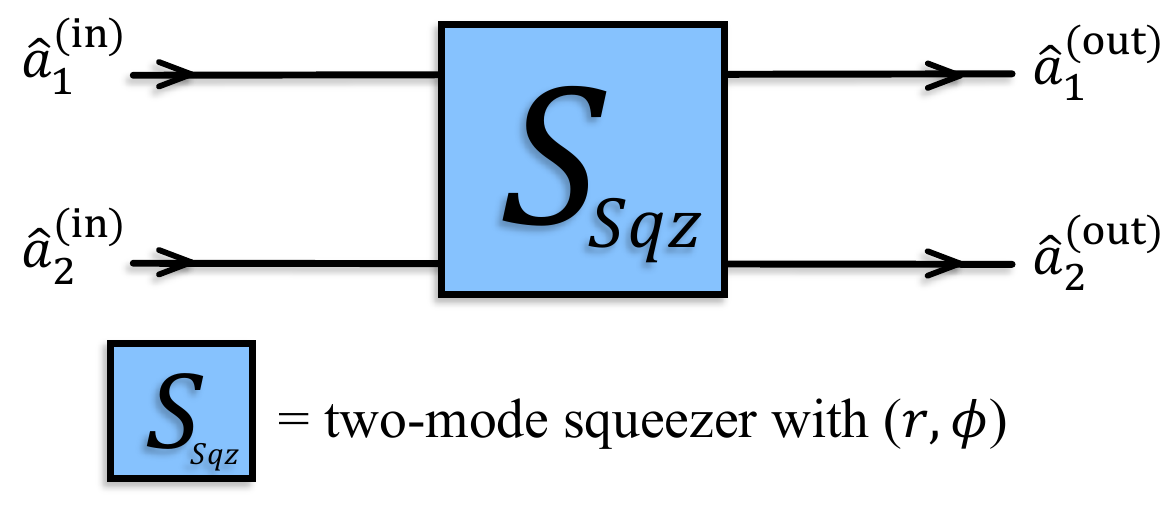}
    \caption{Circuit diagram for two-mode squeezing interaction, with squeezing intensity $r$ and squeezing angle $\phi$, described by symplectic matrix $\bm S_{\rm Sqz}$.}
    \label{fig:squeezer}
\end{figure}

Physically, a two-mode squeezer describes an amplifier (the transformation does not conserve energy; energy must be injected into the system by an external agent for the process to happen). Quantum mechanically, amplification means creation of quanta; as we show below, two-mode squeezers add two genuinely quantum features to the amplification process: (i) The quanta are created in pairs, which are generically entangled, and (ii) a squeezer can create quanta even if the input is the vacuum---i.e., squeezers amplify the vacuum. Two-mode squeezers appear in diverse physical situations: they are responsible for the phenomenon of parametric down conversion, for particle creation in the early universe and, as we argue in this paper, for the physics of the Hawking effect. 

To better understand the action of two-mode squeezers, we now discuss the way they transform various input states.  \\

{\bf Vacuum input.} Acting on the vacuum state, $({\bm \mu}^{(\rm in)}=0, {\bm \sigma}^{(\rm in)}=\bm{I}_4)$ a two-mode squeezer produces another Gaussian state, called a {\em two-mode squeezed vacuum}, and defined by $ {\bm \mu}^{(\rm out)}$ and ${\bm \sigma}^{(\rm out)}$ given by the following expressions
\bea  \label{2msqv} {\bm \mu}^{(\rm out)}&=& {\bm S}_{\rm Sqz}\cdot {\bm \mu}^{(\rm in)} =0\, ; \\  {\bm \sigma}^{(\rm out)}&=& {\bm S}_{\rm Sqz}\cdot  {\bm \sigma}^{(\rm in)}\cdot {\bm S}_{\rm Sqz}^{\top}=\left(
\begin{array}{cccc}
 \cosh 2 r & 0 & \sinh 2 r\,  \cos \phi  & \sinh 2 r\,  \sin \phi  \\ \nonumber 
 0 & \cosh 2 r & \sinh 2 r\,  \sin \phi  & -\sinh 2 r\,  \cos \phi  \\
 \sinh 2 r\,  \cos \phi  & \sinh 2 r\,  \sin \phi  & \cosh 2 r & 0 \\
 \sinh 2 r\,  \sin \phi  & -\sinh 2 r \, \cos \phi & 0 & \cosh 2 r \\
\end{array}
\right) \, ,\eea
This covariance matrix is of the form 
\be {\bm \sigma}^{(\rm out)}=\begin{pmatrix}
{\bm \sigma}^{(\rm red)}_A & {\bm \sigma}_{AB} \\
{\bm \sigma}^{\top}_{AB} &{\bm \sigma}^{(\rm red)}_B\, ,
\end{pmatrix}
\ee
where 
\be {\bm \sigma}^{(\rm red)}_A={\bm \sigma}^{(\rm red)}_B=\begin{pmatrix}\cosh 2 r & 0 \\ 0 & \cosh 2 r \end{pmatrix} \, , \ee
are the reduced covariance matrix for each of the two subsystems, and ${\bm \sigma}_{AB}$ describes the correlations between them. Recall that the covariance matrix of a thermal state [with density matrix $\hat \rho= N\, \exp(-\beta\, \hat H)$, where $\hat H=\frac{1}{2} \omega\, (\hat Q^2+\hat P^2)$] is ${\bm \sigma}=(1+2\, \bar n)\, \bm{I}$ and $\bar n=(\exp(\beta\, \omega)-1)^{-1}$ is the mean number of thermal quanta. Since both $\sigma_A$ and $\sigma_B$ have this form, each mode in a two-mode squeezed vacuum are individually in a thermal state with inverse temperature $\beta=2\ln(\coth r)/\omega$. [The mean number of quanta $\bar n_A=\bar n_B$, when written in terms of $r$, have the simple expression $\bar n_A=\bar n_B=\sinh^2r$.] That each output is in a thermal state implies, in particular, that both subsystems are in a mixed quantum state, and since the total state is pure, the subsystems must be entangled. This entanglement can be quantified either using the von Newmann entropy of either of the subsystems (since the total state is pure, this entropy defines the entanglement entropy) or the logarithmic negativity. Both quantities can be easily computed as follows. 

The von Neumann entropy of an $N$-mode Gaussian state with covariance matrix $\bm\sigma$ is given by 
\be S[\bm\sigma]=\sum_I^N \left( \frac{\nu_I+1}{2}\right) \log_2\left( \frac{\nu_I+1}{2}\right)-\left( \frac{\nu_I-1}{2}\right) \log_2\left( \frac{\nu_I-1}{2}\right) , \ee
where $\nu_I$, $I=1,\cdots,N$ are the absolute value of the eigenvalues of the matrix $\bm\sigma^{ik}\bm\Omega_{kj}$, where $\bm\Omega= \oplus_N \big(\begin{smallmatrix}
    0 & 1\\
    -1 & 0
\end{smallmatrix}\big)$, is the symplectic form. It is common to refer to the $N$ (real and positive) numbers $\nu_I$ as the `symplectic eigenvalues' of the covariance matrix because there exists a symplectic transformation, $\bm S_\sigma$, which diagonalizes $\bm\sigma$ such that the diagonal elements are given by $\nu_I$. 

The symplectic eigenvalue of ${\bm \sigma}^{(\rm red)}_A$ is equal to  $\nu=1+2\sinh^2r$, and produce a von Newmann entropy 
%
\bea S_{\rm ent}=(1+\sinh^2r)\log_2 (1+\sinh^2r)-\sinh^2r\log_2(\sinh^2r). \eea
This is the entanglement entropy of subsystems A and B. 

The logarithmic negativity of a Gaussian state made of two Gaussian sub-systems A and B, is given by 
\be {\rm LN}[{\bm\sigma}]=\sum_I {\rm Max}[0, -\log_2 \tilde \nu_I]\, , \ee
where $\tilde \nu_I$ are the symplectic eigenvalues of the {\em partially transposed} covariance matrix $\bm{\tilde \sigma}$, defined from $\bm \sigma$  by reversing the sign of all components involving momenta $P_i$ of the subsystem B. This is more clearly shown with an example. For a two-mode squeezed vacuum, the partially transposed covariance matrix (with respect to the natural bi-partition) is 

\bea     \tilde {\bm \sigma}^{(\rm out)}=\left(
\begin{array}{cccc}
 \cosh 2 r & 0 & \sinh 2 r\,  \cos \phi  & - \sinh 2 r\,  \sin \phi  \\ \nonumber 
 0 & \cosh 2 r & \sinh 2 r\,  \sin \phi  & 2 \sinh r\,  \cosh r\,  \cos \phi  \\
 \sinh 2 r\,  \cos \phi  & \sinh 2 r\,  \sin \phi  & \cosh 2 r & 0 \\
 -\sinh 2 r\,  \sin \phi  & 2 \sinh r\,  \cosh r \, \cos \phi & 0 & \cosh 2 r \\
\end{array}
\right) \, ,\eea
which differs from ${\bm  \sigma}^{(\rm out)}$ in Eq.~\eqref{2msqv} only in the sign of the components (1,4), (2,4), (3,4), (4,1), (4,2) and (4,3) (note that the sign of the component (4,4) is not reversed; or equivalently, is reversed twice). This is equivalent to applying expression \eqref{PT}. The symplectic eigenvalues of this matrix are $\tilde \nu_1=e^{2r}$ and $\tilde \nu_2=e^{-2r}$. Since $\tilde \nu_2$ is smaller than one for $r>0$, the two subsystems are entangled and ${\rm LN}=-\log_2 e^{-2r}=\frac{2\, r}{\ln 2}$.

Entanglement entropy and LN have different values, but they grow monotonically with the squeezing strength, $r$. Both quantities are explicit entanglement quantifiers for this system. \\

{\bf Coherent state input}. Recall that a coherent state $({\bm \mu}^{(\rm coh)}\neq 0, {\bm \sigma}^{(\rm coh)}=\bm{I}_4)$ differs from the vacuum only in its first moments, while its covariance matrix is the same as for vacuum. Hence, if we send a coherent state through a two-mode squeezer, we get a Gaussian state identical to \eqref{2msqv} except that the first moments are replaced by ${\bm S}_{\rm Sqz}\cdot  {\bm \mu}^{(\rm coh)}$. This implies that the entanglement between the two subsystems in the output state is exactly the same as for vacuum input. Only quantities that depend on the first moments are different. Examples of such quantities are the mean number of quanta in each subsystem,
\bea \langle &&\hat n_A \rangle= \sinh^2r +[\mu^{({\rm in})}_2 \, \cosh r +\ (-\mu^{({\rm in})}_4 \,\cos \phi+\mu^{({\rm in})}_3 \,\sin \phi)\, \sinh r]^2+[ \mu^{({\rm in})}_1\, \cosh r +(\mu^{({\rm in})}_3\, \cos \phi +\mu^{({\rm in})}_4\, \sin \phi) \sinh^2 r]^2\, \nonumber \\
 &&\langle \hat n_B \rangle= \langle \hat n_A \rangle+
 - \mu^{({\rm in})\, 2}_1-\mu^{({\rm in})\, 2}_2+\mu^{({\rm in})\, 2}_3+\mu^{({\rm in})\, 2}_4\, 
. \eea
Note that the last equation tells us that the difference in the mean number of quanta between both subsystems is the same before and after the action of the squeezer. In this sense, quanta are created in pairs. We identify in these last equations the term $\sinh^2r$ that we would have obtained for vacuum input. The other terms are proportional to the components of the first moments of the initial state ${\bm \mu}^{({\rm in})}=(\mu^{({\rm in})}_1,\mu^{({\rm in})}_2,\mu^{({\rm in})}_3,\mu^{({\rm in})}_4)$. The term $\sinh^2r$ corresponds, therefore, to the spontaneous creation of quanta, while the rest account for induced or stimulated creation. 
Interestingly, since the entanglement in the final state is exactly the same as for the case of vacuum input, one can say that the extra quanta created by this stimulated process is, in some sense, not entangled. In other words, there is nothing genuinely quantum in the stimulated radiation. We will see below, however, that this classical character of the stimulated process is a peculiarity of using a coherent state input, and it is not true in general. In particular, this classical interpretation of the stimulated process does not hold if we use a single-mode squeezed vacuum for the initial state. \\ 

\begin{figure}
{\centering   
    \includegraphics[width=0.45\linewidth]{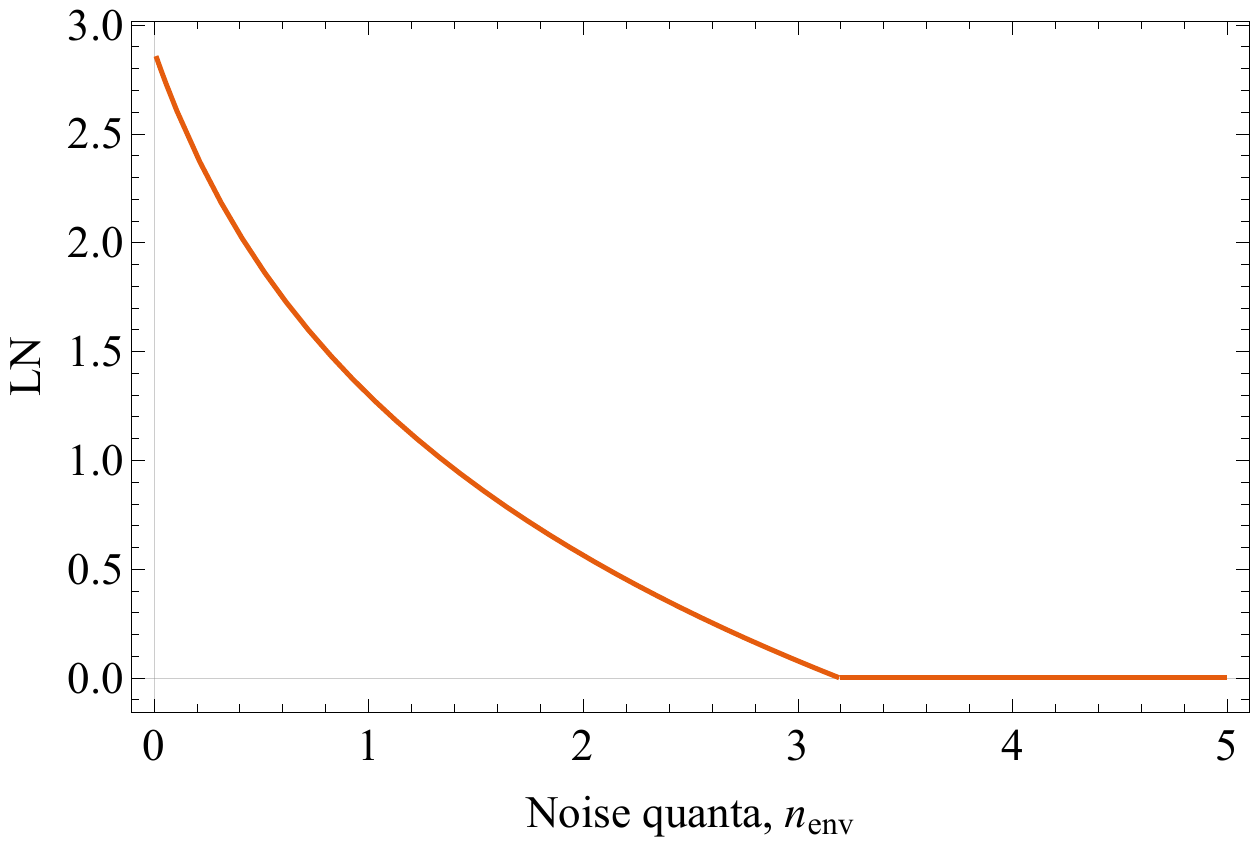}
     \includegraphics[width=0.45\linewidth]{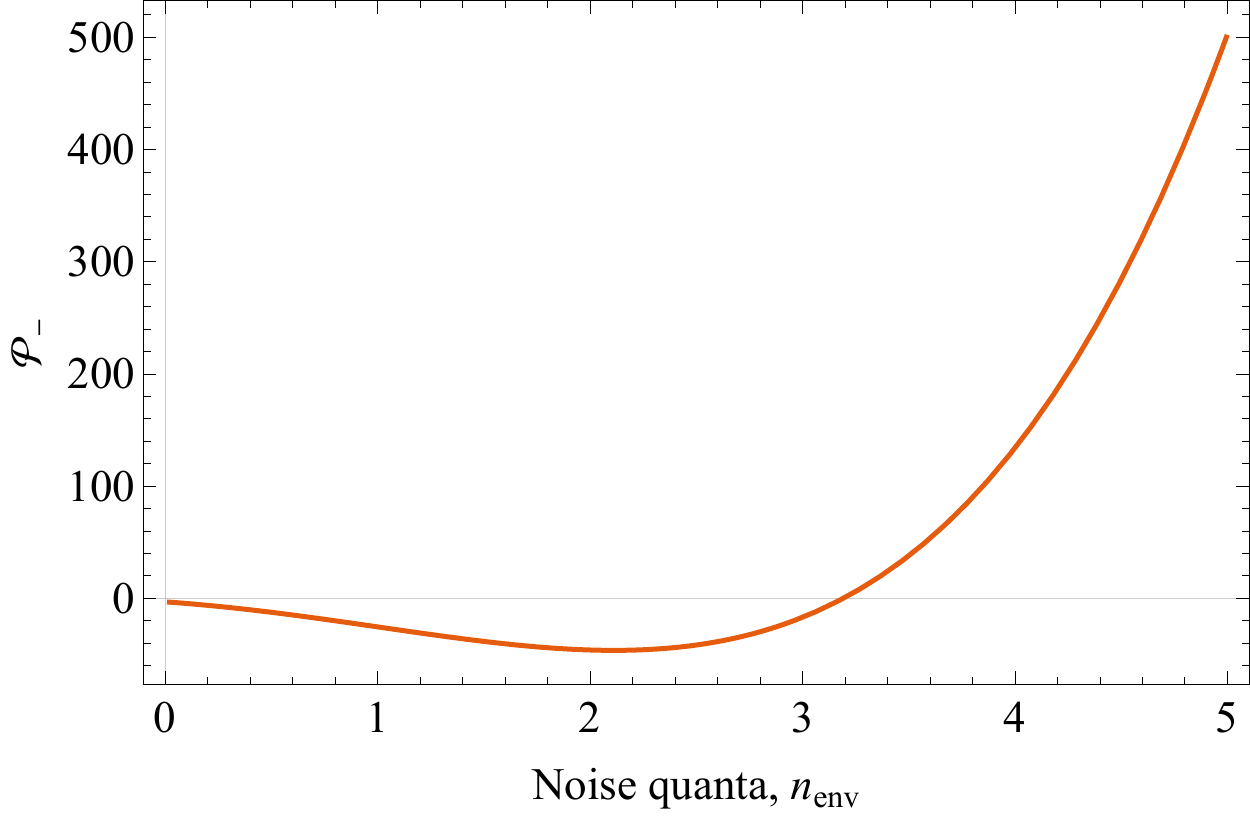}
      \includegraphics[width=0.45\linewidth]{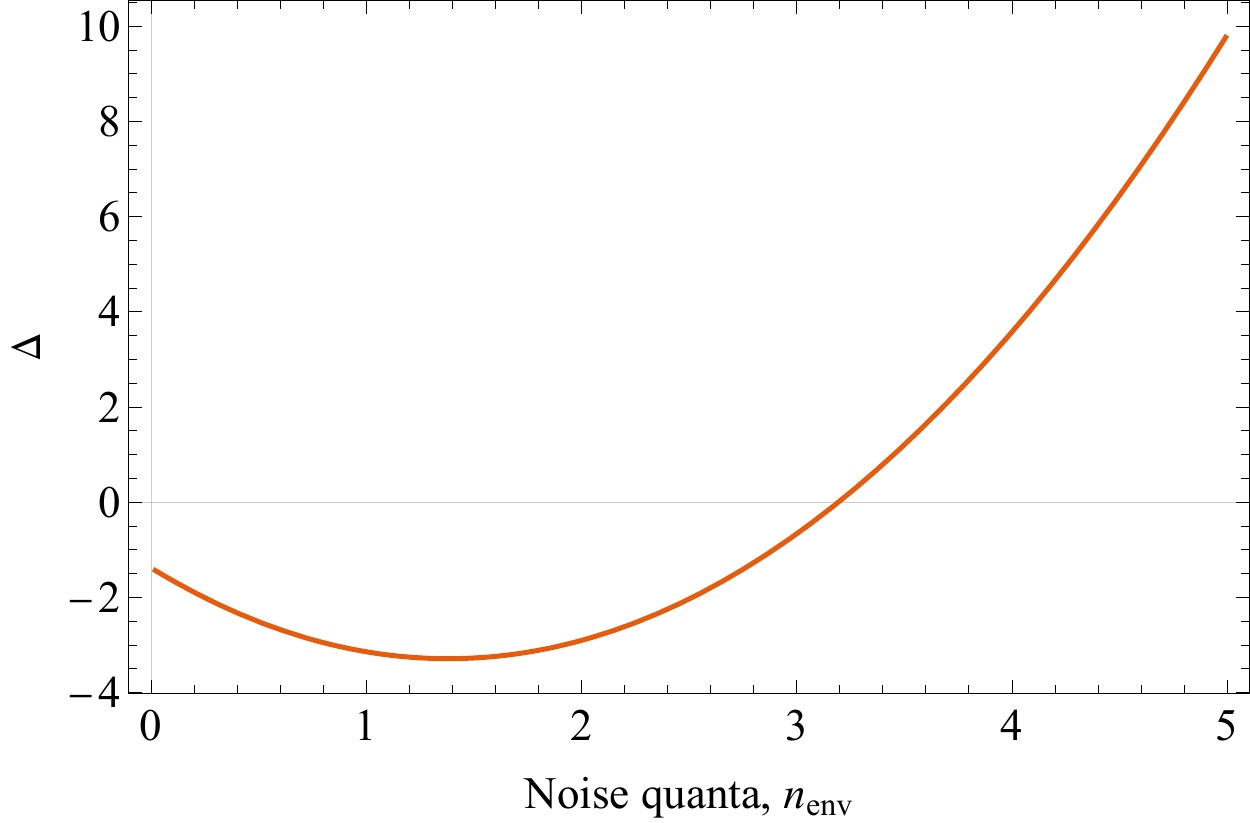}
    \caption{Upper-left panel: Logarithmic Negativity (LN) for the two-outputs of a two-mode squeezer seeded with a thermal state of $ n_{\rm env}$ mean number of quanta in both input modes. Entanglement in the final state decreases monotonically with $ n_{\rm env}$, completely disappearing when $ n_{\rm env}$ is above the threshold value $e^r\, \sinh r$, which depends on the intensity $r$ of the two-mode squeezer. Upper-right panel: Entanglement witness $P_-$ [defined in Eq.~\eqref{eq:pminus}] versus  $ n_{\rm env}$, for the same system. $P_-$ correctly signals the presence of entanglement ($P_-<0$) for values of $ n_{\rm env}$ for which ${\rm LN}$ is different from zero. However, $P_-$ is not an entanglement quantifier and thus does not tell us whether there is more or less entanglement. Lower panel: $\Delta$ [defined in   Eq.~\eqref{Delta}] versus $ n_{\rm env}$. In this example $\Delta$ also signals correctly the range of $ n_{\rm env}$ for which the final state is entangled. Though, it is not an entanglement quantifier either. Plots are computed using a squeezing intensity $r=1$. The three figures are independent of the value of the squeezing angle $\phi$. The effect of increasing $r$ is to shift (horizontally to the right) the value of $ n_{\rm env}$ for which the entanglement vanishes.}
    \label{fig:entanglementthermalinput}
   }
\end{figure}

{\bf Thermal state input}.  Intuitively, we expect thermal noise to degrade quantum coherence and entanglement. This can indeed be shown using the simple example of a thermally seeded two-mode squeezing process. Consider the input state to be a thermal state, and for simplicity let us assume that the two initial modes have the same mean number of thermal quanta, which we will denote by $ n_{\rm env}$ (a generalization is straightforward). The input state is a Gaussian state with ${\bm \mu}^{(\rm th)}=0$, and ${\bm \sigma}^{(\rm th)}=(1+2n_{\rm env})\, \bm{I}_4$. The only difference with vacuum is the multiplicative factor $(1+2n_{\rm env})$ in the covariance matrix, which is easy to carry over through evolution. For instance, the final state is given by Eq.~\eqref{2msqv} times $(1+2n_{\rm env})$. However, this simple factor has an significant impact on entanglement. The expression for LN is now
\be {\rm LN}={\rm Max}\big[0,- \log_2[(1+2\, n_{\rm env})\, e^{-2\,r }]\big]\, , \ee
(rather than ${\rm LN}=-\log_2 e^{-2r}$) which shows that LN \emph{decreases} with $n_{\rm env}$; even more, LN completely vanishes if $(1+2\, n_{\rm env})> e^{2\,r }$. [Note that the entropy no longer quantifies entanglement, since the total state is not pure.] Figure~\ref{fig:entanglementthermalinput} shows how the entanglement between the two output modes changes with $n_{\rm env}$. The upper-left panel traces the Logarithmic Negativity (LN) versus $n_{\rm env}$ and confirms the intuition that thermal noise degrades entanglement---completely removing it beyond the threshold  value  $n_{\rm env}=e^r \sinh r$ (thus leading to a convex combination of separable states for the final state). Figure~\ref{fig:entanglementthermalinput} (upper-right panel) also shows the entanglement witness $P_-$ [defined in Eq.~\eqref{eq:pminus}] versus  $ n_{\rm env}$. Here, $P_-$ correctly signals the presence of entanglement ($P_-<0$),  since $P_-$ takes negative values in exactly the same range of $ n_{\rm env}$ for which LN is different from zero. This is expected, since for a system like the one we are working with (a Gaussian state of a two-mode system), $P_-$ is negative if and only if the state is entangled. On the other hand, the plot also shows that $P_-$ does not change monotonically with $n_{\rm env}$ as the LN does. For instance, for small values $ n_{\rm env}$, $P_-$ becomes \emph{more} negative as $ n_{\rm env}$ \emph{increases}, suggesting that the state becomes more entangled---even though we know from LN that entanglement degrades as we add more thermal noise. This result reminds us that $P_-$ does not quantify entanglement but only signals its presence. The lower panel of Fig.~\ref{fig:entanglementthermalinput} shows the quantity $\Delta$ [defined in Eq.~\eqref{Delta}] versus $ n_{\rm env}$. We see that, for the current situation, $\Delta$ also correctly signals the presence of entanglement for the appropriate values of $ n_{\rm env}$ in accordance with the LN (although $\Delta$ does not quantify entanglement either). We will see below an example for which $\Delta$ is not able to signal entanglement, reminding us that  $\Delta<0$ is only a sufficient condition for entanglement (not necessary and sufficient). \\

\begin{figure}
{\centering   
    \includegraphics[width=0.49\linewidth]{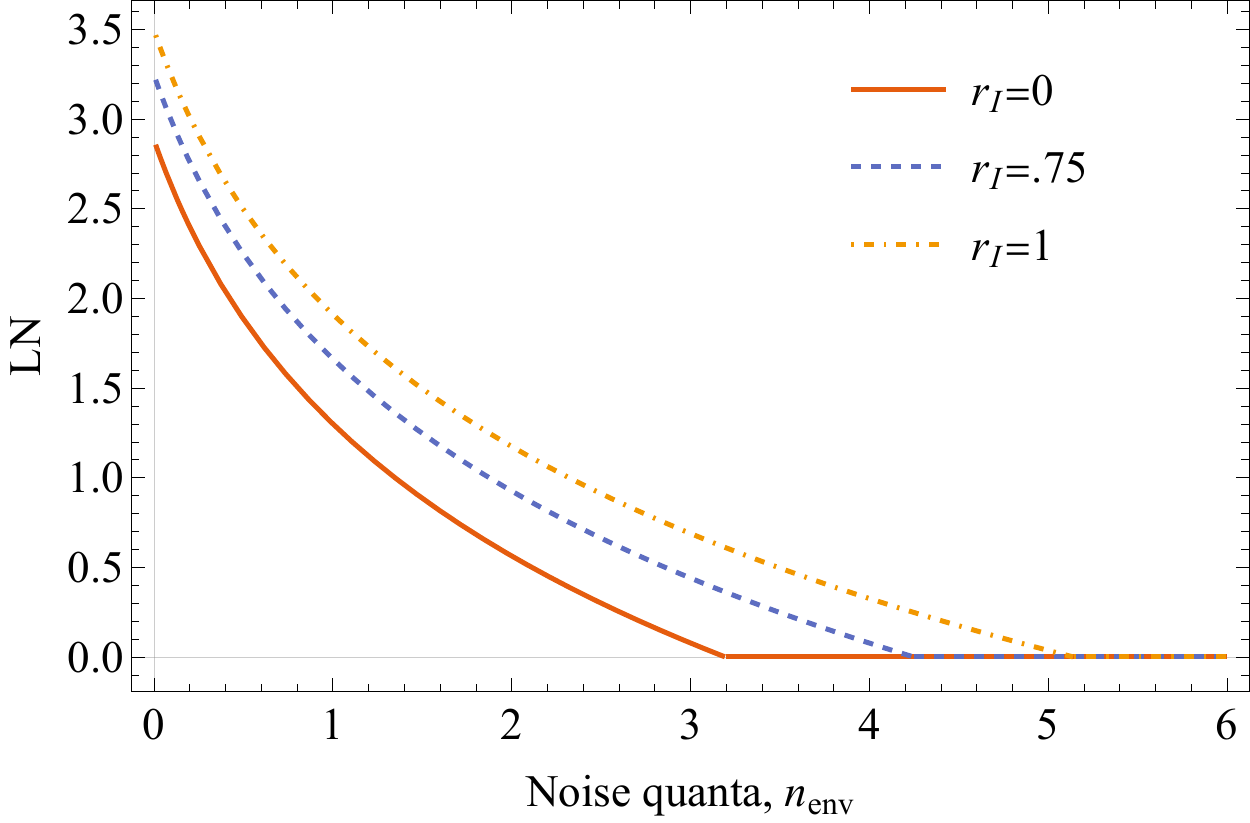}
     \includegraphics[width=0.49\linewidth]{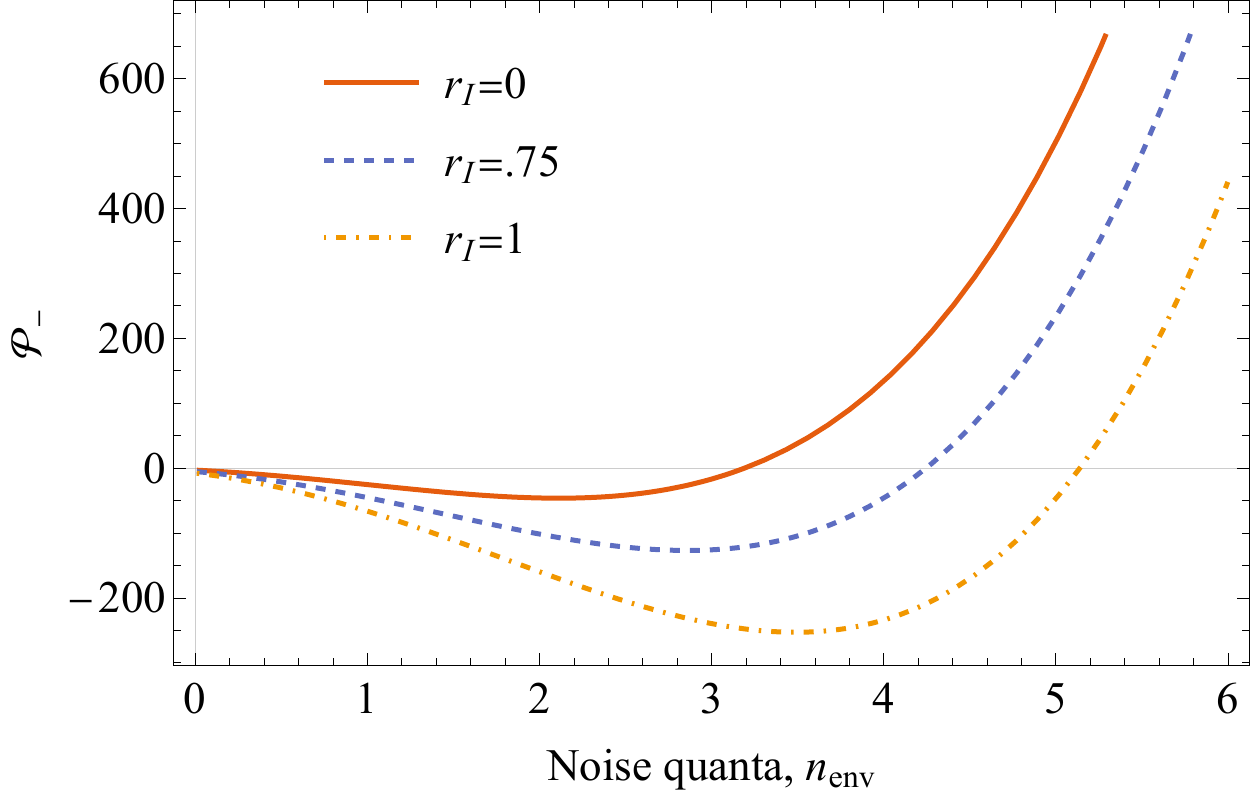}
      \includegraphics[width=0.48\linewidth]{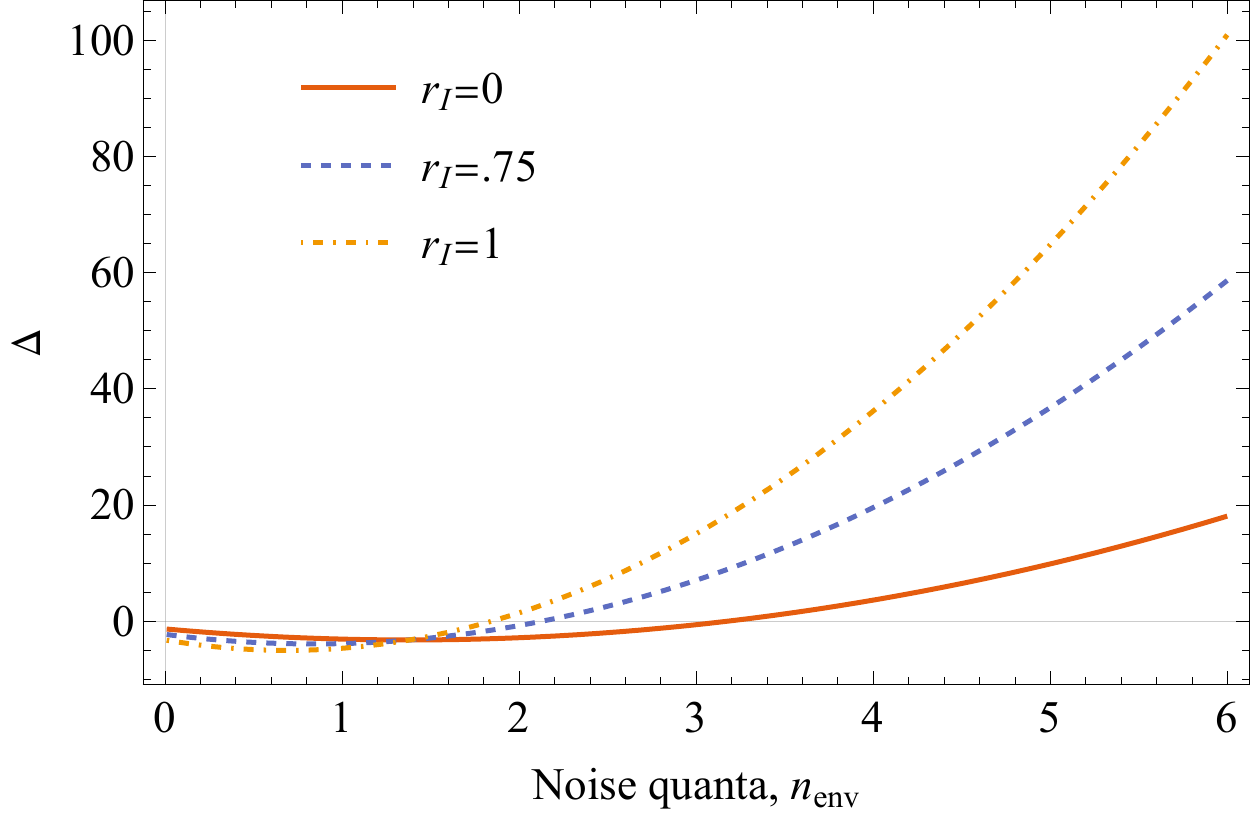}
    \caption{{Upper-left panel:} Logarithmic Negativity (LN) for the two-outputs of a two-mode squeezer (with squeezing intensity $r=1$) seeded with a single-mode-squeezed thermal state with $ n_{\rm env}$  mean number of quanta in both input modes. The horizontal axis represents $ n_{\rm env}$, and the three curves correspond to different values of the initial squeezing intensity, namely $r_I=0$ (solid), $r_I=.75$ (dashed) and $r_I=1$ (dot-dashed). The plots show that, by increasing $r_I$, we can restore entanglement for values of $n_{\rm env}$ for which otherwise entanglement would not exist. {Upper-right panel:} Entanglement witness $P_-$ [defined in Eq.~\eqref{eq:pminus}] versus $ n_{\rm env}$. As in the previous example, the condition $P_- <0$ captures faithfully the values of $ n_{\rm env}$ for which the final state is entangled.  {Lower panel:} $\Delta$ [defined in   Eq.~\eqref{Delta}] versus $ n_{\rm env}$. The Cauchy-Schwarz inequality $\Delta < 0$ misses the existence of entanglement for values of $n_{\rm env}$ at which the state is entangled. As we increase $r_I$, the portion of the horizontal axis for which $\Delta < 0$ decreases, while we know from the LN (upper-left panel) that the region for which the final state is entangled increases. This example reminds us that  $\Delta < 0$ is not a necessary condition for entanglement (in other words, a positive value of $\Delta$ cannot rule out that the state is entangled).}  \label{fig:entanglementSQZthermalinput}
 }
\end{figure}

{\bf Single-mode squeezed vacuum input}. Lastly, we consider a situation in which   one of the inputs of the two-mode squeezer is a single-mode squeezed state. To illustrate the effect of initial single-mode squeezing, we also include thermal noise in the initial state. That is, we take the initial state to be a single-mode squeezed thermal state,
\bea  \label{sqzth} {\bm \mu}^{(\rm in)}&=&0\, ; \\  {\bm \sigma}^{(\rm in)}&=&(1+2\,  n_{\rm env})  \left(
\begin{array}{cccc}
  e^{2 r_I} & 0 & 0  & 0  \\ \nonumber 
 0 & e^{-2 r_I}& 0  & 0  \\
0  & 0  & 1 & 0 \\
 0 &0& 0 &1 \\
\end{array}
\right) \, ,\eea
where $r_I$ is the initial squeezing intensity on the first mode. [Note that we have chosen to squeeze the state in the direction $P^{(\rm in)}_1$. This choice does not alter the following discussion.]  As in previous examples, the final state is computed by simple multiplication with ${\bm S}_{\rm Sqz}(r,\phi)$, from which we can  study how different quantities of interest behave with $r_I$. We focus on the mean number of quanta in each output mode and on the entanglement between the modes. Figure~\ref{fig:numberSQZthermalinput} shows the number of quanta in each subsystem, $\langle \hat n_A\rangle$ and $\langle \hat n_B\rangle$, versus the initial squeezing intensity $r_I$ (the plot uses $r=1$ for the intensity of the two-mode squeezer, and $ n_{\rm env}=2$; the choice of squeezing angle $\phi$ does not change the result). The plot shows that larger values of $r_I$ induce additional creation of quanta (stimulated radiation). Figure~\ref{fig:entanglementSQZthermalinput} shows the entanglement between the two output modes versus the number of noise quanta $n_{\rm env}$, for various initial squeezing intensities $r_I$. This figure also shows (lower panel) an example for which the Cauchy-Schwarz inequality $\Delta < 0$ misses the presence of entanglement between the modes.

\begin{figure}
{\centering   
    \includegraphics[width=0.45\linewidth]{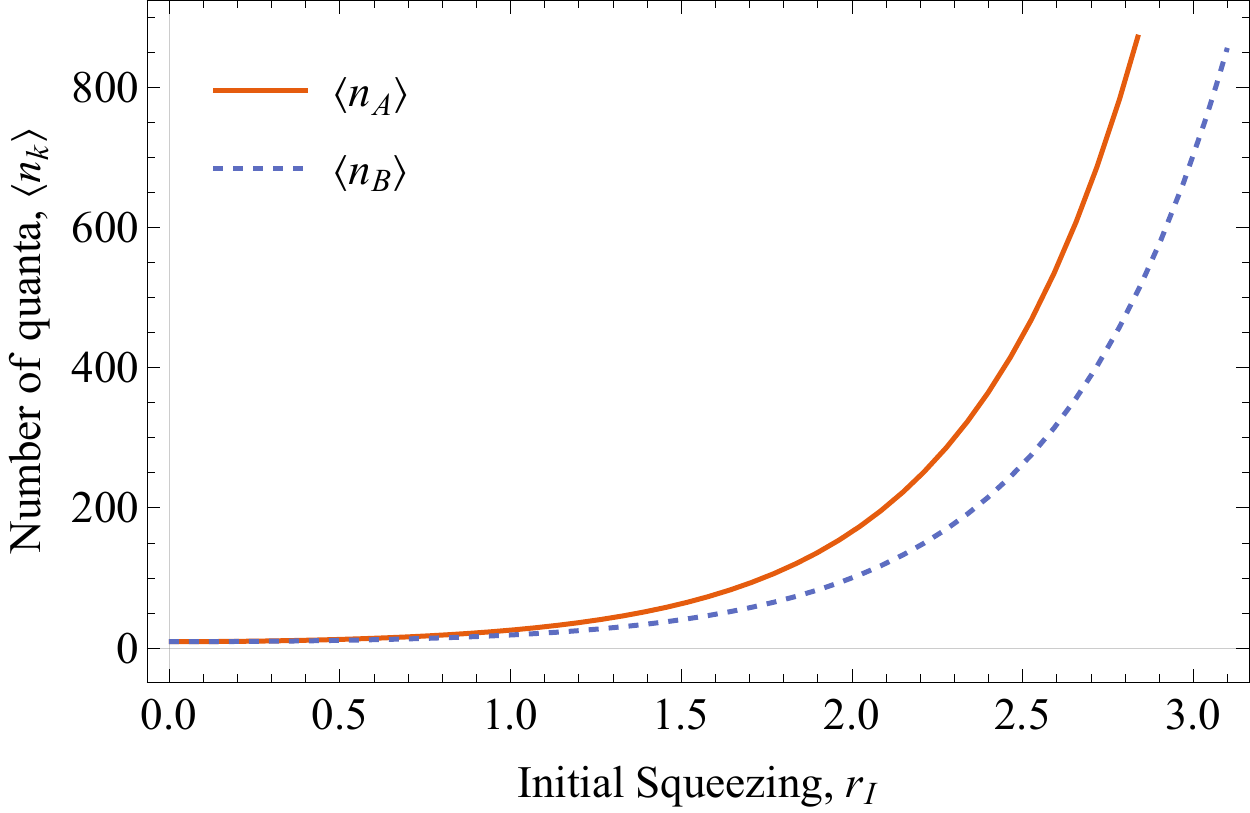}
      \caption{$\langle \hat n_A\rangle$ (solid) and $\langle \hat n_B\rangle$ (dashed), versus  initial squeezing intensity $r_I$.  The plot corresponds to $r=1$ and $n_{\rm env}=2$; the choice of squeezing angle $\phi$ does not change the result.}
     \label{fig:numberSQZthermalinput}
  }
\end{figure}

\subsection{Beam-splitter} 

A beam-splitter is a two-mode transformation defined by the matrix%
\be {\bm S}_{\rm BS}(\theta) =\left(
\begin{array}{cccc}
 \cos \theta  & 0 & \sin \theta  & 0 \\
 0 & \cos \theta & 0 & \sin \theta  \\
 -\sin \theta & 0 & \cos \theta & 0 \\
 0 & -\sin \theta & 0 & \cos \theta  \\
\end{array}
\right)\, .\ee
This is a simple rotation in the two-mode phase space, which splits the amplitudes of the inputs among the outputs. It represents, for instance, the effect of a potential barrier with transmission and reflection probabilities $\cos^2 \theta$ and $\sin^2 \theta$, respectively. [The two in-modes in this case are the waves approaching the barrier from right and left, and similarly, the out-modes are the ones leaving the barrier from both sides.] 
When acting on creation and annihilation operators, the transformations become
\bea \hat a^{\rm(in)}_1 & \to & \hat a^{\rm(out)}_1=\cos \theta \,  \hat a^{\rm(in)}_1+\sin \theta\,  \hat a^{\rm(in)}_2\, , \\
 \hat a^{\rm(in)}_2 &\to& \hat a^{\rm(out)}_2=\cos \theta\,  \hat a^{\rm(in)}_2- \sin \theta\,  \hat a^{\rm(in)}_1\, . 
\eea
We depict the action of a beam-splitter as in Fig.~\ref{fig:beamsplitter}.
The matrix ${\bm S}_{\rm BS}(\theta)$ satisfies, ${\bm S}_{\rm BS}(\theta){\bm S}^{\top}_{\rm BS}(\theta)=\bm{I}_4$, hence it belongs to the orthogonal subgroup of the symplectic group. This automatically implies that it leaves the vacuum invariant
\bea
{\bm \mu}^{(\rm in)}&=&0\ \longrightarrow {\bm \mu}^{(\rm out)}={\bm S}_{\rm BS}\cdot {\bm \mu}^{(\rm in)}=0 \\ \nonumber 
 {\bm \sigma}^{(\rm in)}&=&\bm{I}_4 \longrightarrow   {\bm \sigma}^{(\rm out)}= {\bm S}_{\rm BS}\cdot {\bm \sigma}^{(\rm in)}\cdot {\bm S}^{\top}_{\rm BS}= {\bm S}_{\rm BS}\cdot {\bm S}^{\top}_{\rm BS}=\bm{I}_4,\eea
which, in turn, implies that  the total number of quanta is left invariant via Eq.~\eqref{eq:mean_quanta}. For this reason, a beam-splitter is often referred to as a passive transformation.

When acting on a coherent state, the covariance matrix of the transformed state is again the identity,  and the first moments  ${\bm \mu}^{({\rm in})}=(\mu^{({\rm in})}_1,\mu^{({\rm in})}_2,\mu^{({\rm in})}_3,\mu^{({\rm in})}_4)$ transform to
\be  {\bm \mu}^{(\rm out)}={\bm S}_{\rm BS}\cdot {\bm \mu}^{(\rm in)}= (\mu^{({\rm in})}_1\, \cos \theta+\mu^{({\rm in})}_2\, \sin \theta,\mu^{({\rm in})}_2\, \cos \theta+\mu^{({\rm in})}_4\, \cos \theta,\mu^{({\rm in})}_3\, \cos \theta-\mu^{({\rm in})}_1\, \sin \theta,\mu^{({\rm in})}_4\, \cos \theta-\mu^{({\rm in})}_2\, \sin \theta)\, . \nonumber \ee

\begin{figure}
    \centering
    \includegraphics[width=.5\linewidth]{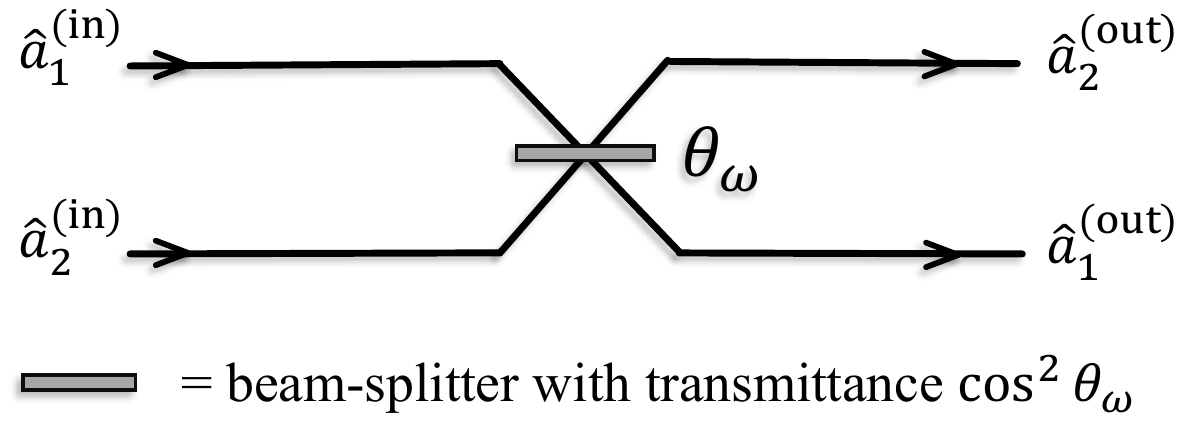}
    \caption{Schematic of a beam-splitter transformation with transmission probability $\cos^2\theta_\omega$.}
    \label{fig:beamsplitter}
\end{figure}

\section{Further comments on symplectic-circuit construction}\label{app:counting}
We are analyzing linear relationships between inputs and outputs of $N$ bosonic modes, due to symplectic transformations. Hence, the scattering matrix, $\bm S$, uniquely encoding such relationships is a $2N\times2N$ matrix and an element of the real symplectic group, ${\rm Sp}(2N,\mathbb{R})$. Generally, $\bm S$ has $(2N)^2$ free parameters. However the symplectic condition, $\bm S\bm\Omega\bm S^\top=\bm\Omega$, which is the only condition for $\bm S\in{\rm Sp}(2N,\mathbb{R})$, introduces $N(2N-1)$ constraints.\footnote{$\bm S\bm\Omega\bm S^\top$ is a $2N\times2N$ real anti-symmetric matrix, which has $2N(2N-1)/2=N(2N-1)$ free parameters corresponding to the number of constraints imposed on any $\bm S\in{\rm Sp}(2N,\mathbb{R})$.} Therefore, the number of parameters needed to describe any $\bm S\in{\rm Sp}(2N,\mathbb{R})$ is $\abs{{\rm Sp}(2N,\mathbb{R})}=2N^2+N$, which is just the dimension of $\bm S$ minus the number of constraints imposed on $\bm S$ by the symplectic condition. Without further information about, e.g., symmetries, allowed interactions etc., the dimension of the symplectic group sets an ultimate upper bound on the number of free parameters (i.e., circuit elements) required to fully describe a symplectic transformation on $N$ bosonic modes.

In analog-gravity models (dispersive theories), information about permissible interactions may be found from the dispersion relation for the modes~\cite{Corley:1996ar,linder16}. Information about the symplectic norm of interacting modes, in turn, tells us which modes transform passively (i.e., via beam-splitter-like transformations) or actively (i.e., via squeezing-like transformations) during the scattering process. For instance, negative-norm modes mix with positive-norm modes via active transformations only, whereas positive-norm (negative-norm) modes mix positive-norm (negative-norm) modes via passive transformations only. This information can substantially reduce the number of free parameters needed to describe interactions between the $N$ modes. Furthermore, dynamical symmetries (such as, e.g., space-time reversal symmetry in the white-black hole scattering; see Fig.~\ref{fig:wbh_circuit}) play a role here as well. We implicitly have such notions in mind when devising circuit-diagrams to explain multi-mode scattering events.


\section{Aspects of the entanglement witness $\Delta$ for analog horizons}\label{app:ent_compare}
Here we further illustrate the general differences between the LN and entanglement witness of $\Delta$ from Eq.~\eqref{eq:cs_inequality} with two examples. In Fig.~\ref{fig:Delta_eta_T}, we plot $\Delta$ for the correlated pair $a_{\omega}^{(\rm out)}$ and $a_{\omega}^{(\rm int)}$ of a black hole, assuming $\Gamma\approx1$ and $\omega/T_H=1$, which is the same parameter setting as in Fig.~\ref{fig:logneg_bh} for the LN between this pair of modes. A few observations can be made. First, the boundary curve in Fig.~\ref{fig:Delta_eta_T}, for which $\Delta=0$, demarcates the boundary between entangled (colorful region) and not entangled (null region),  agrees with the boundary curve in the LN of Fig.~\ref{fig:logneg_bh}. Hence, $\Delta$ is a good indicator of entanglement in this simple setting. However, it is not a quantifier of quantum correlations since, for instance, there are regions in Fig.~\ref{fig:Delta_eta_T} of higher environment temperature $T_{\rm env}$ where $\Delta$ is more negative, which is inconsistent with the LN. Incorrectly interpreting $\Delta$ as a quantifier of entanglement would thus have us deduce that increasing thermal fluctuations can increase entanglement between Hawking pairs in some regions of parameter space, which is unreasonable and, more importantly, disagrees with the behavior of the LN.

\begin{figure}[t]
    \centering
    \includegraphics[width=.5\linewidth]{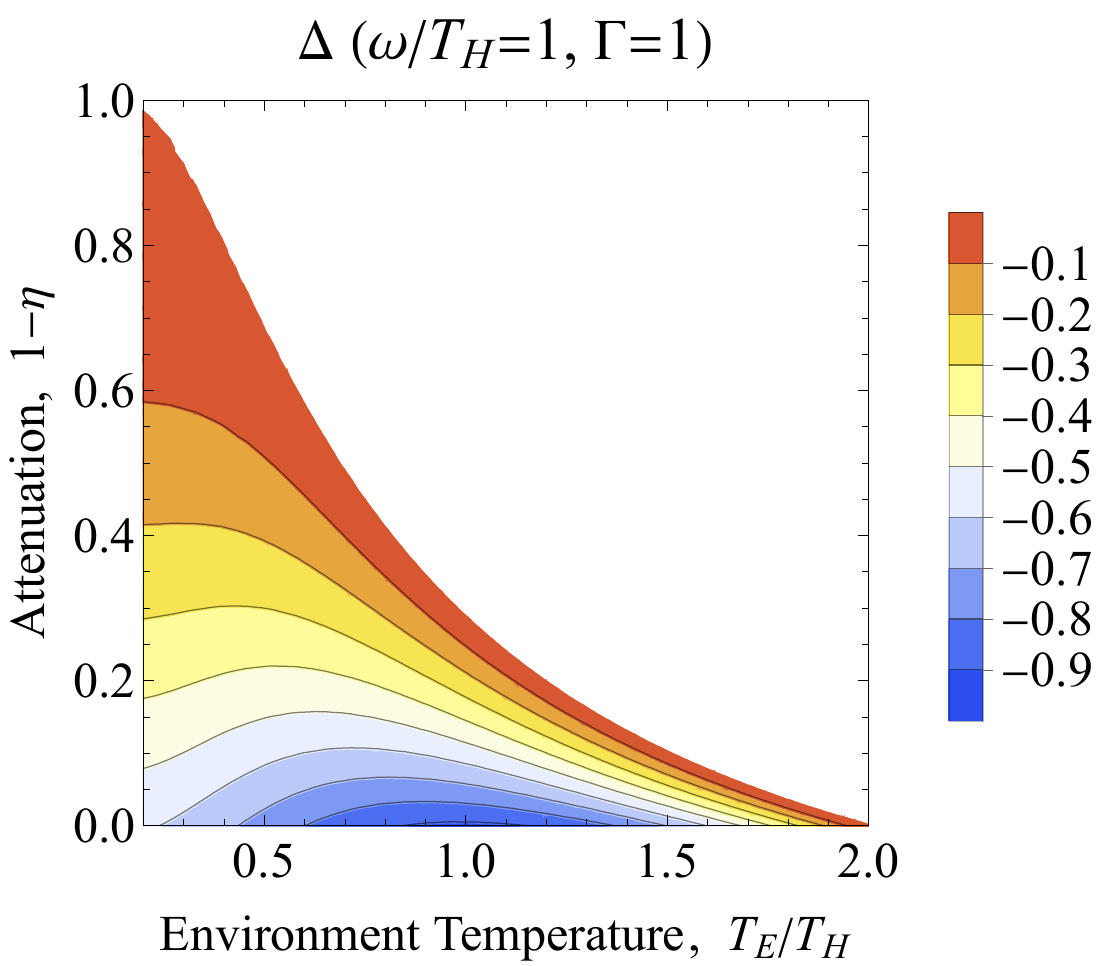}
    \caption{Plot of the entanglement witness $\Delta$ of Eq.~\eqref{eq:cs_inequality} ($\Delta<0$ signifies entanglement) for the output modes, $a_{\omega}^{(\rm out)}$ and $a_{\omega}^{(\rm int)}$, of a black hole. Boundary curve at $\Delta=0$ correctly demarcates the entanglement and no-entanglement regions, consistent with the boundary curve of the LN in Fig.~\ref{fig:logneg_bh} of the main text. Observe, however, that $\Delta$ is non-monotonic in $T_{\rm env}$, and hence non-monotonic with LN or any other entanglement quantifier.}
    \label{fig:Delta_eta_T}
\end{figure}

In Fig.~\ref{fig:delta_sq} (left panel), we plot $\Delta$ for the output pairs emitted by a white hole, $a_{k_4}^{(\rm out)}$ and $a_{k_1}^{(\rm out)}$, in a white-black hole setup, when the white hole is stimulated with a squeezed vacuum state. This is the setup described in Section~\ref{sec:sq_wbh}, but here, we take an isotropic comoving temperature $T_{\rm env}=T_H$ to highlight discrepancies between $\Delta$ and the LN. We see that $\Delta$ does not witness entanglement for squeezing levels $\e^{2s}\gtrsim3$, in contrast with the LN shown in Fig.~\ref{fig:delta_sq} (right panel), which shows a monotonic increase in the amount of entanglement with input squeezing (depending on the efficiency $\eta$). Thus, $\Delta$ does not serve as a good proxy of entanglement for stimulated Hawking radiation when initial squeezing is present in all regions of parameter space.

\section{Formulae} \label{app:formulae}
For completeness, we quote the y-intercepts, $b_i$, for the squeezing-enhanced setup discussed in Section~\ref{sec:sq_wbh},
\bea  \label{bs}
b_1&=&-\frac{\eta }{2}+\frac{1}{2} \eta  \sinh ^2r_H \cosh ^2r_H \left( N_{k_4}\, \cos ^4\theta -2  \,  N_{k_1}\, \sin ^2\theta
   \right)+\frac{1}{2} \eta    N_{k_1}\, \sin ^4\theta  \sinh ^4r_H
    \nonumber \\  \nonumber &+& \frac{1}{2} \eta    N_{k_2}\,  \sin ^2\theta  \cos
   ^2\theta  \sinh ^2r_H+\frac{1}{2} \eta    N_{k_3}\, \cos^2\theta  \sinh ^2r_H+\frac{1}{2} \eta    N_{k_1}\,  \cosh ^4r_H, \\
b_2&=&\frac{1}{2} \eta   N_{k_2}\,  \cos ^4 \theta +\frac{1}{2} \eta   N_{k_3} \, \sin ^2\theta -\frac{\eta }{2}+\frac{1}{8} \eta 
    N_{k_1}\, \sin ^2(2 \theta ) \sinh ^2r_H+\frac{1}{8} \eta  \, N_{k_4}\,  \sin ^2(2 \theta ) \cosh ^2r_H, \nonumber \\
b_3&=&  \frac{1}{2} \eta  \left(N_{k_2}\,   \sin ^2\theta +N_{k_1}\,   \cos ^2\theta \, \sinh ^2r_H+ N_{k_3}\,   \cos ^2\theta
   \cosh ^2r_H-1\right), \nonumber \\
b_4&=&-\frac{\eta}{2}+\frac{1}{2} \eta  \sinh ^2r_H \cosh^2r_H \left( N_{k_1}\, \, \cos^4 \theta -2  N_{k_4}\,  \sin^2\theta
   \right)+\frac{1}{2} \eta   N_{k_2}\,  \sin^2\theta  \cos^2\theta  \cosh ^2r_H \nonumber \\  \nonumber &+&\frac{1}{2} \eta   N_{k_3}\,  \cos^2\theta  \cosh ^2r_H+\frac{1}{2} \eta   N_{k_4}\,  \sin^4\theta  \cosh ^4r_H+\frac{1}{2} \eta   N_{k_4}\, \sinh^4r_H,
\eea
 where $N_{k_i}=1+2n_{k_i}$ and $n_{k_i}$ is the number of noisy quanta in the $i$th in mode.

\begin{figure}[t]
    \centering
    \includegraphics[width=.49\linewidth]{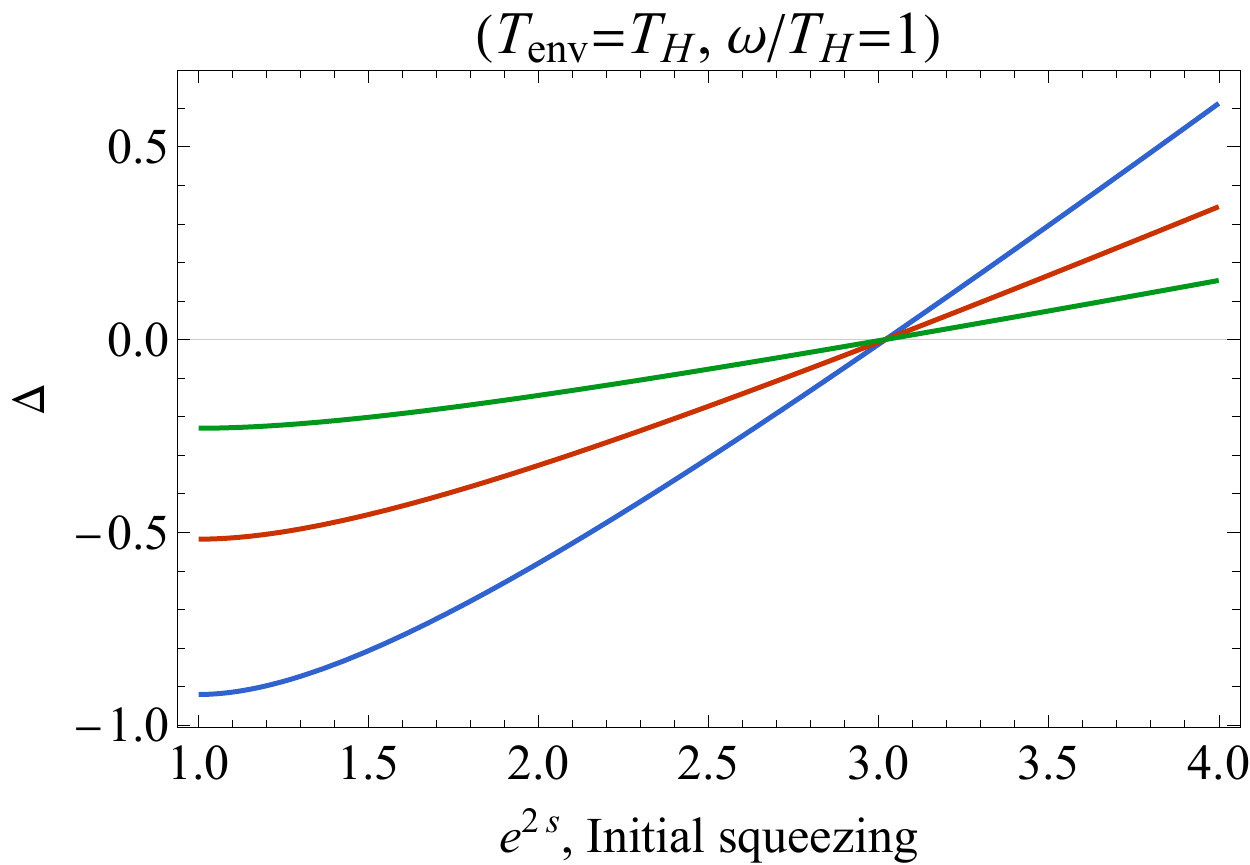}
    \includegraphics[width=.47\linewidth]{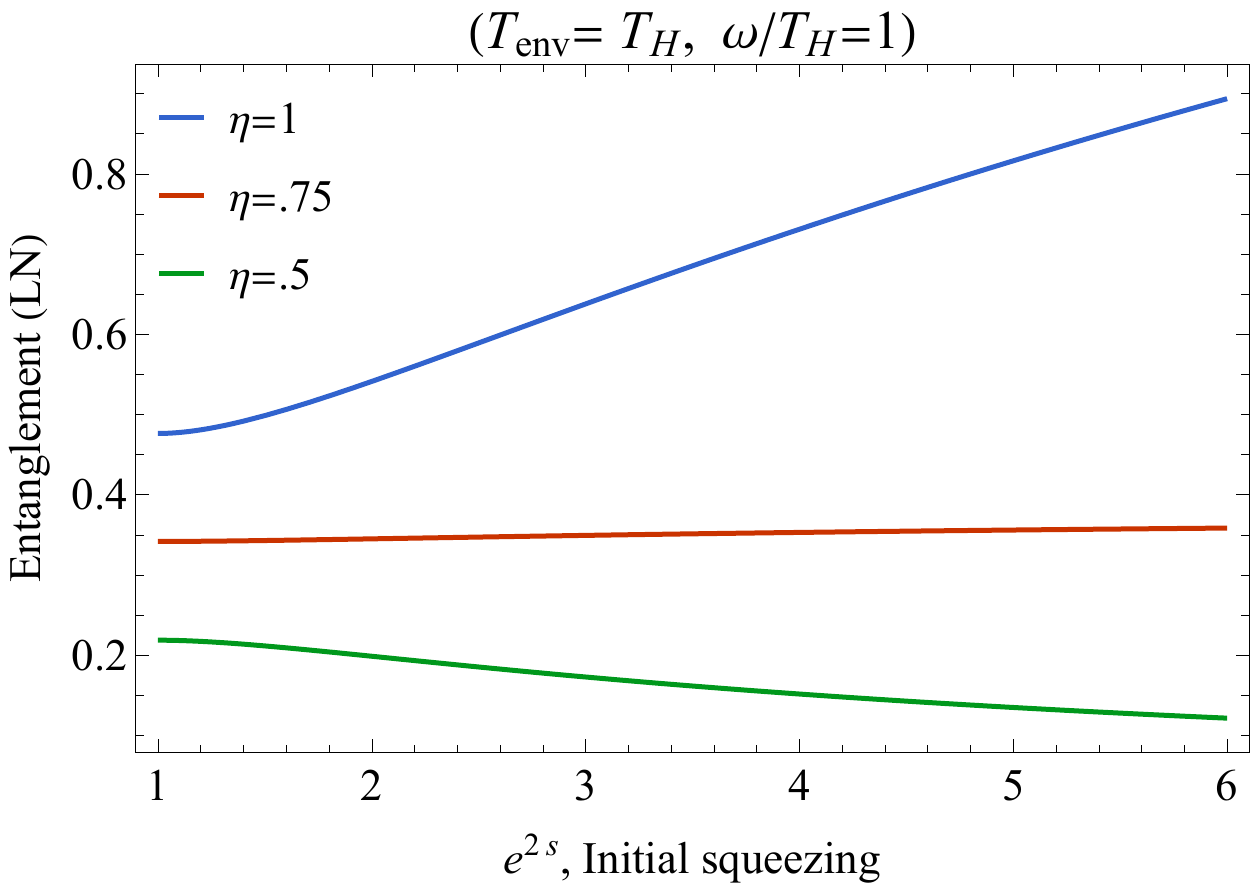}
    \caption{(Left panel) $\Delta$ between output modes, $a_{k_4}^{(\rm out)}$ and $a_{k_1}^{(\rm out)}$, of a stimulated white-black hole pair. The white hole is stimulated with a (noisy) single-mode squeezed vacuum of squeezing intensity $\e^{2s}$ and initial thermal fluctuations at temperature $T_{\rm env}=T_H$ in the comoving frame. Curves from bottom to top indicate increasing levels of attenuation ($1-\eta=0,.5,.25$). $\Delta$ does not faithfully indicate the presence of entanglement for $\e^{2s}>3$ since $\Delta$ is not in correspondence with the LN in this regime (right panel).}
    \label{fig:delta_sq}
\end{figure}
\end{widetext}


\pagebreak

\begin{thebibliography}{72}%
\makeatletter
\providecommand \@ifxundefined [1]{%
 \@ifx{#1\undefined}
}%
\providecommand \@ifnum [1]{%
 \ifnum #1\expandafter \@firstoftwo
 \else \expandafter \@secondoftwo
 \fi
}%
\providecommand \@ifx [1]{%
 \ifx #1\expandafter \@firstoftwo
 \else \expandafter \@secondoftwo
 \fi
}%
\providecommand \natexlab [1]{#1}%
\providecommand \enquote  [1]{``#1''}%
\providecommand \bibnamefont  [1]{#1}%
\providecommand \bibfnamefont [1]{#1}%
\providecommand \citenamefont [1]{#1}%
\providecommand \href@noop [0]{\@secondoftwo}%
\providecommand \href [0]{\begingroup \@sanitize@url \@href}%
\providecommand \@href[1]{\@@startlink{#1}\@@href}%
\providecommand \@@href[1]{\endgroup#1\@@endlink}%
\providecommand \@sanitize@url [0]{\catcode `\\12\catcode `\$12\catcode
  `\&12\catcode `\#12\catcode `\^12\catcode `\_12\catcode `\%12\relax}%
\providecommand \@@startlink[1]{}%
\providecommand \@@endlink[0]{}%
\providecommand \url  [0]{\begingroup\@sanitize@url \@url }%
\providecommand \@url [1]{\endgroup\@href {#1}{\urlprefix }}%
\providecommand \urlprefix  [0]{URL }%
\providecommand \Eprint [0]{\href }%
\providecommand \doibase [0]{https://doi.org/}%
\providecommand \selectlanguage [0]{\@gobble}%
\providecommand \bibinfo  [0]{\@secondoftwo}%
\providecommand \bibfield  [0]{\@secondoftwo}%
\providecommand \translation [1]{[#1]}%
\providecommand \BibitemOpen [0]{}%
\providecommand \bibitemStop [0]{}%
\providecommand \bibitemNoStop [0]{.\EOS\space}%
\providecommand \EOS [0]{\spacefactor3000\relax}%
\providecommand \BibitemShut  [1]{\csname bibitem#1\endcsname}%
\let\auto@bib@innerbib\@empty
\bibitem [{\citenamefont {{Hawking, Stephen W}}(1974)}]{Hawking74}%
  \BibitemOpen
  \bibfield  {author} {\bibinfo {author} {\bibnamefont {{Hawking, Stephen
  W}}},\ }\bibfield  {title} {\bibinfo {title} {Black hole explosions?},\
  }\href {https://doi.org/10.1038/248030a0} {\bibfield  {journal} {\bibinfo
  {journal} {Nature}\ }\textbf {\bibinfo {volume} {248}},\ \bibinfo {pages}
  {30} (\bibinfo {year} {1974})}\BibitemShut {NoStop}%
\bibitem [{\citenamefont {{Hawking, Stephen W}}(1975)}]{Hawking75}%
  \BibitemOpen
  \bibfield  {author} {\bibinfo {author} {\bibnamefont {{Hawking, Stephen
  W}}},\ }\bibfield  {title} {\bibinfo {title} {Particle creation by black
  holes},\ }\href {https://doi.org/10.1007/BF02345020} {\bibfield  {journal}
  {\bibinfo  {journal} {Communications in Mathematical Physics}\ }\textbf
  {\bibinfo {volume} {43}},\ \bibinfo {pages} {199} (\bibinfo {year}
  {1975})}\BibitemShut {NoStop}%
\bibitem [{\citenamefont {Hawking}(1976)}]{Hawking:1976ra}%
  \BibitemOpen
  \bibfield  {author} {\bibinfo {author} {\bibfnamefont {S.~W.}\ \bibnamefont
  {Hawking}},\ }\bibfield  {title} {\bibinfo {title} {{Breakdown of
  Predictability in Gravitational Collapse}},\ }\href
  {https://doi.org/10.1103/PhysRevD.14.2460} {\bibfield  {journal} {\bibinfo
  {journal} {Phys. Rev. D}\ }\textbf {\bibinfo {volume} {14}},\ \bibinfo
  {pages} {2460} (\bibinfo {year} {1976})}\BibitemShut {NoStop}%
\bibitem [{\citenamefont {Israel}(1976)}]{israel1976thermo}%
  \BibitemOpen
  \bibfield  {author} {\bibinfo {author} {\bibfnamefont {W.}~\bibnamefont
  {Israel}},\ }\bibfield  {title} {\bibinfo {title} {Thermo-field dynamics of
  black holes},\ }\href {https://doi.org/10.1016/0375-9601(76)90178-X}
  {\bibfield  {journal} {\bibinfo  {journal} {Physics Letters A}\ }\textbf
  {\bibinfo {volume} {57}},\ \bibinfo {pages} {107} (\bibinfo {year}
  {1976})}\BibitemShut {NoStop}%
\bibitem [{\citenamefont {{Unruh, William George}}(1981)}]{unruh81}%
  \BibitemOpen
  \bibfield  {author} {\bibinfo {author} {\bibnamefont {{Unruh, William
  George}}},\ }\bibfield  {title} {\bibinfo {title} {Experimental black-hole
  evaporation?},\ }\href {https://doi.org/10.1103/PhysRevLett.46.1351}
  {\bibfield  {journal} {\bibinfo  {journal} {Physical Review Letters}\
  }\textbf {\bibinfo {volume} {46}},\ \bibinfo {pages} {1351} (\bibinfo {year}
  {1981})}\BibitemShut {NoStop}%
\bibitem [{\citenamefont {Michel}\ and\ \citenamefont
  {Parentani}(2014)}]{Michel:2014zsa}%
  \BibitemOpen
  \bibfield  {author} {\bibinfo {author} {\bibfnamefont {F.}~\bibnamefont
  {Michel}}\ and\ \bibinfo {author} {\bibfnamefont {R.}~\bibnamefont
  {Parentani}},\ }\bibfield  {title} {\bibinfo {title} {{Probing the thermal
  character of analogue Hawking radiation for shallow water waves?}},\ }\href
  {https://doi.org/10.1103/PhysRevD.90.044033} {\bibfield  {journal} {\bibinfo
  {journal} {Phys. Rev. D}\ }\textbf {\bibinfo {volume} {90}},\ \bibinfo
  {pages} {044033} (\bibinfo {year} {2014})},\ \Eprint
  {https://arxiv.org/abs/1404.7482} {arXiv:1404.7482 [gr-qc]} \BibitemShut
  {NoStop}%
\bibitem [{\citenamefont {{Demircan, A and Amiranashvili, Sh and Steinmeyer,
  G}}(2011)}]{demircan11TRANSISTOR}%
  \BibitemOpen
  \bibfield  {author} {\bibinfo {author} {\bibnamefont {{Demircan, A and
  Amiranashvili, Sh and Steinmeyer, G}}},\ }\bibfield  {title} {\bibinfo
  {title} {Controlling light by light with an optical event horizon},\ }\href
  {https://doi.org/10.1103/PhysRevLett.106.163901} {\bibfield  {journal}
  {\bibinfo  {journal} {Physical Review Letters}\ }\textbf {\bibinfo {volume}
  {106}},\ \bibinfo {pages} {163901} (\bibinfo {year} {2011})}\BibitemShut
  {NoStop}%
\bibitem [{\citenamefont {{Rubino, Elenora and Lotti, A and Belgiorno, F and
  Cacciatori, SL and Couairon, Arnaud and Leonhardt, Ulf and Faccio,
  D}}(2012)}]{rubino2012soliton}%
  \BibitemOpen
  \bibfield  {author} {\bibinfo {author} {\bibnamefont {{Rubino, Elenora and
  Lotti, A and Belgiorno, F and Cacciatori, SL and Couairon, Arnaud and
  Leonhardt, Ulf and Faccio, D}}},\ }\bibfield  {title} {\bibinfo {title}
  {Soliton-induced relativistic-scattering and amplification},\ }\href
  {https://doi.org/10.1038/srep00932} {\bibfield  {journal} {\bibinfo
  {journal} {Scientific Reports}\ }\textbf {\bibinfo {volume} {2}},\ \bibinfo
  {pages} {1} (\bibinfo {year} {2012})}\BibitemShut {NoStop}%
\bibitem [{\citenamefont {{Petev, Mike and Westerberg, Niclas and Moss, Daniel
  and Rubino, Elenora and Rimoldi, C and Cacciatori, SL and Belgiorno, F and
  Faccio, D}}(2013)}]{petev2013blackbody}%
  \BibitemOpen
  \bibfield  {author} {\bibinfo {author} {\bibnamefont {{Petev, Mike and
  Westerberg, Niclas and Moss, Daniel and Rubino, Elenora and Rimoldi, C and
  Cacciatori, SL and Belgiorno, F and Faccio, D}}},\ }\bibfield  {title}
  {\bibinfo {title} {Blackbody emission from light interacting with an
  effective moving dispersive medium},\ }\href
  {https://doi.org/10.1103/PhysRevLett.111.043902} {\bibfield  {journal}
  {\bibinfo  {journal} {Physical Review Letters}\ }\textbf {\bibinfo {volume}
  {111}},\ \bibinfo {pages} {043902} (\bibinfo {year} {2013})}\BibitemShut
  {NoStop}%
\bibitem [{\citenamefont {Finazzi}\ and\ \citenamefont
  {Carusotto}(2013)}]{finazzi13}%
  \BibitemOpen
  \bibfield  {author} {\bibinfo {author} {\bibfnamefont {S.}~\bibnamefont
  {Finazzi}}\ and\ \bibinfo {author} {\bibfnamefont {I.}~\bibnamefont
  {Carusotto}},\ }\bibfield  {title} {\bibinfo {title} {Quantum vacuum emission
  in a nonlinear optical medium illuminated by a strong laser pulse},\ }\href
  {https://doi.org/10.1103/PhysRevA.87.023803} {\bibfield  {journal} {\bibinfo
  {journal} {Physical Review A}\ }\textbf {\bibinfo {volume} {87}},\ \bibinfo
  {pages} {023803} (\bibinfo {year} {2013})}\BibitemShut {NoStop}%
\bibitem [{\citenamefont {Belgiorno}\ \emph {et~al.}(2015)\citenamefont
  {Belgiorno}, \citenamefont {Cacciatori},\ and\ \citenamefont
  {Dalla~Piazza}}]{Belgiorno:2014ana}%
  \BibitemOpen
  \bibfield  {author} {\bibinfo {author} {\bibfnamefont {F.}~\bibnamefont
  {Belgiorno}}, \bibinfo {author} {\bibfnamefont {S.~L.}\ \bibnamefont
  {Cacciatori}},\ and\ \bibinfo {author} {\bibfnamefont {F.}~\bibnamefont
  {Dalla~Piazza}},\ }\bibfield  {title} {\bibinfo {title} {{Hawking effect in
  dielectric media and the Hopfield model}},\ }\href
  {https://doi.org/10.1103/PhysRevD.91.124063} {\bibfield  {journal} {\bibinfo
  {journal} {Phys. Rev. D}\ }\textbf {\bibinfo {volume} {91}},\ \bibinfo
  {pages} {124063} (\bibinfo {year} {2015})},\ \Eprint
  {https://arxiv.org/abs/1411.7870} {arXiv:1411.7870 [gr-qc]} \BibitemShut
  {NoStop}%
\bibitem [{\citenamefont {{Linder, Malte F and Sch{\"u}tzhold, Ralf and Unruh,
  William G}}(2016)}]{linder16}%
  \BibitemOpen
  \bibfield  {author} {\bibinfo {author} {\bibnamefont {{Linder, Malte F and
  Sch{\"u}tzhold, Ralf and Unruh, William G}}},\ }\bibfield  {title} {\bibinfo
  {title} {{Derivation of Hawking radiation in dispersive dielectric media}},\
  }\href {https://doi.org/10.1103/PhysRevD.93.104010} {\bibfield  {journal}
  {\bibinfo  {journal} {Physical Review D}\ }\textbf {\bibinfo {volume} {93}},\
  \bibinfo {pages} {104010} (\bibinfo {year} {2016})}\BibitemShut {NoStop}%
\bibitem [{\citenamefont {Bermudez}\ and\ \citenamefont
  {Leonhardt}(2016)}]{Bermudez:2016hbl}%
  \BibitemOpen
  \bibfield  {author} {\bibinfo {author} {\bibfnamefont {D.}~\bibnamefont
  {Bermudez}}\ and\ \bibinfo {author} {\bibfnamefont {U.}~\bibnamefont
  {Leonhardt}},\ }\bibfield  {title} {\bibinfo {title} {{Hawking spectrum for a
  fiber-optical analog of the event horizon}},\ }\href
  {https://doi.org/10.1103/PhysRevA.93.053820} {\bibfield  {journal} {\bibinfo
  {journal} {Phys. Rev. A}\ }\textbf {\bibinfo {volume} {93}},\ \bibinfo
  {pages} {053820} (\bibinfo {year} {2016})},\ \Eprint
  {https://arxiv.org/abs/1601.06816} {arXiv:1601.06816 [gr-qc]} \BibitemShut
  {NoStop}%
\bibitem [{\citenamefont {Belgiorno}\ \emph {et~al.}(2017)\citenamefont
  {Belgiorno}, \citenamefont {Cacciatori}, \citenamefont {Dalla~Piazza},\ and\
  \citenamefont {Doronzo}}]{Belgiorno:2017glw}%
  \BibitemOpen
  \bibfield  {author} {\bibinfo {author} {\bibfnamefont {F.}~\bibnamefont
  {Belgiorno}}, \bibinfo {author} {\bibfnamefont {S.~L.}\ \bibnamefont
  {Cacciatori}}, \bibinfo {author} {\bibfnamefont {F.}~\bibnamefont
  {Dalla~Piazza}},\ and\ \bibinfo {author} {\bibfnamefont {M.}~\bibnamefont
  {Doronzo}},\ }\bibfield  {title} {\bibinfo {title} {{Hopfield-Kerr model and
  analogue black hole radiation in dielectrics}},\ }\href
  {https://doi.org/10.1103/PhysRevD.96.096024} {\bibfield  {journal} {\bibinfo
  {journal} {Phys. Rev. D}\ }\textbf {\bibinfo {volume} {96}},\ \bibinfo
  {pages} {096024} (\bibinfo {year} {2017})},\ \Eprint
  {https://arxiv.org/abs/1707.01663} {arXiv:1707.01663 [hep-th]} \BibitemShut
  {NoStop}%
\bibitem [{\citenamefont {{Jacquet, Maxime J and K{\"o}nig,
  Friedrich}}(2020)}]{jacquet20emission}%
  \BibitemOpen
  \bibfield  {author} {\bibinfo {author} {\bibnamefont {{Jacquet, Maxime J and
  K{\"o}nig, Friedrich}}},\ }\bibfield  {title} {\bibinfo {title} {Analytical
  description of quantum emission in optical analogs to gravity},\ }\href
  {https://doi.org/10.1103/PhysRevA.102.013725} {\bibfield  {journal} {\bibinfo
   {journal} {Physical Review A}\ }\textbf {\bibinfo {volume} {102}},\ \bibinfo
  {pages} {013725} (\bibinfo {year} {2020})}\BibitemShut {NoStop}%
\bibitem [{\citenamefont {Jacquet}\ and\ \citenamefont
  {Koenig}(2020)}]{Jacquet:2020jpj}%
  \BibitemOpen
  \bibfield  {author} {\bibinfo {author} {\bibfnamefont {M.}~\bibnamefont
  {Jacquet}}\ and\ \bibinfo {author} {\bibfnamefont {F.}~\bibnamefont
  {Koenig}},\ }\bibfield  {title} {\bibinfo {title} {The influence of spacetime
  curvature on quantum emission in optical analogues to gravity},\ }\bibfield
  {journal} {\bibinfo  {journal} {SciPost Physics Core}\ }\textbf {\bibinfo
  {volume} {3}},\ \href {https://doi.org/10.21468/scipostphyscore.3.1.005}
  {10.21468/scipostphyscore.3.1.005} (\bibinfo {year} {2020})\BibitemShut
  {NoStop}%
\bibitem [{\citenamefont {{Rosenberg, Yuval}}(2020)}]{rosenberg2020optical}%
  \BibitemOpen
  \bibfield  {author} {\bibinfo {author} {\bibnamefont {{Rosenberg, Yuval}}},\
  }\bibfield  {title} {\bibinfo {title} {Optical analogues of black-hole
  horizons},\ }\href {https://doi.org/10.1098/rsta.2019.0232} {\bibfield
  {journal} {\bibinfo  {journal} {Philosophical Transactions of the Royal
  Society A}\ }\textbf {\bibinfo {volume} {378}},\ \bibinfo {pages} {20190232}
  (\bibinfo {year} {2020})}\BibitemShut {NoStop}%
\bibitem [{\citenamefont {Aguero-Santacruz}\ and\ \citenamefont
  {Bermudez}(2020)}]{Aguero-Santacruz:2020krw}%
  \BibitemOpen
  \bibfield  {author} {\bibinfo {author} {\bibfnamefont {R.}~\bibnamefont
  {Aguero-Santacruz}}\ and\ \bibinfo {author} {\bibfnamefont {D.}~\bibnamefont
  {Bermudez}},\ }\bibfield  {title} {\bibinfo {title} {{Hawking radiation in
  optics and beyond}},\ }\href {https://doi.org/10.1098/rsta.2019.0223}
  {\bibfield  {journal} {\bibinfo  {journal} {Phil. Trans. Roy. Soc. Lond. A}\
  }\textbf {\bibinfo {volume} {378}},\ \bibinfo {pages} {20190223} (\bibinfo
  {year} {2020})},\ \Eprint {https://arxiv.org/abs/2002.07907}
  {arXiv:2002.07907 [gr-qc]} \BibitemShut {NoStop}%
\bibitem [{\citenamefont {Garay}\ \emph {et~al.}(2000)\citenamefont {Garay},
  \citenamefont {Anglin}, \citenamefont {Cirac},\ and\ \citenamefont
  {Zoller}}]{cirac2000}%
  \BibitemOpen
  \bibfield  {author} {\bibinfo {author} {\bibfnamefont {L.~J.}\ \bibnamefont
  {Garay}}, \bibinfo {author} {\bibfnamefont {J.}~\bibnamefont {Anglin}},
  \bibinfo {author} {\bibfnamefont {J.~I.}\ \bibnamefont {Cirac}},\ and\
  \bibinfo {author} {\bibfnamefont {P.}~\bibnamefont {Zoller}},\ }\bibfield
  {title} {\bibinfo {title} {Sonic analog of gravitational black holes in
  bose-einstein condensates},\ }\href
  {https://doi.org/10.1103/PhysRevLett.85.4643} {\bibfield  {journal} {\bibinfo
   {journal} {Physical Review Letters}\ }\textbf {\bibinfo {volume} {85}},\
  \bibinfo {pages} {4643} (\bibinfo {year} {2000})}\BibitemShut {NoStop}%
\bibitem [{\citenamefont {Macher}\ and\ \citenamefont
  {Parentani}(2009{\natexlab{a}})}]{Parentani09}%
  \BibitemOpen
  \bibfield  {author} {\bibinfo {author} {\bibfnamefont {J.}~\bibnamefont
  {Macher}}\ and\ \bibinfo {author} {\bibfnamefont {R.}~\bibnamefont
  {Parentani}},\ }\bibfield  {title} {\bibinfo {title} {Black-hole radiation in
  bose-einstein condensates},\ }\href
  {https://doi.org/10.1103/PhysRevA.80.043601} {\bibfield  {journal} {\bibinfo
  {journal} {Physical Review A}\ }\textbf {\bibinfo {volume} {80}},\ \bibinfo
  {pages} {043601} (\bibinfo {year} {2009}{\natexlab{a}})}\BibitemShut
  {NoStop}%
\bibitem [{\citenamefont {Macher}\ and\ \citenamefont
  {Parentani}(2009{\natexlab{b}})}]{macher2009}%
  \BibitemOpen
  \bibfield  {author} {\bibinfo {author} {\bibfnamefont {J.}~\bibnamefont
  {Macher}}\ and\ \bibinfo {author} {\bibfnamefont {R.}~\bibnamefont
  {Parentani}},\ }\bibfield  {title} {\bibinfo {title} {Black/white hole
  radiation from dispersive theories},\ }\href
  {https://doi.org/10.1103/PhysRevD.79.124008} {\bibfield  {journal} {\bibinfo
  {journal} {Physical Review D}\ }\textbf {\bibinfo {volume} {79}},\ \bibinfo
  {pages} {124008} (\bibinfo {year} {2009}{\natexlab{b}})}\BibitemShut
  {NoStop}%
\bibitem [{\citenamefont {Finazzi}\ and\ \citenamefont
  {Parentani}(2011{\natexlab{a}})}]{Finazzi:2010yq}%
  \BibitemOpen
  \bibfield  {author} {\bibinfo {author} {\bibfnamefont {S.}~\bibnamefont
  {Finazzi}}\ and\ \bibinfo {author} {\bibfnamefont {R.}~\bibnamefont
  {Parentani}},\ }\bibfield  {title} {\bibinfo {title} {{Spectral properties of
  acoustic black hole radiation: broadening the horizon}},\ }\href
  {https://doi.org/10.1103/PhysRevD.83.084010} {\bibfield  {journal} {\bibinfo
  {journal} {Phys. Rev. D}\ }\textbf {\bibinfo {volume} {83}},\ \bibinfo
  {pages} {084010} (\bibinfo {year} {2011}{\natexlab{a}})},\ \Eprint
  {https://arxiv.org/abs/1012.1556} {arXiv:1012.1556 [gr-qc]} \BibitemShut
  {NoStop}%
\bibitem [{\citenamefont {Finazzi}\ and\ \citenamefont
  {Parentani}(2011{\natexlab{b}})}]{Finazzi:2011jd}%
  \BibitemOpen
  \bibfield  {author} {\bibinfo {author} {\bibfnamefont {S.}~\bibnamefont
  {Finazzi}}\ and\ \bibinfo {author} {\bibfnamefont {R.}~\bibnamefont
  {Parentani}},\ }\bibfield  {title} {\bibinfo {title} {{On the robustness of
  acoustic black hole spectra}},\ }\href
  {https://doi.org/10.1088/1742-6596/314/1/012030} {\bibfield  {journal}
  {\bibinfo  {journal} {J. Phys. Conf. Ser.}\ }\textbf {\bibinfo {volume}
  {314}},\ \bibinfo {pages} {012030} (\bibinfo {year} {2011}{\natexlab{b}})},\
  \Eprint {https://arxiv.org/abs/1102.1452} {arXiv:1102.1452 [gr-qc]}
  \BibitemShut {NoStop}%
\bibitem [{\citenamefont {Finazzi}\ and\ \citenamefont
  {Parentani}(2012)}]{Finazzi:2012iu}%
  \BibitemOpen
  \bibfield  {author} {\bibinfo {author} {\bibfnamefont {S.}~\bibnamefont
  {Finazzi}}\ and\ \bibinfo {author} {\bibfnamefont {R.}~\bibnamefont
  {Parentani}},\ }\bibfield  {title} {\bibinfo {title} {{Hawking radiation in
  dispersive theories, the two regimes}},\ }\href
  {https://doi.org/10.1103/PhysRevD.85.124027} {\bibfield  {journal} {\bibinfo
  {journal} {Phys. Rev. D}\ }\textbf {\bibinfo {volume} {85}},\ \bibinfo
  {pages} {124027} (\bibinfo {year} {2012})},\ \Eprint
  {https://arxiv.org/abs/1202.6015} {arXiv:1202.6015 [gr-qc]} \BibitemShut
  {NoStop}%
\bibitem [{\citenamefont {Busch}\ \emph {et~al.}(2014)\citenamefont {Busch},
  \citenamefont {Carusotto},\ and\ \citenamefont {Parentani}}]{Busch:2013gna}%
  \BibitemOpen
  \bibfield  {author} {\bibinfo {author} {\bibfnamefont {X.}~\bibnamefont
  {Busch}}, \bibinfo {author} {\bibfnamefont {I.}~\bibnamefont {Carusotto}},\
  and\ \bibinfo {author} {\bibfnamefont {R.}~\bibnamefont {Parentani}},\
  }\bibfield  {title} {\bibinfo {title} {{Spectrum and entanglement of phonons
  in quantum fluids of light}},\ }\href
  {https://doi.org/10.1103/PhysRevA.89.043819} {\bibfield  {journal} {\bibinfo
  {journal} {Phys. Rev. A}\ }\textbf {\bibinfo {volume} {89}},\ \bibinfo
  {pages} {043819} (\bibinfo {year} {2014})},\ \Eprint
  {https://arxiv.org/abs/1311.3507} {arXiv:1311.3507 [cond-mat.quant-gas]}
  \BibitemShut {NoStop}%
\bibitem [{\citenamefont {Busch}\ and\ \citenamefont
  {Parentani}(2014)}]{busch14}%
  \BibitemOpen
  \bibfield  {author} {\bibinfo {author} {\bibfnamefont {X.}~\bibnamefont
  {Busch}}\ and\ \bibinfo {author} {\bibfnamefont {R.}~\bibnamefont
  {Parentani}},\ }\bibfield  {title} {\bibinfo {title} {{Quantum entanglement
  in analogue Hawking radiation: When is the final state nonseparable?}},\
  }\href {https://doi.org/10.1103/PhysRevD.89.105024} {\bibfield  {journal}
  {\bibinfo  {journal} {Physical Review D}\ }\textbf {\bibinfo {volume} {89}},\
  \bibinfo {pages} {105024} (\bibinfo {year} {2014})}\BibitemShut {NoStop}%
\bibitem [{\citenamefont {Michel}\ \emph {et~al.}(2016)\citenamefont {Michel},
  \citenamefont {Coupechoux},\ and\ \citenamefont
  {Parentani}}]{Michel:2016tog}%
  \BibitemOpen
  \bibfield  {author} {\bibinfo {author} {\bibfnamefont {F.}~\bibnamefont
  {Michel}}, \bibinfo {author} {\bibfnamefont {J.-F.}\ \bibnamefont
  {Coupechoux}},\ and\ \bibinfo {author} {\bibfnamefont {R.}~\bibnamefont
  {Parentani}},\ }\bibfield  {title} {\bibinfo {title} {{Phonon spectrum and
  correlations in a transonic flow of an atomic Bose gas}},\ }\href
  {https://doi.org/10.1103/PhysRevD.94.084027} {\bibfield  {journal} {\bibinfo
  {journal} {Phys. Rev. D}\ }\textbf {\bibinfo {volume} {94}},\ \bibinfo
  {pages} {084027} (\bibinfo {year} {2016})},\ \Eprint
  {https://arxiv.org/abs/1605.09752} {arXiv:1605.09752 [cond-mat.quant-gas]}
  \BibitemShut {NoStop}%
\bibitem [{\citenamefont {Nambu}\ and\ \citenamefont {Osawa}(2021)}]{nambu21}%
  \BibitemOpen
  \bibfield  {author} {\bibinfo {author} {\bibfnamefont {Y.}~\bibnamefont
  {Nambu}}\ and\ \bibinfo {author} {\bibfnamefont {Y.}~\bibnamefont {Osawa}},\
  }\bibfield  {title} {\bibinfo {title} {Tripartite entanglement of hawking
  radiation in dispersive model},\ }\href
  {https://doi.org/10.1103/PhysRevD.103.125007} {\bibfield  {journal} {\bibinfo
   {journal} {Phys. Rev. D}\ }\textbf {\bibinfo {volume} {103}},\ \bibinfo
  {pages} {125007} (\bibinfo {year} {2021})}\BibitemShut {NoStop}%
\bibitem [{\citenamefont {{Visser, Matt}}(2003)}]{Visser2003}%
  \BibitemOpen
  \bibfield  {author} {\bibinfo {author} {\bibnamefont {{Visser, Matt}}},\
  }\bibfield  {title} {\bibinfo {title} {{Essential and inessential features of
  Hawking radiation}},\ }\href {https://doi.org/10.1142/S0218271803003190}
  {\bibfield  {journal} {\bibinfo  {journal} {International Journal of Modern
  Physics D}\ }\textbf {\bibinfo {volume} {12}},\ \bibinfo {pages} {649}
  (\bibinfo {year} {2003})}\BibitemShut {NoStop}%
\bibitem [{\citenamefont {{Novello, M{\'a}rio and Visser, Matt and Volovik,
  Grigory E}}(2002)}]{novello02ArtificialBHs}%
  \BibitemOpen
  \bibfield  {author} {\bibinfo {author} {\bibnamefont {{Novello, M{\'a}rio and
  Visser, Matt and Volovik, Grigory E}}},\ }\href@noop {} {\emph {\bibinfo
  {title} {Artificial black holes}}}\ (\bibinfo  {publisher} {World
  Scientific},\ \bibinfo {year} {2002})\BibitemShut {NoStop}%
\bibitem [{\citenamefont {{Barcel{\'o}, Carlos and Liberati, Stefano and
  Visser, Matt}}(2011)}]{barcelo11}%
  \BibitemOpen
  \bibfield  {author} {\bibinfo {author} {\bibnamefont {{Barcel{\'o}, Carlos
  and Liberati, Stefano and Visser, Matt}}},\ }\bibfield  {title} {\bibinfo
  {title} {Analogue gravity},\ }\href {https://doi.org/10.12942/lrr-2005-12}
  {\bibfield  {journal} {\bibinfo  {journal} {Living Reviews in Relativity}\
  }\textbf {\bibinfo {volume} {14}},\ \bibinfo {pages} {1} (\bibinfo {year}
  {2011})}\BibitemShut {NoStop}%
\bibitem [{\citenamefont {{Barcel{\'o}, Carlos}}(2019)}]{barcelo2019analogue}%
  \BibitemOpen
  \bibfield  {author} {\bibinfo {author} {\bibnamefont {{Barcel{\'o},
  Carlos}}},\ }\bibfield  {title} {\bibinfo {title} {Analogue black-hole
  horizons},\ }\href {https://doi.org/10.1038/s41567-018-0367-6} {\bibfield
  {journal} {\bibinfo  {journal} {Nature Physics}\ }\textbf {\bibinfo {volume}
  {15}},\ \bibinfo {pages} {210} (\bibinfo {year} {2019})}\BibitemShut
  {NoStop}%
\bibitem [{\citenamefont {{Jacquet, Maxime J and Weinfurtner, Silke and Koenig,
  Friedrich}}(2020)}]{jacquet20}%
  \BibitemOpen
  \bibfield  {author} {\bibinfo {author} {\bibnamefont {{Jacquet, Maxime J and
  Weinfurtner, Silke and Koenig, Friedrich}}},\ }\href
  {https://doi.org/10.1098/rsta.2019.0239} {\bibinfo {title} {The next
  generation of analogue gravity experiments}} (\bibinfo {year}
  {2020})\BibitemShut {NoStop}%
\bibitem [{\citenamefont {{Philbin, Thomas G and Kuklewicz, Chris and
  Robertson, Scott and Hill, Stephen and K{\"o}nig, Friedrich and Leonhardt,
  Ulf}}(2008)}]{philbin08}%
  \BibitemOpen
  \bibfield  {author} {\bibinfo {author} {\bibnamefont {{Philbin, Thomas G and
  Kuklewicz, Chris and Robertson, Scott and Hill, Stephen and K{\"o}nig,
  Friedrich and Leonhardt, Ulf}}},\ }\bibfield  {title} {\bibinfo {title}
  {Fiber-optical analog of the event horizon},\ }\href
  {https://doi.org/10.1126/science.1153625} {\bibfield  {journal} {\bibinfo
  {journal} {Science}\ }\textbf {\bibinfo {volume} {319}},\ \bibinfo {pages}
  {1367} (\bibinfo {year} {2008})}\BibitemShut {NoStop}%
\bibitem [{\citenamefont {{Weinfurtner, Silke and Tedford, Edmund W and
  Penrice, Matthew CJ and Unruh, William G and Lawrence, Gregory
  A}}(2011)}]{weinfurtner2011}%
  \BibitemOpen
  \bibfield  {author} {\bibinfo {author} {\bibnamefont {{Weinfurtner, Silke and
  Tedford, Edmund W and Penrice, Matthew CJ and Unruh, William G and Lawrence,
  Gregory A}}},\ }\bibfield  {title} {\bibinfo {title} {{Measurement of
  stimulated Hawking emission in an analogue system}},\ }\href
  {https://doi.org/10.1103/PhysRevLett.106.021302} {\bibfield  {journal}
  {\bibinfo  {journal} {Physical Review Letters}\ }\textbf {\bibinfo {volume}
  {106}},\ \bibinfo {pages} {021302} (\bibinfo {year} {2011})}\BibitemShut
  {NoStop}%
\bibitem [{\citenamefont {Euv{\'e}}\ \emph {et~al.}(2016)\citenamefont
  {Euv{\'e}}, \citenamefont {Michel}, \citenamefont {Parentani}, \citenamefont
  {Philbin},\ and\ \citenamefont {Rousseaux}}]{euve2016}%
  \BibitemOpen
  \bibfield  {author} {\bibinfo {author} {\bibfnamefont {L.-P.}\ \bibnamefont
  {Euv{\'e}}}, \bibinfo {author} {\bibfnamefont {F.}~\bibnamefont {Michel}},
  \bibinfo {author} {\bibfnamefont {R.}~\bibnamefont {Parentani}}, \bibinfo
  {author} {\bibfnamefont {T.~G.}\ \bibnamefont {Philbin}},\ and\ \bibinfo
  {author} {\bibfnamefont {G.}~\bibnamefont {Rousseaux}},\ }\bibfield  {title}
  {\bibinfo {title} {{Observation of noise correlated by the Hawking effect in
  a water tank}},\ }\href {https://doi.org/10.1103/PhysRevLett.117.121301}
  {\bibfield  {journal} {\bibinfo  {journal} {Physical Review Letters}\
  }\textbf {\bibinfo {volume} {117}},\ \bibinfo {pages} {121301} (\bibinfo
  {year} {2016})}\BibitemShut {NoStop}%
\bibitem [{\citenamefont {{Steinhauer, Jeff}}(2016)}]{steinhauer2016}%
  \BibitemOpen
  \bibfield  {author} {\bibinfo {author} {\bibnamefont {{Steinhauer, Jeff}}},\
  }\bibfield  {title} {\bibinfo {title} {{Observation of quantum Hawking
  radiation and its entanglement in an analogue black hole}},\ }\href
  {https://doi.org/10.1038/nphys3863} {\bibfield  {journal} {\bibinfo
  {journal} {Nature Physics}\ }\textbf {\bibinfo {volume} {12}},\ \bibinfo
  {pages} {959} (\bibinfo {year} {2016})}\BibitemShut {NoStop}%
\bibitem [{\citenamefont {{De Nova, Juan Ramon Munoz and Golubkov, Katrine and
  Kolobov, Victor I and Steinhauer, Jeff}}(2019)}]{de19BEC}%
  \BibitemOpen
  \bibfield  {author} {\bibinfo {author} {\bibnamefont {{De Nova, Juan Ramon
  Munoz and Golubkov, Katrine and Kolobov, Victor I and Steinhauer, Jeff}}},\
  }\bibfield  {title} {\bibinfo {title} {{Observation of thermal Hawking
  radiation and its temperature in an analogue black hole}},\ }\href
  {https://doi.org/10.1038/s41586-019-1241-0} {\bibfield  {journal} {\bibinfo
  {journal} {Nature}\ }\textbf {\bibinfo {volume} {569}},\ \bibinfo {pages}
  {688} (\bibinfo {year} {2019})}\BibitemShut {NoStop}%
\bibitem [{\citenamefont {{Drori, Jonathan and Rosenberg, Yuval and Bermudez,
  David and Silberberg, Yaron and Leonhardt, Ulf}}(2019)}]{drori19}%
  \BibitemOpen
  \bibfield  {author} {\bibinfo {author} {\bibnamefont {{Drori, Jonathan and
  Rosenberg, Yuval and Bermudez, David and Silberberg, Yaron and Leonhardt,
  Ulf}}},\ }\bibfield  {title} {\bibinfo {title} {{Observation of stimulated
  Hawking radiation in an optical analogue}},\ }\href
  {https://doi.org/10.1103/PhysRevLett.122.010404} {\bibfield  {journal}
  {\bibinfo  {journal} {Physical Review Letters}\ }\textbf {\bibinfo {volume}
  {122}},\ \bibinfo {pages} {010404} (\bibinfo {year} {2019})}\BibitemShut
  {NoStop}%
\bibitem [{\citenamefont {{Kolobov, Victor I and Golubkov, Katrine and de Nova,
  Juan Ram{\'o}n Mu{\~n}oz and Steinhauer, Jeff}}(2021)}]{kolobov2021BEC}%
  \BibitemOpen
  \bibfield  {author} {\bibinfo {author} {\bibnamefont {{Kolobov, Victor I and
  Golubkov, Katrine and de Nova, Juan Ram{\'o}n Mu{\~n}oz and Steinhauer,
  Jeff}}},\ }\bibfield  {title} {\bibinfo {title} {{Observation of stationary
  spontaneous Hawking radiation and the time evolution of an analogue black
  hole}},\ }\href {https://doi.org/10.1038/s41567-020-01076-0} {\bibfield
  {journal} {\bibinfo  {journal} {Nature Physics}\ ,\ \bibinfo {pages} {1}}
  (\bibinfo {year} {2021})}\BibitemShut {NoStop}%
\bibitem [{\citenamefont {Peres}(1996)}]{peres96}%
  \BibitemOpen
  \bibfield  {author} {\bibinfo {author} {\bibfnamefont {A.}~\bibnamefont
  {Peres}},\ }\bibfield  {title} {\bibinfo {title} {Separability criterion for
  density matrices},\ }\href {https://doi.org/10.1103/PhysRevLett.77.1413}
  {\bibfield  {journal} {\bibinfo  {journal} {Physical Review Letters}\
  }\textbf {\bibinfo {volume} {77}},\ \bibinfo {pages} {1413} (\bibinfo {year}
  {1996})}\BibitemShut {NoStop}%
\bibitem [{\citenamefont {Plenio}(2005)}]{plenio05}%
  \BibitemOpen
  \bibfield  {author} {\bibinfo {author} {\bibfnamefont {M.~B.}\ \bibnamefont
  {Plenio}},\ }\bibfield  {title} {\bibinfo {title} {Logarithmic negativity: A
  full entanglement monotone that is not convex},\ }\href
  {https://doi.org/10.1103/PhysRevLett.95.090503} {\bibfield  {journal}
  {\bibinfo  {journal} {Physical Review Letters}\ }\textbf {\bibinfo {volume}
  {95}},\ \bibinfo {pages} {090503} (\bibinfo {year} {2005})}\BibitemShut
  {NoStop}%
\bibitem [{\citenamefont {Agullo}\ \emph {et~al.}(2022)\citenamefont {Agullo},
  \citenamefont {Brady},\ and\ \citenamefont {Kranas}}]{agullo2022prl}%
  \BibitemOpen
  \bibfield  {author} {\bibinfo {author} {\bibfnamefont {I.}~\bibnamefont
  {Agullo}}, \bibinfo {author} {\bibfnamefont {A.~J.}\ \bibnamefont {Brady}},\
  and\ \bibinfo {author} {\bibfnamefont {D.}~\bibnamefont {Kranas}},\
  }\bibfield  {title} {\bibinfo {title} {{Quantum Aspects of Stimulated Hawking
  Radiation in an Optical Analog White-Black Hole Pair}},\ }\href
  {https://doi.org/10.1103/PhysRevLett.128.091301} {\bibfield  {journal}
  {\bibinfo  {journal} {Physical Review Letters}\ }\textbf {\bibinfo {volume}
  {128}},\ \bibinfo {pages} {091301} (\bibinfo {year} {2022})}\BibitemShut
  {NoStop}%
\bibitem [{\citenamefont {Weedbrook}\ \emph {et~al.}(2012)\citenamefont
  {Weedbrook}, \citenamefont {Pirandola}, \citenamefont
  {Garc{\'\i}a-Patr{\'o}n}, \citenamefont {Cerf}, \citenamefont {Ralph},
  \citenamefont {Shapiro},\ and\ \citenamefont {Lloyd}}]{weedbrook2012}%
  \BibitemOpen
  \bibfield  {author} {\bibinfo {author} {\bibfnamefont {C.}~\bibnamefont
  {Weedbrook}}, \bibinfo {author} {\bibfnamefont {S.}~\bibnamefont
  {Pirandola}}, \bibinfo {author} {\bibfnamefont {R.}~\bibnamefont
  {Garc{\'\i}a-Patr{\'o}n}}, \bibinfo {author} {\bibfnamefont {N.~J.}\
  \bibnamefont {Cerf}}, \bibinfo {author} {\bibfnamefont {T.~C.}\ \bibnamefont
  {Ralph}}, \bibinfo {author} {\bibfnamefont {J.~H.}\ \bibnamefont {Shapiro}},\
  and\ \bibinfo {author} {\bibfnamefont {S.}~\bibnamefont {Lloyd}},\ }\bibfield
   {title} {\bibinfo {title} {Gaussian quantum information},\ }\href
  {https://doi.org/10.1103/RevModPhys.84.621} {\bibfield  {journal} {\bibinfo
  {journal} {Reviews of Modern Physics}\ }\textbf {\bibinfo {volume} {84}},\
  \bibinfo {pages} {621} (\bibinfo {year} {2012})}\BibitemShut {NoStop}%
\bibitem [{\citenamefont {Serafini}(2017)}]{serafini17QCV}%
  \BibitemOpen
  \bibfield  {author} {\bibinfo {author} {\bibfnamefont {A.}~\bibnamefont
  {Serafini}},\ }\href@noop {} {\emph {\bibinfo {title} {Quantum continuous
  variables: a primer of theoretical methods}}}\ (\bibinfo  {publisher} {CRC
  press},\ \bibinfo {year} {2017})\BibitemShut {NoStop}%
\bibitem [{\citenamefont {Wang}\ and\ \citenamefont
  {Wilde}(2020{\natexlab{a}})}]{wilde2020ent_cost}%
  \BibitemOpen
  \bibfield  {author} {\bibinfo {author} {\bibfnamefont {X.}~\bibnamefont
  {Wang}}\ and\ \bibinfo {author} {\bibfnamefont {M.~M.}\ \bibnamefont
  {Wilde}},\ }\bibfield  {title} {\bibinfo {title} {Cost of quantum
  entanglement simplified},\ }\href
  {https://doi.org/10.1103/PhysRevLett.125.040502} {\bibfield  {journal}
  {\bibinfo  {journal} {Phys. Rev. Lett.}\ }\textbf {\bibinfo {volume} {125}},\
  \bibinfo {pages} {040502} (\bibinfo {year} {2020}{\natexlab{a}})}\BibitemShut
  {NoStop}%
\bibitem [{\citenamefont {Wang}\ and\ \citenamefont
  {Wilde}(2020{\natexlab{b}})}]{wilde2020alpha_ln}%
  \BibitemOpen
  \bibfield  {author} {\bibinfo {author} {\bibfnamefont {X.}~\bibnamefont
  {Wang}}\ and\ \bibinfo {author} {\bibfnamefont {M.~M.}\ \bibnamefont
  {Wilde}},\ }\bibfield  {title} {\bibinfo {title}
  {$\ensuremath{\alpha}$-logarithmic negativity},\ }\href
  {https://doi.org/10.1103/PhysRevA.102.032416} {\bibfield  {journal} {\bibinfo
   {journal} {Phys. Rev. A}\ }\textbf {\bibinfo {volume} {102}},\ \bibinfo
  {pages} {032416} (\bibinfo {year} {2020}{\natexlab{b}})}\BibitemShut
  {NoStop}%
\bibitem [{\citenamefont {Steinhauer}(2015)}]{steinhauer2015PRD}%
  \BibitemOpen
  \bibfield  {author} {\bibinfo {author} {\bibfnamefont {J.}~\bibnamefont
  {Steinhauer}},\ }\bibfield  {title} {\bibinfo {title} {Measuring the
  entanglement of analogue {H}awking radiation by the density-density
  correlation function},\ }\href {https://doi.org/10.1103/PhysRevD.92.024043}
  {\bibfield  {journal} {\bibinfo  {journal} {Phys. Rev. D}\ }\textbf {\bibinfo
  {volume} {92}},\ \bibinfo {pages} {024043} (\bibinfo {year}
  {2015})}\BibitemShut {NoStop}%
\bibitem [{\citenamefont {de~Nova}\ \emph {et~al.}(2015)\citenamefont
  {de~Nova}, \citenamefont {Sols},\ and\ \citenamefont {Zapata}}]{de2015iop}%
  \BibitemOpen
  \bibfield  {author} {\bibinfo {author} {\bibfnamefont {J.~M.}\ \bibnamefont
  {de~Nova}}, \bibinfo {author} {\bibfnamefont {F.}~\bibnamefont {Sols}},\ and\
  \bibinfo {author} {\bibfnamefont {I.}~\bibnamefont {Zapata}},\ }\bibfield
  {title} {\bibinfo {title} {Entanglement and violation of classical
  inequalities in the {H}awking radiation of flowing atom condensates},\ }\href
  {https://doi.org/10.1088/1367-2630/17/10/105003} {\bibfield  {journal}
  {\bibinfo  {journal} {New journal of physics}\ }\textbf {\bibinfo {volume}
  {17}},\ \bibinfo {pages} {105003} (\bibinfo {year} {2015})}\BibitemShut
  {NoStop}%
\bibitem [{\citenamefont {Simon}(2000)}]{simon2000criterion}%
  \BibitemOpen
  \bibfield  {author} {\bibinfo {author} {\bibfnamefont {R.}~\bibnamefont
  {Simon}},\ }\bibfield  {title} {\bibinfo {title} {Peres-{H}orodecki
  separability criterion for continuous variable systems},\ }\href
  {https://doi.org/10.1103/PhysRevLett.84.2726} {\bibfield  {journal} {\bibinfo
   {journal} {Phys. Rev. Lett.}\ }\textbf {\bibinfo {volume} {84}},\ \bibinfo
  {pages} {2726} (\bibinfo {year} {2000})}\BibitemShut {NoStop}%
\bibitem [{\citenamefont {Lvovsky}\ and\ \citenamefont
  {Raymer}(2009)}]{raymer2009}%
  \BibitemOpen
  \bibfield  {author} {\bibinfo {author} {\bibfnamefont {A.~I.}\ \bibnamefont
  {Lvovsky}}\ and\ \bibinfo {author} {\bibfnamefont {M.~G.}\ \bibnamefont
  {Raymer}},\ }\bibfield  {title} {\bibinfo {title} {Continuous-variable
  optical quantum-state tomography},\ }\href
  {https://doi.org/10.1103/RevModPhys.81.299} {\bibfield  {journal} {\bibinfo
  {journal} {Reviews of Modern Physics}\ }\textbf {\bibinfo {volume} {81}},\
  \bibinfo {pages} {299} (\bibinfo {year} {2009})}\BibitemShut {NoStop}%
\bibitem [{\citenamefont {{Bruschi, David Edward and Friis, Nicolai and
  Fuentes, Ivette and Weinfurtner, Silke}}(2013)}]{bruschi2013}%
  \BibitemOpen
  \bibfield  {author} {\bibinfo {author} {\bibnamefont {{Bruschi, David Edward
  and Friis, Nicolai and Fuentes, Ivette and Weinfurtner, Silke}}},\ }\bibfield
   {title} {\bibinfo {title} {On the robustness of entanglement in analogue
  gravity systems},\ }\href {https://doi.org/10.1088/1367-2630/15/11/113016}
  {\bibfield  {journal} {\bibinfo  {journal} {New Journal of Physics}\ }\textbf
  {\bibinfo {volume} {15}},\ \bibinfo {pages} {113016} (\bibinfo {year}
  {2013})}\BibitemShut {NoStop}%
\bibitem [{\citenamefont {Page}(1976)}]{Page:1976df}%
  \BibitemOpen
  \bibfield  {author} {\bibinfo {author} {\bibfnamefont {D.~N.}\ \bibnamefont
  {Page}},\ }\bibfield  {title} {\bibinfo {title} {{Particle Emission Rates
  from a Black Hole: Massless Particles from an Uncharged, Nonrotating Hole}},\
  }\href {https://doi.org/10.1103/PhysRevD.13.198} {\bibfield  {journal}
  {\bibinfo  {journal} {Phys. Rev. D}\ }\textbf {\bibinfo {volume} {13}},\
  \bibinfo {pages} {198} (\bibinfo {year} {1976})}\BibitemShut {NoStop}%
\bibitem [{\citenamefont {Wald}(1975)}]{Wald:1975kc}%
  \BibitemOpen
  \bibfield  {author} {\bibinfo {author} {\bibfnamefont {R.~M.}\ \bibnamefont
  {Wald}},\ }\bibfield  {title} {\bibinfo {title} {{On Particle Creation by
  Black Holes}},\ }\href {https://doi.org/10.1007/BF01609863} {\bibfield
  {journal} {\bibinfo  {journal} {Commun. Math. Phys.}\ }\textbf {\bibinfo
  {volume} {45}},\ \bibinfo {pages} {9} (\bibinfo {year} {1975})}\BibitemShut
  {NoStop}%
\bibitem [{\citenamefont {Frolov}\ and\ \citenamefont
  {Novikov}(2012)}]{Frolov:1998wf}%
  \BibitemOpen
  \bibfield  {author} {\bibinfo {author} {\bibfnamefont {V.}~\bibnamefont
  {Frolov}}\ and\ \bibinfo {author} {\bibfnamefont {I.}~\bibnamefont
  {Novikov}},\ }\href@noop {} {\emph {\bibinfo {title} {Black hole physics:
  Basic concepts and new developments}}},\ Vol.~\bibinfo {volume} {96}\
  (\bibinfo  {publisher} {Springer Science \& Business Media},\ \bibinfo {year}
  {2012})\BibitemShut {NoStop}%
\bibitem [{\citenamefont {Wald}(1995)}]{Wald:1995yp}%
  \BibitemOpen
  \bibfield  {author} {\bibinfo {author} {\bibfnamefont {R.~M.}\ \bibnamefont
  {Wald}},\ }\href@noop {} {\emph {\bibinfo {title} {{Quantum Field Theory in
  Curved Space-Time and Black Hole Thermodynamics}}}},\ Chicago Lectures in
  Physics\ (\bibinfo  {publisher} {University of Chicago Press},\ \bibinfo
  {address} {Chicago, IL},\ \bibinfo {year} {1995})\BibitemShut {NoStop}%
\bibitem [{\citenamefont {{Fabbri, Alessandro and Navarro-Salas,
  Jos\'e}}(2005)}]{fabbri05}%
  \BibitemOpen
  \bibfield  {author} {\bibinfo {author} {\bibnamefont {{Fabbri, Alessandro and
  Navarro-Salas, Jos\'e}}},\ }\href@noop {} {\emph {\bibinfo {title} {Modeling
  black hole evaporation}}}\ (\bibinfo  {publisher} {World Scientific},\
  \bibinfo {year} {2005})\BibitemShut {NoStop}%
\bibitem [{\citenamefont {Corley}\ and\ \citenamefont
  {Jacobson}(1999)}]{corley1999lasers}%
  \BibitemOpen
  \bibfield  {author} {\bibinfo {author} {\bibfnamefont {S.}~\bibnamefont
  {Corley}}\ and\ \bibinfo {author} {\bibfnamefont {T.}~\bibnamefont
  {Jacobson}},\ }\bibfield  {title} {\bibinfo {title} {Black hole lasers},\
  }\href {https://doi.org/0.1103/PhysRevD.59.124011} {\bibfield  {journal}
  {\bibinfo  {journal} {Physical Review D}\ }\textbf {\bibinfo {volume} {59}},\
  \bibinfo {pages} {124011} (\bibinfo {year} {1999})}\BibitemShut {NoStop}%
\bibitem [{\citenamefont {Katayama}(2021)}]{katayama2021circuit}%
  \BibitemOpen
  \bibfield  {author} {\bibinfo {author} {\bibfnamefont {H.}~\bibnamefont
  {Katayama}},\ }\bibfield  {title} {\bibinfo {title} {Quantum-circuit black
  hole lasers},\ }\href {https://doi.org/10.1038/s41598-021-98456-0} {\bibfield
   {journal} {\bibinfo  {journal} {Scientific Reports}\ }\textbf {\bibinfo
  {volume} {11}},\ \bibinfo {pages} {1} (\bibinfo {year} {2021})}\BibitemShut
  {NoStop}%
\bibitem [{\citenamefont {Corley}\ and\ \citenamefont
  {Jacobson}(1996)}]{Corley:1996ar}%
  \BibitemOpen
  \bibfield  {author} {\bibinfo {author} {\bibfnamefont {S.}~\bibnamefont
  {Corley}}\ and\ \bibinfo {author} {\bibfnamefont {T.}~\bibnamefont
  {Jacobson}},\ }\bibfield  {title} {\bibinfo {title} {{Hawking spectrum and
  high frequency dispersion}},\ }\href
  {https://doi.org/10.1103/PhysRevD.54.1568} {\bibfield  {journal} {\bibinfo
  {journal} {Phys. Rev. D}\ }\textbf {\bibinfo {volume} {54}},\ \bibinfo
  {pages} {1568} (\bibinfo {year} {1996})}\BibitemShut {NoStop}%
\bibitem [{\citenamefont {Gaona-Reyes}\ and\ \citenamefont
  {Bermudez}(2017)}]{Gaona-Reyes:2017mks}%
  \BibitemOpen
  \bibfield  {author} {\bibinfo {author} {\bibfnamefont {J.~L.}\ \bibnamefont
  {Gaona-Reyes}}\ and\ \bibinfo {author} {\bibfnamefont {D.}~\bibnamefont
  {Bermudez}},\ }\bibfield  {title} {\bibinfo {title} {{The theory of optical
  black hole lasers}},\ }\href {https://doi.org/10.1016/j.aop.2017.03.005}
  {\bibfield  {journal} {\bibinfo  {journal} {Annals Phys.}\ }\textbf {\bibinfo
  {volume} {380}},\ \bibinfo {pages} {41} (\bibinfo {year} {2017})},\ \Eprint
  {https://arxiv.org/abs/1701.05655} {arXiv:1701.05655 [gr-qc]} \BibitemShut
  {NoStop}%
\bibitem [{\citenamefont {Moreno-Ruiz}\ and\ \citenamefont
  {Bermudez}(2020)}]{Moreno-Ruiz:2019lgn}%
  \BibitemOpen
  \bibfield  {author} {\bibinfo {author} {\bibfnamefont {A.}~\bibnamefont
  {Moreno-Ruiz}}\ and\ \bibinfo {author} {\bibfnamefont {D.}~\bibnamefont
  {Bermudez}},\ }\bibfield  {title} {\bibinfo {title} {{Hawking temperature in
  dispersive media: Analytics and numerics}},\ }\href
  {https://doi.org/10.1016/j.aop.2020.168268} {\bibfield  {journal} {\bibinfo
  {journal} {Annals Phys.}\ }\textbf {\bibinfo {volume} {420}},\ \bibinfo
  {pages} {168268} (\bibinfo {year} {2020})},\ \Eprint
  {https://arxiv.org/abs/1908.02368} {arXiv:1908.02368 [gr-qc]} \BibitemShut
  {NoStop}%
\bibitem [{\citenamefont {{Belgiorno, F and Cacciatori, SL and Dalla Piazza,
  F}}(2015)}]{belgiorno15}%
  \BibitemOpen
  \bibfield  {author} {\bibinfo {author} {\bibnamefont {{Belgiorno, F and
  Cacciatori, SL and Dalla Piazza, F}}},\ }\bibfield  {title} {\bibinfo {title}
  {{Hawking effect in dielectric media and the Hopfield model}},\ }\href
  {https://doi.org/10.1103/PhysRevD.91.124063} {\bibfield  {journal} {\bibinfo
  {journal} {Physical Review D}\ }\textbf {\bibinfo {volume} {91}},\ \bibinfo
  {pages} {124063} (\bibinfo {year} {2015})}\BibitemShut {NoStop}%
\bibitem [{\citenamefont {Hopfield}(1958)}]{hopfield1958}%
  \BibitemOpen
  \bibfield  {author} {\bibinfo {author} {\bibfnamefont {J.}~\bibnamefont
  {Hopfield}},\ }\bibfield  {title} {\bibinfo {title} {Theory of the
  contribution of excitons to the complex dielectric constant of crystals},\
  }\href {https://doi.org/10.1103/PhysRev.112.1555} {\bibfield  {journal}
  {\bibinfo  {journal} {Physical Review}\ }\textbf {\bibinfo {volume} {112}},\
  \bibinfo {pages} {1555} (\bibinfo {year} {1958})}\BibitemShut {NoStop}%
\bibitem [{\citenamefont {Kranas}\ \emph {et~al.}()\citenamefont {Kranas},
  \citenamefont {Agullo},\ and\ \citenamefont {Brady}}]{paper3}%
  \BibitemOpen
  \bibfield  {author} {\bibinfo {author} {\bibfnamefont {D.}~\bibnamefont
  {Kranas}}, \bibinfo {author} {\bibfnamefont {I.}~\bibnamefont {Agullo}},\
  and\ \bibinfo {author} {\bibfnamefont {A.~J.}\ \bibnamefont {Brady}},\
  }\href@noop {} {\bibinfo {title} {In preparation}}\BibitemShut {NoStop}%
\bibitem [{\citenamefont {Kim}\ \emph {et~al.}(2002)\citenamefont {Kim},
  \citenamefont {Son}, \citenamefont {Bu{\v{z}}ek},\ and\ \citenamefont
  {Knight}}]{kim02}%
  \BibitemOpen
  \bibfield  {author} {\bibinfo {author} {\bibfnamefont {M.}~\bibnamefont
  {Kim}}, \bibinfo {author} {\bibfnamefont {W.}~\bibnamefont {Son}}, \bibinfo
  {author} {\bibfnamefont {V.}~\bibnamefont {Bu{\v{z}}ek}},\ and\ \bibinfo
  {author} {\bibfnamefont {P.}~\bibnamefont {Knight}},\ }\bibfield  {title}
  {\bibinfo {title} {Entanglement by a beam splitter: Nonclassicality as a
  prerequisite for entanglement},\ }\href
  {https://doi.org/10.1103/PhysRevA.65.032323} {\bibfield  {journal} {\bibinfo
  {journal} {Physical Review A}\ }\textbf {\bibinfo {volume} {65}},\ \bibinfo
  {pages} {032323} (\bibinfo {year} {2002})}\BibitemShut {NoStop}%
\bibitem [{\citenamefont {Jiang}\ \emph {et~al.}(2013)\citenamefont {Jiang},
  \citenamefont {Lang},\ and\ \citenamefont {Caves}}]{jiang13}%
  \BibitemOpen
  \bibfield  {author} {\bibinfo {author} {\bibfnamefont {Z.}~\bibnamefont
  {Jiang}}, \bibinfo {author} {\bibfnamefont {M.~D.}\ \bibnamefont {Lang}},\
  and\ \bibinfo {author} {\bibfnamefont {C.~M.}\ \bibnamefont {Caves}},\
  }\bibfield  {title} {\bibinfo {title} {{Mixing nonclassical pure states in a
  linear-optical network almost always generates modal entanglement}},\ }\href
  {https://doi.org/10.1103/PhysRevA.88.044301} {\bibfield  {journal} {\bibinfo
  {journal} {Physical Review A}\ }\textbf {\bibinfo {volume} {88}},\ \bibinfo
  {pages} {044301} (\bibinfo {year} {2013})}\BibitemShut {NoStop}%
\bibitem [{\citenamefont {Isoard}\ \emph {et~al.}(2021)\citenamefont {Isoard},
  \citenamefont {Milazzo}, \citenamefont {Pavloff},\ and\ \citenamefont
  {Giraud}}]{Isoard:2021peb}%
  \BibitemOpen
  \bibfield  {author} {\bibinfo {author} {\bibfnamefont {M.}~\bibnamefont
  {Isoard}}, \bibinfo {author} {\bibfnamefont {N.}~\bibnamefont {Milazzo}},
  \bibinfo {author} {\bibfnamefont {N.}~\bibnamefont {Pavloff}},\ and\ \bibinfo
  {author} {\bibfnamefont {O.}~\bibnamefont {Giraud}},\ }\bibfield  {title}
  {\bibinfo {title} {Bipartite and tripartite entanglement in a bose-einstein
  acoustic black hole},\ }\href {https://doi.org/10.1103/PhysRevA.104.063302}
  {\bibfield  {journal} {\bibinfo  {journal} {Phys. Rev. A}\ }\textbf {\bibinfo
  {volume} {104}},\ \bibinfo {pages} {063302} (\bibinfo {year}
  {2021})}\BibitemShut {NoStop}%
\bibitem [{\citenamefont {Coutant}\ and\ \citenamefont
  {Parentani}(2010)}]{Coutant:2009cu}%
  \BibitemOpen
  \bibfield  {author} {\bibinfo {author} {\bibfnamefont {A.}~\bibnamefont
  {Coutant}}\ and\ \bibinfo {author} {\bibfnamefont {R.}~\bibnamefont
  {Parentani}},\ }\bibfield  {title} {\bibinfo {title} {Black hole lasers, a
  mode analysis},\ }\href {https://doi.org/10.1103/PhysRevD.81.084042}
  {\bibfield  {journal} {\bibinfo  {journal} {Phys. Rev. D}\ }\textbf {\bibinfo
  {volume} {81}},\ \bibinfo {pages} {084042} (\bibinfo {year}
  {2010})}\BibitemShut {NoStop}%
\bibitem [{\citenamefont {Finazzi}\ and\ \citenamefont
  {Parentani}(2010)}]{Finazzi:2010nc}%
  \BibitemOpen
  \bibfield  {author} {\bibinfo {author} {\bibfnamefont {S.}~\bibnamefont
  {Finazzi}}\ and\ \bibinfo {author} {\bibfnamefont {R.}~\bibnamefont
  {Parentani}},\ }\bibfield  {title} {\bibinfo {title} {{Black hole lasers in
  Bose--Einstein condensates}},\ }\href
  {https://doi.org/10.1088/1367-2630/12/9/095015} {\bibfield  {journal}
  {\bibinfo  {journal} {New Journal of Physics}\ }\textbf {\bibinfo {volume}
  {12}},\ \bibinfo {pages} {095015} (\bibinfo {year} {2010})}\BibitemShut
  {NoStop}%
\bibitem [{\citenamefont {Jacquet}\ \emph {et~al.}(2022)\citenamefont
  {Jacquet}, \citenamefont {Joly}, \citenamefont {Giacomelli}, \citenamefont
  {Claude}, \citenamefont {Glorieux}, \citenamefont {Bramati}, \citenamefont
  {Carusotto},\ and\ \citenamefont {Giacobino}}]{Jacquet:2022vak}%
  \BibitemOpen
  \bibfield  {author} {\bibinfo {author} {\bibfnamefont {M.~J.}\ \bibnamefont
  {Jacquet}}, \bibinfo {author} {\bibfnamefont {M.}~\bibnamefont {Joly}},
  \bibinfo {author} {\bibfnamefont {L.}~\bibnamefont {Giacomelli}}, \bibinfo
  {author} {\bibfnamefont {F.}~\bibnamefont {Claude}}, \bibinfo {author}
  {\bibfnamefont {Q.}~\bibnamefont {Glorieux}}, \bibinfo {author}
  {\bibfnamefont {A.}~\bibnamefont {Bramati}}, \bibinfo {author} {\bibfnamefont
  {I.}~\bibnamefont {Carusotto}},\ and\ \bibinfo {author} {\bibfnamefont
  {E.}~\bibnamefont {Giacobino}},\ }\href@noop {} {\bibinfo {title} {{Analogue
  quantum simulation of the Hawking effect in a polariton superfluid}}}
  (\bibinfo {year} {2022}),\ \Eprint {https://arxiv.org/abs/2201.02038}
  {arXiv:2201.02038 [quant-ph]} \BibitemShut {NoStop}%
\bibitem [{\citenamefont {Chelpanova}\ \emph {et~al.}(2021)\citenamefont
  {Chelpanova}, \citenamefont {Lerose}, \citenamefont {Zhang}, \citenamefont
  {Carusotto}, \citenamefont {Tserkovnyak},\ and\ \citenamefont
  {Marino}}]{Chelpanova:2021gnm}%
  \BibitemOpen
  \bibfield  {author} {\bibinfo {author} {\bibfnamefont {O.}~\bibnamefont
  {Chelpanova}}, \bibinfo {author} {\bibfnamefont {A.}~\bibnamefont {Lerose}},
  \bibinfo {author} {\bibfnamefont {S.}~\bibnamefont {Zhang}}, \bibinfo
  {author} {\bibfnamefont {I.}~\bibnamefont {Carusotto}}, \bibinfo {author}
  {\bibfnamefont {Y.}~\bibnamefont {Tserkovnyak}},\ and\ \bibinfo {author}
  {\bibfnamefont {J.}~\bibnamefont {Marino}},\ }\href@noop {} {\bibinfo {title}
  {Competition between lasing and superradiance under spintronic pumping}}
  (\bibinfo {year} {2021}),\ \Eprint {https://arxiv.org/abs/2112.04509}
  {arXiv:2112.04509 [cond-mat.stat-mech]} \BibitemShut {NoStop}%
\end{thebibliography}

%

\end{document}